\documentclass[a4paper,fleqn,usenatbib]{mnras}

\usepackage{newtxtext,newtxmath}

\usepackage[T1]{fontenc}
\usepackage{ae,aecompl}

\usepackage{graphicx}	
\usepackage{amsmath}	
\usepackage{amssymb}	

\title[Spectra quality and nebular abundances]{The impact of spectra quality on nebular abundances}

\author[M. Rodr\'iguez]{
M\'onica Rodr\'iguez,\thanks{E-mail: mrodri@inaoep.mx}
\\
Instituto Nacional de Astrof\'isica \'Optica y Electr\'onica, Luis Enrique Erro 1, Tonantzintla 72840, Puebla, Mexico
}

\date{Accepted XXX. Received YYY; in original form ZZZ}

\pubyear{2020}

\begin{document}
\label{firstpage}
\pagerange{\pageref{firstpage}--\pageref{lastpage}}
\maketitle

\begin{abstract}
I explore the effects of observational errors on nebular chemical abundances using a sample of 179 optical spectra of 42 planetary nebulae (PNe) observed by different authors. The spectra are analysed in a homogeneous way to derive physical conditions and ionic and total abundances. The effects of recombination on the [\ion{O}{ii}] and [\ion{N}{ii}] emission lines are estimated by including the effective recombination coefficients in the statistical equilibrium equations that are solved for O$^+$ and N$^+$. The results are shown to be significantly different than those derived using previous approaches. The O$^+$ abundances derived with the blue and red lines of [\ion{O}{ii}] differ by up to a factor of 6, indicating that the relative intensities of lines widely separated in wavelength can be highly uncertain. In fact, the \ion{He}{i} lines in the range 4000--6800~\AA\ imply that most of the spectra are bluer than expected. Scores are assigned to the spectra using different criteria and the spectrum with the highest score for each PN is taken as the reference spectrum. The differences between the abundances derived with the reference spectrum and those derived with the other spectra available for each object are used to estimate the one-sigma observational uncertainties in the final abundances: 0.11~dex for O/H and Ar/H, 0.14~dex for N/H, Ne/H, and Cl/H, and 0.16~dex for S/H.
\end{abstract}




\section{Introduction}

The chemical composition of the ionized gas in planetary nebulae (PNe) provides information on the composition of the interstellar medium when their progenitor stars were formed and on the nucleosynthesis processes that take place in these low and intermediate mass stars. This information can be used to constrain the models of galactic chemical evolution and of stellar evolution. Chemical abundances have been calculated for hundreds of Galactic PNe \citep[see, e.g.,][]{Qui07, Hen10, Sta18, Mol18} and for dozens of extragalactic PNe in nearby galaxies \citep[e.g.,][and references therein]{Kwi14, Mag16}. When the PN abundances are plotted as a function of their galactocentric distances, they show a large scatter, and it is not clear how much of that scatter is real. This might explain why different authors reach different conclusions regarding the slopes of the metallicity gradients and their time variation. A much better understanding of the uncertainties that affect our abundance determinations is clearly needed.

PN chemical abundances are usually derived using optical spectra, which in general cover the wavelength range 3700--6800~\AA. The relative intensities of the different emission lines that appear in this range can be used to determine the physical conditions (electron density and electron temperature) and the ionic abundances of several elements, like He, O, N, S, Ne, Ar, and Cl \citep{OF06}. Total abundances are calculated by adding the abundances of all relevant ions or using ionization correction factors to estimate the contribution of unobserved ions. Hence, three sources of uncertainty in the final results can be identified: observational uncertainties, uncertainties in the atomic data used in the calculations, and uncertainties resulting from departures from our assumptions about the structure of the objects (in density, temperature, metallicity, or ionization structure).

I will focus here on the effects of observational uncertainties in the line intensity ratios used in our abundance determinations. Observational uncertainties can arise from a variety of sources: the effects of atmospheric differential refraction \citep{Fil82}, contamination by telluric absorption or emission \citep[e.g.,][]{Sme15}, blends with weaker nebular lines \citep[e.g.,][]{Rod11}, a poor signal to noise ratio, problems with the flux calibration, the extinction correction or any other part of the reduction process, and even typos at the time of reporting the measurements. Some authors report estimates of the uncertainties affecting their measurements but, these being just estimates, they are likely to be somewhat subjective, and it is unrealistic to expect that all possible sources of uncertainty are included. This can be illustrated with the results of \citet{Per04}, who compared the abundances obtained in a homogeneous way with spectra of the same PNe observed by different authors. \citeauthor{Per04} propagated the reported observational errors to the final abundances, and found that in 40 per cent of the objects the differences in the oxygen abundances implied by the different spectra exceed the estimated errors. \citeauthor{Per04} conclude that the reported errors are probably underestimated.

Since many of the brightest Galactic PNe have been observed more than once, one can use the spectra of the same object obtained by different authors to explore and constrain the effect of observational errors. Most of these objects are extended, and the different spectra are likely to cover different regions of the nebula. This means that the observed intensities cannot be directly compared to infer their reliability, but if one makes the assumption that the element abundances do not vary significantly across each nebula (at least in the regions sampled by collisionally-excited line emission), a comparison of the abundances derived from different spectra will provide an idea of the observational uncertainties. This is the approach taken here, which is applied to 179 optical spectra of 42 PNe. Besides, different features of the observed spectra are used to explore their quality and to assign them a score on this account. The data obtained from this analysis and the procedure described below can be used in the future to assign a quality to any observed PN spectrum and to relate this quality to the expected observational uncertainties.

\section{Observational sample}

The sample consists of 179 optical spectra of 42 PNe, each of them with 3--7 spectra. The spectra were compiled from the literature by looking for PNe that have three or more available spectra that contain: at least one working density diagnostic (from among [\ion{S}{ii}]~$\lambda6716/\lambda6731$, [\ion{O}{ii}]~$\lambda3726/\lambda3729$, [\ion{Cl}{iii}]~$\lambda5518/\lambda5538$, and [\ion{Ar}{iv}]~$\lambda4711/\lambda4740$), the two most used temperature diagnostics ([\ion{O}{iii}]~$\lambda4363/(\lambda4959+\lambda5007)$ and [\ion{N}{ii}]~$\lambda5755/(\lambda6548+\lambda6583)$), and the [\ion{O}{ii}]~$\lambda3727$ lines.

The [\ion{O}{ii}] lines are required for the calculation of the total abundances not only of oxygen, but of most of the other elements through the use of ionization correction factors \citep[see, e.g.,][]{DI14}. The O$^+$ abundance can be calculated using either the [\ion{O}{ii}]~$\lambda3727$ lines or the [\ion{O}{ii}]~$\lambda7325$ lines. The red [\ion{O}{ii}] lines are weaker than their blue counterparts and can be severely contaminated by telluric emission lines. The blue [\ion{O}{ii}] lines are not without problems, since they lie in a spectral region where instruments can have a low efficiency and the effects of atmospheric differential refraction might be important \citep{Fil82}, but they seem a safer bet. This is why the presence of the blue [\ion{O}{ii}] lines in the measured spectrum is used as a selection criterion here. However, since 153 of the 179 compiled spectra have measurements of both the red and blue [\ion{O}{ii}] lines, it is possible to compare their performance, as will be shown below.

The sample objects are Galactic PNe, although the halo PN BoBn~1 might be located in the Sagittarius dwarf spheroidal galaxy \citep{Zij06}. These objects and the spectra used for their analysis are listed in Tables~\ref{tab:neTe} to \ref{tab:scores}.

\vskip2truecm

\section{Analysis}

The analysis is based on the extinction-corrected line ratios provided by the different authors. Most of the calculations of physical conditions and ionic abundances use PyNeb, a python-based package for the analysis of emission lines developed by \citet{Lur15}. Electron densities are derived from all available diagnostics ([\ion{S}{ii}]~$\lambda6716/\lambda6731$, [\ion{O}{ii}]~$\lambda3726/\lambda3729$, [\ion{Cl}{iii}]~$\lambda5518/\lambda5538$, and/or [\ion{Ar}{iv}]~$\lambda4711/\lambda4740$), and their geometric mean is used as the electron density, $n_{\rm{e}}$, in all calculations. The geometric mean, i.e., the average of the logarithmic values, is used because it is more robust to the large variations in $n_{\rm{e}}$ introduced by observational or atomic-data uncertainties when the diagnostics are near the limits of their validity.

The temperature values of $T_{\rm{e}}$[\ion{O}{iii}] and $T_{\rm{e}}$[\ion{N}{ii}], obtained from the diagnostics [\ion{O}{iii}]~$\lambda4363/(\lambda4959+\lambda5007)$ and [\ion{N}{ii}]~$\lambda5755/(\lambda6548+\lambda6583)$, are used to characterize a two-zone temperature structure. The abundances derived for He$^+$, He$^{++}$, O$^{++}$, Ne$^{++}$, S$^{++}$, and Ar$^{++}$ are based on the value found for $T_{\rm{e}}$[\ion{O}{iii}]; whereas the abundances of O$^+$, N$^+$, S$^+$, and Cl$^{++}$ are based on $T_{\rm{e}}$[\ion{N}{ii}]. The lines used in the abundance determination, when available, are the following: [\ion{O}{ii}]~$\lambda3726+29$, [\ion{O}{ii}]~$\lambda7319+30$, [\ion{O}{iii}]~$\lambda\lambda4959$, 5007, [\ion{N}{ii}]~$\lambda6548+84$, [\ion{S}{ii}]~$\lambda6716+31$, [\ion{S}{iii}]~$\lambda6312$, [\ion{Cl}{iii}]~$\lambda5518+38$, [\ion{Ar}{iii}]~$\lambda7135$, [\ion{Ar}{iv}]~$\lambda4711+40$, [\ion{Ne}{iii}]~$\lambda\lambda3868$, 3967, and \ion{He}{i}~$\lambda4471$, $\lambda5876$, $\lambda6678$.

The calculations use the atomic data listed in Table~\ref{tab:atdata}. The choice of atomic data is based on different considerations discussed by \citet[][and 2020, in preparation]{JdD17}. The ionization correction factors used to derive total abundances are the ones calculated by \cite{DI14}, except for nitrogen, where N/O=N$^+/$O$^+$ is used \citep[see][]{DI15}. In the case of helium, the final abundance is given as He/H~$\simeq$~He$^+$/H$^+ +$~He$^{++}$/H$^+$. This will be a lower limit to the total abundance whenever He$^0$ has a significant concentration in the volume of ionized gas, but it can also be an upper limit to the total abundance if He$^+$ extends beyond the region where H is ionized.

\begin{table}
	\centering
	\caption{Atomic data used in the calculations.}
	\label{tab:atdata}
	\begin{tabular}{lll}
		\hline
		Ion & Transition probabilities & Collision strengths \\
		\hline
		O$^+$ & \citeauthor{FFT04} & \citet{Kal09} \\
		      & \citeyearpar{FFT04} &  \\
		O$^{++}$ & \citet{SZ00} & \citet{SSB14} \\
		         & \citet{WFD96} &  \\
		N$^+$ & \citeauthor{FFT04} & \citet{T11} \\
		      & \citeyearpar{FFT04} &  \\
		S$^+$ & \citet{PKW09} & \citeauthor{TZ10} \\
		      &               & \citeyearpar{TZ10} \\
		S$^{++}$ & \citet{PKW09} & \citet{GRHK14} \\
		Ne$^{++}$ & \citet{McL11} & \citeauthor{McL11} \\
		      &                  & \citeyearpar{McL11} \\
		Ar$^{++}$ & \citet{M83} & \citet{GMZ95} \\
		          & \citet{KS86} &  \\
		Ar$^{3+}$ & \citet{MZ82} & \citeauthor{RB97} \\
		      &                  & \citeyearpar{RB97} \\
		Cl$^{++}$ & \citet{Fal99} & \citet{GMZ95} \\
		\hline
		Ion & Recombination coefficients & \\
		\hline
		H$^+$ & \citet{SH95} & \\
		He$^+$ & \citet{Por12, Por13} & \\
		He$^{++}$ & \citet{SH95} & \\
		O$^{++}$ & \citet{Sto17} & \\
		N$^+$ & \citet{Kis02} & \\
		\hline
	\end{tabular}
\end{table}

\subsection{Corrections for the effects of recombination}\label{rec_corr}

The calculations performed with the [\ion{O}{ii}] lines have been corrected for the effects of recombination using the recombination coefficients of \citet{Sto17}. This is done by adding to the statistical equilibrium equations for the four levels that lie above the ground level (2s$^2$2p$^3$~$^2$D$_{5/2}^{\rm{o}}$, $^2$D$_{3/2}^{\rm{o}}$, $^2$P$_{3/2}^{\rm{o}}$, and  $^2$P$_{1/2}^{\rm{o}}$) a term that includes the population of each level $i$ due to the recombination of O$^{++}$:
\begin{displaymath}
    n_i\sum_{j\ne{i}}n_{\rm{e}}\, q_{ij} + n_i\sum_{j\prec{i}}A_{ij} = 
\end{displaymath}
\begin{equation}
    \qquad = \sum_{j\ne{i}}n_{\rm{e}}\, n_j\, q_{ji} + \sum_{j\succ{i}} n_j\, A_{ji} + \, n_{\rm{e}}\, n(\mbox{O}^{++})\, \alpha_i(T_{\rm{e}}, n_{\rm{e}}),
	\label{eq:stateq}
\end{equation}
where $n_i$ is the population of level $i$ of O$^+$, $A_{ij}$ and $q_{ij}$ are the transition probability and the collisional excitation or de-excitation rate between levels $i$ and $j$, $n(\mbox{O}^{++})$ is the O$^{++}$ abundance, and $\alpha_i$ is the effective recombination coefficient to level $i$. The recombination coefficients are interpolated linearly from the tables provided by \citet{Sto17} for the values of $n_{\rm{e}}$ and $T_{\rm{e}}$[\ion{O}{iii}] found for each object. I use the case B values of the recombination coefficients since, according to \citet{Sto17}, this case is a good approximation for PNe. Case B considers that the nebula is optically thick to permitted transitions from the O$^+$ ground level.

The four equations resulting from equation~(\ref{eq:stateq}) above for $i=2$--5, with $n(\mbox{O}^{++})=(\mbox{O}^{++}/\mbox{O}^{+})\sum_{i}n_i$, along with the usual fifth equation, $\sum_{i}n_i=1$, can be easily solved to derive the level populations. However, since the recombination term of the equation includes a dependence on the relative abundances of O$^+$ and O$^{++}$, and since the derived O$^+$ abundance changes with the correction for recombination, it is necessary to iterate. Using the condition that the O$^+$ abundance derived from either the blue or red [\ion{O}{ii}] lines should change by less than 0.02~dex in the last iteration, only up to three iterations are needed in all cases but one. The exception is the analysis performed with the spectrum of NGC~7009 obtained by \citet{KH98}, where five iterations are needed. The effects of the correction for recombination in the O$^+$ abundances, in $n_{\rm{e}}$[\ion{O}{ii}], and in the final abundance ratios are discussed below in Section~\ref{rec}.

In the case of [\ion{N}{ii}], the recombination coefficients of \citet{Kis02} for case B can be used to introduce the effects of recombination in the statistical equilibrium equations of levels 4 and 5 of N$^+$. The effects of recombination will be proportional in this case to the N$^{++}$/N$^+$ abundance ratio, which, to a first approximation, can be considered equal to O$^{++}$/O$^+$. Since the corrections for recombination for both [\ion{O}{ii}] and [\ion{N}{ii}] emission depend on O$^{++}$/O$^+$ and $T_{\rm{e}}$[\ion{N}{ii}], in four cases it was necessary to iterate these corrections, and for the spectrum of NGC~7009 obtained by \citet{KH98}, the value of N$^{++}$/N$^+$ had to be reduced to 80 per cent that of O$^{++}$/O$^+$ in order to find a solution. The corrections for recombination lead to lower values of $T_{\rm{e}}$[\ion{N}{ii}], with the differences being higher that 5 percent in 23/179 spectra. The corrections start to be important at O$^{++}$/O$^+>30$, especially for the lower values of $T_{\rm{e}}$[\ion{N}{ii}]. The largest correction, 3000~K or $\sim20$ per cent of the initial value, is found for the spectrum of NGC~7009 of \citet{KH98}, where, after the corrections for recombination have been performed, O$^{++}$/O$^+\simeq210$ and $T_{\rm{e}}\mbox{[\ion{N}{ii}]}=9700$~K.

The effects of the corrections for recombination of [\ion{O}{ii}] and [\ion{N}{ii}] emission go in opposite directions since the corrections decrease the values of $T_{\rm{e}}$[\ion{N}{ii}] and O$^+$/H$^+$, but the lower values of $T_{\rm{e}}$[\ion{N}{ii}] lead to higher values of O$^+$/H$^+$, which in turn imply lower corrections because of the dependence of the corrections on O$^{++}$/O$^+$. The net effect of the corrections is that the values of $T_{\rm{e}}$[\ion{N}{ii}] always decrease, whereas in most cases the O$^+$ abundances derived with the blue [\ion{O}{ii}] lines increase and those derived with the red [\ion{O}{ii}] lines decrease. This implies that if only one of the corrections is performed (either the one for [\ion{O}{ii}] or the one for [\ion{N}{ii}]), the effects of this correction on the derived abundances will be higher than the ones reported here.

The [\ion{O}{iii}]~$\lambda4363$ line can also be affected by recombination, but this correction was not performed, since the formula provided by \citet{Liu00} implies that the change in the intensity of this line is below three per cent in all the spectra analysed here.

The physical conditions and ionic and total abundances derived from each spectra are presented in Tables~\ref{tab:neTe}, \ref{tab:ion}, \ref{tab:He}, and \ref{tab:XH}, whose full versions are available online.

\section{Effects of the corrections for recombination in [O~II] and [N~II] emission}
\label{rec}

\begin{figure*} 
\begin{center}
\includegraphics[width=0.75\textwidth, trim=30 10 30 10, clip=yes]{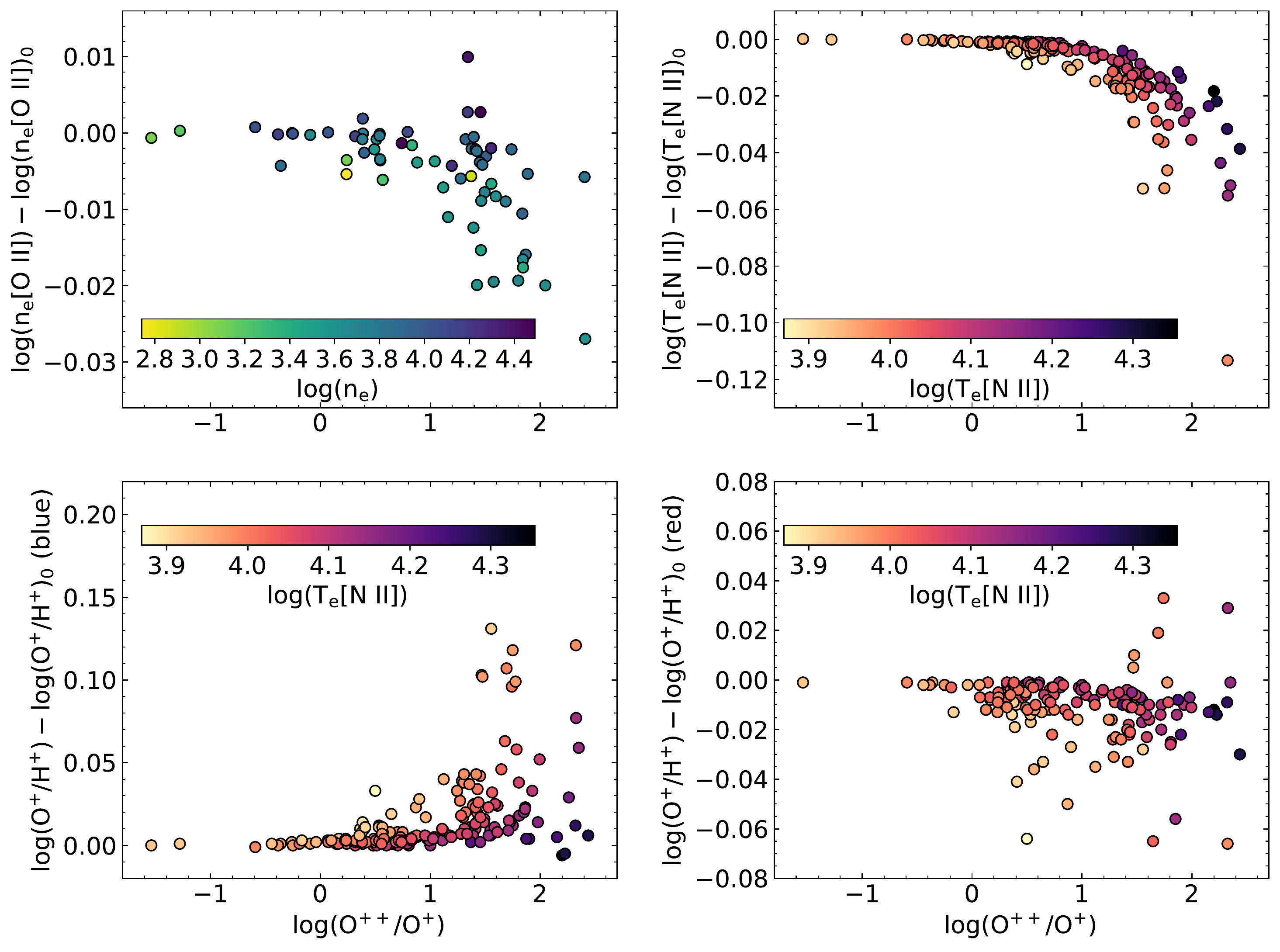}
\caption{Differences introduced by the corrections for recombination to the values of $n_{\rm{e}}$[\ion{O}{ii}] (in units of cm$^{-3}$), $T_{\rm{e}}$[\ion{N}{ii}] (in K), and the O$^+$ abundances derived using the blue and red lines of [\ion{O}{ii}], presented as a function of the degree of ionization, O$^{++}$/O$^+$. The color/grey code shows the dependence of the corrections on either $n_{\rm{e}}$ or $T_{\rm{e}}$[\ion{N}{ii}]. A subindex of 0 is used for the uncorrected values.} 
\label{figrec}
\end{center}
\end{figure*}

Fig.~\ref{figrec} shows the effects of the correction for recombination in the derived values of $n_{\rm{e}}$[\ion{O}{ii}], $T_{\rm{e}}$[\ion{N}{ii}], and in the O$^+$ abundances obtained from the blue and red [\ion{O}{ii}] lines. The differences between the corrected values and the uncorrected ones are shown as a function of the degree of ionization and are colour/grey-coded to illustrate their dependence on $n_{\rm{e}}$ and $T_{\rm{e}}$[\ion{N}{ii}]. The corrections are very small for $n_{\rm{e}}$[\ion{O}{ii}], and smaller that 0.05~dex for $T_{\rm{e}}$[\ion{N}{ii}] and O$^+$/H$^+$ in most cases, especially when O$^{++}$/O$^+<10$. Note that the size of the corrections is larger for lower values of $n_{\rm{e}}$ and $T_{\rm{e}}$[\ion{N}{ii}]. This reflects the larger impact that recombinations have on the level populations when collisions are less important. The corrections are larger for the O$^+$ abundances obtained from the blue [\ion{O}{ii}] lines than for those obtained from the red lines. This is due to the lower values of $T_{\rm{e}}$[\ion{N}{ii}] introduced by the recombination corrections. If the recombination correction of [\ion{N}{ii}] is not performed, the opposite effect is observed: the corrections are larger for the O$^+$ abundances obtained from the red [\ion{O}{ii}] lines and the O$^+$/H$^+$ abundance ratio decreases for both the blue and the red [\ion{O}{ii}] lines.

\begin{figure*} 
\begin{center}
\includegraphics[width=0.9\textwidth, trim=30 10 20 10, clip=yes]{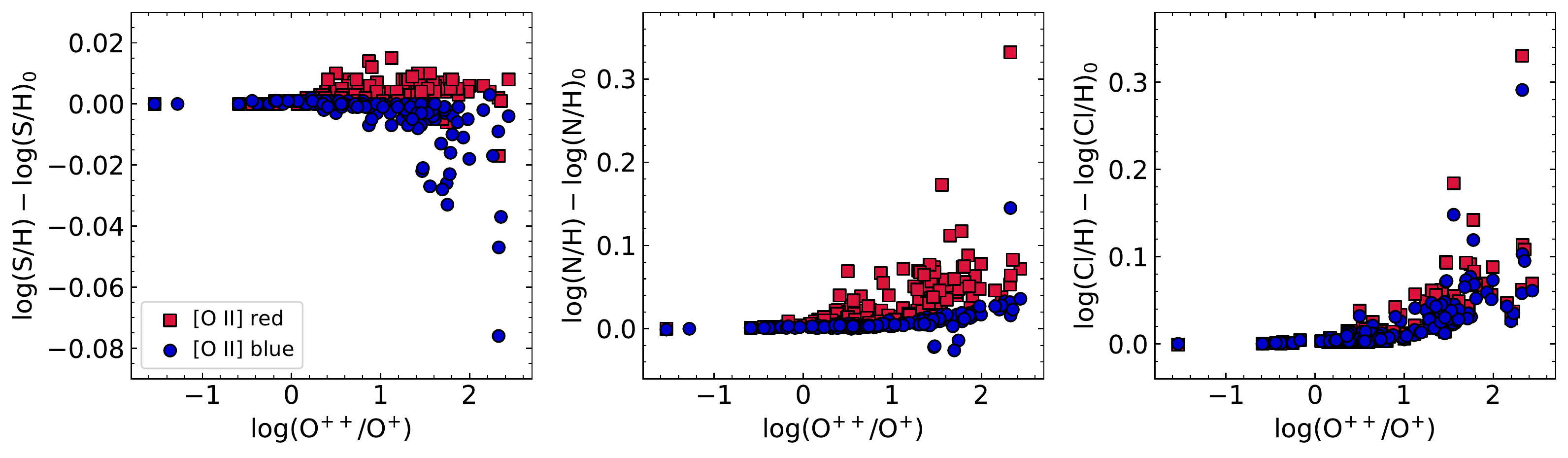}
\caption{Differences introduced by the correction for recombination in the total abundances of sulfur, nitrogen and chlorine, with subindex 0 indicating the results obtained without the correction. The differences are plotted as a function of O$^{++}$/O$^+$. The circles and squares show the results obtained with the blue and red [\ion{O}{ii}] lines, respectively.} 
\label{figrec2}
\end{center}
\end{figure*}

Since the largest changes in O$^+$/H$^+$ occur when O$^{++}/$O$^+>>1$, the corrections for recombination do not introduce any significant difference in O/H. On the other hand, since the ionization correction factors for all elements but O and He depend on O$^{++}$/O$^+$, the recombination corrections introduce changes in the total abundances of N, Ne, S, Cl, and Ar. These changes are below 0.05~dex in most cases, but are larger, up to $\sim0.3$~dex, in the N, S, and Cl abundances derived for some spectra. This is illustrated in Fig.~\ref{figrec2}, where the differences between the total abundances of N, S, and Cl derived with and without the recombinations corrections are plotted as a function of O$^{++}$/O$^+$. The circles and squares show the results obtained with the blue and red [\ion{O}{ii}] lines, respectively. The results obtained for N and Cl are less sensitive to the recombination correction when the blue [\ion{O}{ii}] lines are used to derive the O$^+$ abundances; whereas the opposite is true for the S abundances. This reflects the different dependences on the physical conditions of the emissivities of the lines used in the calculations. Since the effects for the S abundance are smaller than those for N and Cl, and since the recombination corrections introduce some uncertainty, these results indicate that the blue [\ion{O}{ii}] lines will in general lead to better estimates of the total abundances.

\subsection{A comparison between the two approaches used to correct for the effects of recombination}

Previous works have used the formulas provided by \citet{Liu00} to correct for the effects of recombination the intensities of [\ion{O}{ii}]~$\lambda7325$, [\ion{N}{ii}]~$\lambda5755$, and [\ion{O}{iii}]~$\lambda4363$ relative to H$\beta$. These formulas estimate the contribution of recombination to the intensities of these lines as a function of the value of the electron temperature and the abundance of the recombining ion. In order to derive these formulas, \citet{Liu00} calculate how recombinations populate the upper levels of the lines. This approach, which is much simpler than the one used here, will be a good approximation in the limit of low density, when collisions do not change the population balance introduced by recombinations.

I have compared the two approaches, using the same atomic data in both cases, and find that the alternative approach of \citet{Liu00} does not lead to the same results. In order to perform a more meaningful comparison, the calculations that explore this low-density approach were performed using as initial values the temperatures, densities, and ionic abundances derived from the analysis described above in Section~\ref{rec_corr}, which includes the correction for recombination in the equations of statistical equilibrium for O$^+$ and N$^+$. The case B recombination coefficients of \citet{Sto17} and \citet{Kis02} were interpolated from their tables and used to estimate what fraction of the [\ion{O}{ii}] and [\ion{N}{ii}] line intensities arise from direct recombination to their upper levels and from cascades involving electric-quadrupole and magnetic dipole transitions.

\begin{figure*} 
\begin{center}
\includegraphics[width=0.9\textwidth, trim=30 10 20 10, clip=yes]{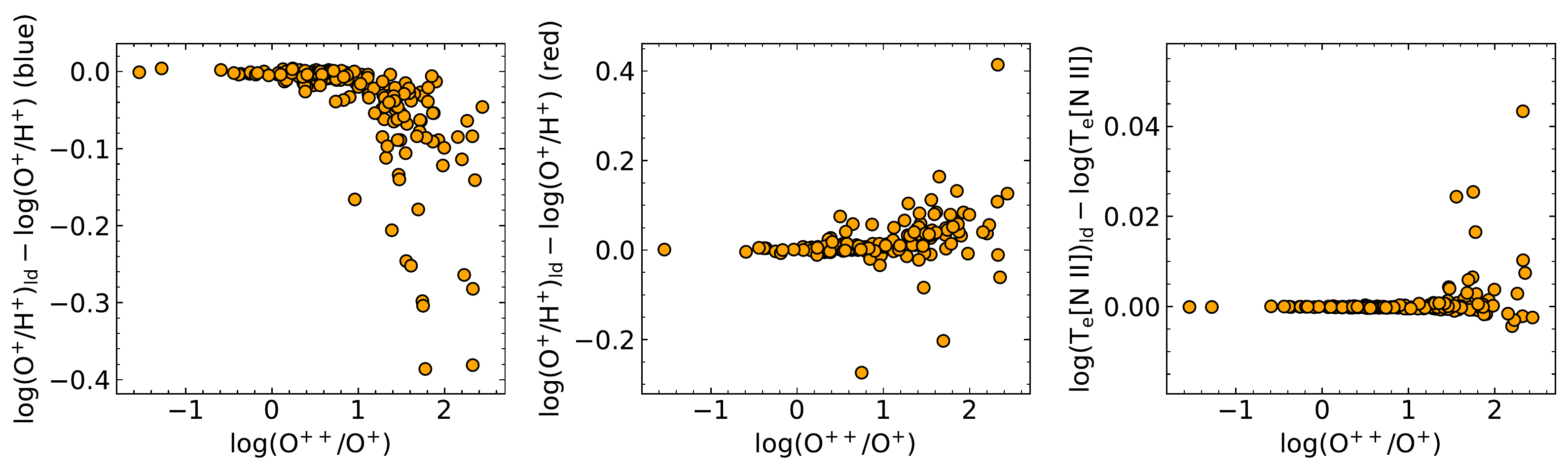}
\caption{Differences introduced by ignoring the effects of collisions in the corrections for recombination of the O$^+$ abundances calculated with the blue (left panel) and red (middle panel) [\ion{O}{ii}] lines and of $T_{\rm{e}}$[\ion{N}{ii}] (right panel, in K). The differences are plotted as a function of O$^{++}$/O$^+$. The subscript `ld' is used for the values calculated with the low-density approximation.} 
\label{figrec1b}
\end{center}
\end{figure*}

Fig.~\ref{figrec1b} compares the values of O$^+$/H$^+$ and $T_{\rm{e}}$[\ion{N}{ii}] estimated with the two approaches. The O$^+$ abundances can be seen to differ by up to $\sim0.4$~dex for both the blue and [\ion{O}{ii}] lines, although the results implied by the blue [\ion{O}{ii}] lines are in general more affected. In fact, the corrections to the O$^+$ abundance derived with the blue lines are mostly negative, instead of positive, when the low-density approximation is used. The results for N$^+$ are slightly better, since the values of $T_{\rm{e}}$[\ion{N}{ii}] estimated with the two approaches differ by 5--10 per cent only in a few cases. These non-negligible changes imply that recombination and collisions cannot be treated separately for these ions.

\section{[O~II]~$\lambda3727$ versus [O~II]~$\lambda7325$}
\label{oii}

\begin{figure} 
\begin{center}
\includegraphics[width=0.5\textwidth, trim=-5 35 -35 -5, clip=yes]{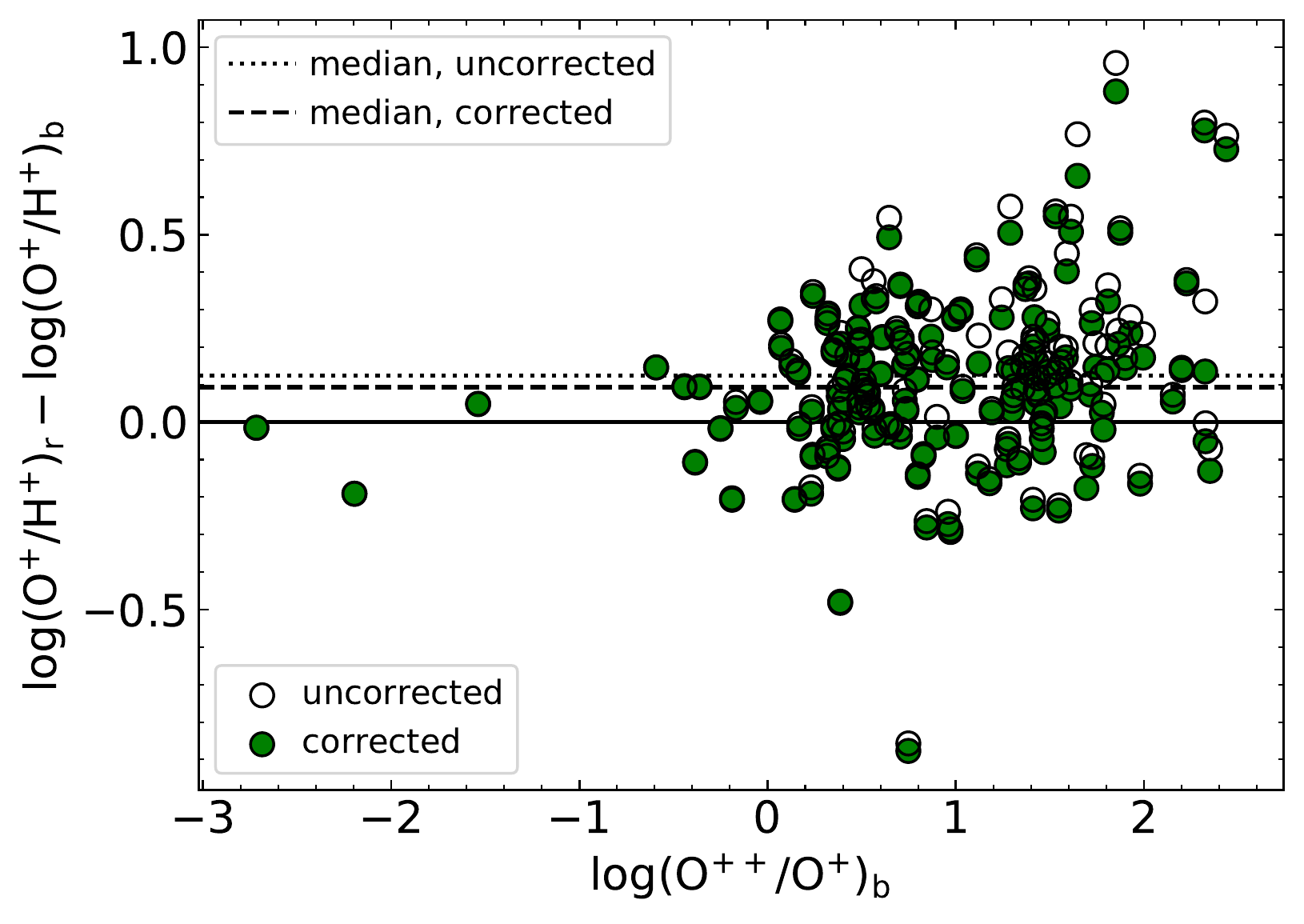}
\caption{Differences between the O$^+$ abundances derived with the blue (subscript `b') and red (subscript `r') [\ion{O}{ii}] lines as a function of the degree of ionization. Empty and solid circles present the results before and after the correction for recombination, and the dotted and dashed lines show the median values.} 
\label{figbr1}
\end{center}
\end{figure}

Fig.~\ref{figbr1} shows the differences between the O$^+$ abundances implied by the red and blue [\ion{O}{ii}] lines, O$^+_{\mathrm{r}}$/O$^+_{\mathrm{b}}$, before and after the recombination correction, as a function of the degree of ionization. The results are presented for the 153 spectra of the sample that include measurements of both the blue and red [\ion{O}{ii}] lines. It can be seen from this figure that the correction improves the agreement between the blue and red [\ion{O}{ii}] lines, with the median value of O$^+_{\mathrm{r}}$/O$^+_{\mathrm{b}}$ changing from 0.12 to 0.09~dex after the correction. 

However, the red lines are leading to systematically larger O$^+$ abundances, a result previously found by \citet{Sta98} and \citet{Esc04}. This disagreement might indicate that the corrections for recombination should be higher \citep[see, e.g.][]{Nemer19}, but it could also imply that the uncertainties in the measurement of the intensities of the blue and red [\ion{O}{ii}] lines relative to H$\beta$ are not symmetrical, with the intensities of the red lines being more easily overestimated (because of contamination with sky emission?) or with the blue lines being more easily underestimated (because of the effects of atmospheric differential refraction?). As discussed below in Section~\ref{o2hhe}, the behaviour of the \ion{H}{i} and \ion{He}{i} lines suggests that the differences in O$^+$ are not due to systematic effects introduced by the flux calibration or the extinction correction.

I explored other possibilities: changing the atomic data used for O$^+$ (the transition probabilities and collision strengths), increasing by 2000~K the temperature used to analyze [\ion{O}{ii}] emission, and decreasing the temperature used for the correction for recombination to 4000~K (if this temperature is reduced further, no solution to the statistical equilibrium equations is found for several spectra). These effects might be important for some objects, but only an increase in the temperature used to analyze [\ion{O}{ii}] emission leads to a decrease in the median value of O$^+_{\mathrm{r}}$/O$^+_{\mathrm{b}}$ (from 0.09~dex to 0.04~dex), and only at the expense of introducing very low values of  O$^+_{\mathrm{r}}$/O$^+_{\mathrm{b}}$ in several objects.

Much better results are obtained when using case C for the recombination corrections, with the median difference between the red- and blue-based O$^+$ abundances decreasing to 0.03 dex. However case C requires that permitted transitions arising from levels 2 and 3 of O$^+$ are optically thick and, in order to explain the results, this should happen in objects with large values of O$^{++}$/O$^+$, which seems unlikely.

I used one of the models described below in Section~\ref{hei} to get a tentative estimate of the optical depth of the transitions arising from levels 2 and 3 of O$^+$. The model is calculated with the photoionization code {\sc cloudy}, version C17.01 \citep{Fer17}. It is a sphere of constant hydrogen density of $10^4$~cm$^{-3}$, an inner radius of $10^{16.5}$~cm, and typical PN abundances and dust grains. The source of ionization is a blackbody with $10^5$~K and a luminosity of $10^{37}$~erg s$^{-1}$. The model is matter-bounded, with a thickness of $10^{16.6}$~cm, and reaches O$^{++}$/O$^+\simeq170$ with a column density of O$^+$ of $10^{14.7}$~cm$^{-2}$ (radiation-bounded models with similar characteristics reach higher O$^+$ column densities, but do not reproduce the required high value of O$^{++}$/O$^+$). One of the strongest transitions linking levels 2 or 3 of O$^+$ to the upper levels is 2s$^2$ 2p$^3$~$^2$D$^{\rm{o}}_{5/2}$~--~2s$^2$ 2p$^4$~$^2$D$_{5/2}$, with a transition probability of $A_{ul}=1.9\times10^9$~s$^{-1}$ and $\lambda=718.5$~\AA\ \citep{WFD96}. The absorption coefficient at the line centre can be calculated using:
\begin{equation}
k_{0l} \simeq \frac{\lambda^2}{8\pi^{3/2}} \frac{g_u}{g_l} n_l \frac{A_{ul}}{\Delta{\nu_D}},
\end{equation} 
where $g_u$ and $g_l$ are the statistical weights of the upper and lower levels and $\Delta{\nu_D}$ is the line width, related to the full width at half-maximum by $\mathrm{FWHM} = 2 \sqrt{\ln2}\, \Delta{\nu_D}$ \citep[see, e.g.,][]{OF06}. With 0.3~\AA\ as a typical $\mathrm{FWHM}$ for \ion{O}{ii} recombination lines in the optical \citep{Pena17}, and taking a population for the lower level of $n_2\simeq10^{-2}$, I find that the optical depth at the centre of this transition is $\tau\simeq0.07$. This seems to rule out case~C as the explanation for the overabundances calculated with the red [\ion{O}{ii}] lines.

Another possible explanation for the high values of O$^+_{\mathrm{r}}$/O$^+_{\mathrm{b}}$ is an underestimate of the electron density. Indeed, this is clearly affecting several spectra, since eight of the ten spectra that imply $\log(\mathrm{O}^+_{\mathrm{r}}/\mathrm{O}^+_{\mathrm{b}})>0.4$ also lead to the lowest density values for their corresponding PNe. In fact, for those spectra that allowed the calculation of $n_{\rm{e}}$[\ion{O}{ii}], the median difference between this density and the adopted one is equal to 0.08~dex. If all the spectra are reanalyzed increasing their adopted densities by 0.08~dex, the median value of O$^+_{\mathrm{r}}$/O$^+_{\mathrm{b}}$ decreases from 0.09~dex to 0.045~dex. Since in those spectra where $n_{\rm{e}}$[\ion{O}{ii}] can be measured, the value enters in the determination of the adopted electron density, the effect could be higher for other cases. In fact, with the set of atomic data used here, $n_{\rm{e}}$[\ion{O}{ii}] is systematically higher than the other diagnostics, with median differences with respect to $n_{\rm{e}}$[\ion{S}{ii}], $n_{\rm{e}}$[\ion{Cl}{iii}], and $n_{\rm{e}}$[\ion{Ar}{iv}] of 0.17~dex, 0.12~dex, and 0.07~dex. With an increase of 0.2~dex in density, the median value of O$^+_{\mathrm{r}}$/O$^+_{\mathrm{b}}$ becomes negative, $-0.024$~dex. Hence, it seems possible to explain the higher than expected values of O$^+_{\mathrm{r}}$/O$^+_{\mathrm{b}}$ as arising from density underestimates.

This does not necessarily imply that the electron density in the [\ion{O}{ii}] emitting region is larger than the one characterizing other diagnostics, since the choice of atomic data determines the differences in density that are found with the different diagnostics (Juan de Dios \& Rodr\'iguez, in prep.). Besides, if the transition probabilities used for O$^+$ are changed to those identified in PyNeb as Z82-WFD96 (which use the data of \citealt{Z82} and \citealt{WFD96} for different transitions), the median differences between $n_{\rm{e}}$[\ion{O}{ii}] and $n_{\rm{e}}$[\ion{S}{ii}], $n_{\rm{e}}$[\ion{Cl}{iii}], and $n_{\rm{e}}$[\ion{Ar}{iv}] change to 0.03, -0.06, and -0.05~dex, but the median of O$^+_{\mathrm{r}}$/O$^+_{\mathrm{b}}$ does not change in a significant way. In order to get the median value of O$^+_{\mathrm{r}}$/O$^+_{\mathrm{b}}$ closer to 0, one would need to change the atomic data for S$^+$, Cl$^{++}$, and Ar$^{+3}$ so that their density diagnostics all lead to higher values. As mentioned above, there might be other effects involved, like systematic uncertainties in the measurement of the relative intensities of the blue and red [\ion{O}{ii}] lines. A selection of a large number of high quality spectra might shed more light on this issue.

\begin{figure*} 
\begin{center}
\includegraphics[width=0.9\textwidth, trim=30 10 20 10, clip=yes]{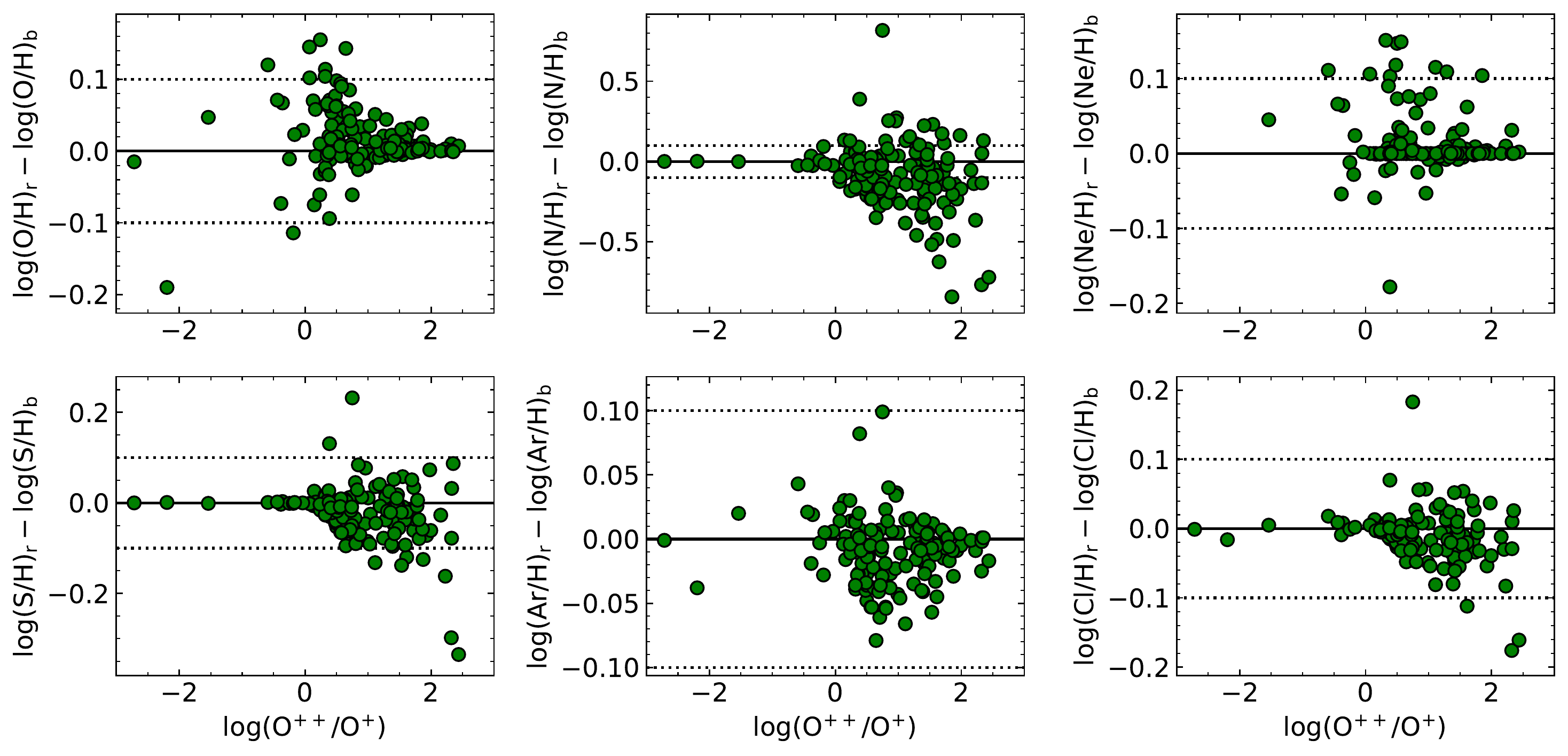}
\caption{Differences between the total abundances derived with the blue and red [\ion{O}{ii}] lines (identified with the subscripts `b' and `r', respectively) presented as a function of the degree of ionization. Dotted lines show differences of $\pm0.1$~dex.} 
\label{figbr2}
\end{center}
\end{figure*}

On the other hand, note that the differences between the O$^+$ abundances implied by the blue and red [\ion{O}{ii}] lines shown in Fig.~\ref{figbr1} have a large dispersion, with discrepancies that easily reach 0.3~dex or more. This suggest that the measurement of the relative intensities of the blue and red [\ion{O}{ii}] lines is very uncertain in many observed spectra. This also implies that the total abundances for all the elements that are based on the O$^+$ abundance (directly or through the use of ionization correction factors), will be different when derived with either the blue or the red [\ion{O}{ii}] lines. The differences in the total abundances implied by the blue and red [\ion{O}{ii}] lines are shown for the sample studied here in Fig.~\ref{figbr2} as a function of the degree of ionization. Differences of $\pm0.1$~dex are indicated by horizontal dotted lines in the figure. It can be seen that in some extreme cases the differences are higher than 0.1~dex for all elements excepting Ar. The effects on N/H are the most serious, with more than half of the spectra leading to differences that are larger than 0.1~dex and can surpass 0.5~dex. 

Note that the blue and red [\ion{O}{ii}] lines, at $\lambda=3727$ and 7325~\AA, lie at the edges of the wavelength range that contains all the lines generally used to determine physical conditions and chemical abundances. The fact that their relative intensities are difficult to measure properly suggest that other important line ratios will also have smaller but still important uncertainties.

So, which lines are better suited to calculate abundances? The blue or the red [\ion{O}{ii}] lines?  Both sets of lines have advantages and disadvantages. The blue lines are very sensitive to the effects of atmospheric differential refraction \citep{Fil82} and are in a region of the spectrum where our instruments are generally not very efficient. On the other hand, the red lines can be severely affected by telluric emission and are more sensitive to uncertainties in the electron temperature. As shown above in Section~\ref{rec}, the total abundances derived with the red lines are also more sensitive in general to the effects of the correction for recombination. The disadvantages of the red lines seem more serious than those affecting the blue lines, suggesting that the blue lines should be preferred, especially when problems with atmospheric differential refraction are not an issue.\footnote{This can be a difficult problem to tackle, since for most of the published spectra no information is provided on the airmass at the time of observation.}

\section{Spectra quality}

The simplest approach one can use to get an estimate of the chemical abundances and their observational uncertainties in each PN is to get an average of the results derived with different spectra and to calculate the deviations from this average. However, there are only 3--7 spectra per object, which means that poor-quality spectra will have a large effect in the results. In fact, with so few determinations per PN, the estimates of the mean values and their uncertainties can be seriously biased. A second approach that can be used is to get a weighted average of the results, using some estimate of the observational errors or the spectra quality in order to decide the contribution of each spectra to the final results. The disadvantage of this approach is that the weighting procedure is unlikely to be perfect and might assign some weight to spectra that should be discarded. A good alternative is to identify for each object the spectrum that can be considered more reliable, and to use this spectrum as the reference for the chemical abundances of the object. The differences between the abundances derived with the other available spectra and the ones provided by the reference spectrum can then be used to explore the uncertainties in the results.

Note that since there are at least three spectra per object, a comparison of the results obtained with each of them makes it possible to trace the causes of some of the differences in the final abundances. For example, the spectrum of \citet{GKA07} for NGC~6369 leads to the lowest values of electron temperature for this object: $T_{\rm{e}}\mbox{[\ion{N}{ii}]}=8900$~K and $T_{\rm{e}}\mbox{[\ion{O}{iii}]}=8000$~K versus $T_{\rm{e}}\mbox{[\ion{N}{ii}]}=10700$--$13000$~K and $T_{\rm{e}}\mbox{[\ion{O}{iii}]}=9700$--$11600$~K with the other four spectra available for this PN. All the derived ionic abundances, excepting He$^+$/H$^+$ and S$^{++}$/H$^+$, are systematically higher for this spectrum. The final abundances are also higher, with the exception of N/H and S/H, which illustrates not only the interdependence of the derived parameters but also the fact that several different sources of error are affecting the results. A similar effect happens with the spectra of \citet{AHF96} and \citet{HAF97} for NGC~6790 and NGC~6884, respectively, which also lead to the lowest temperatures for these objects. In the case of NGC~6884, all the total abundances, excepting He/H, are systematically higher, by 0.3~dex in O/H, for the spectrum observed by \citet{HAF97}. However, many of the differences in total abundances cannot be traced back to differences in temperature or any other single parameter. This implies that the quality of the spectra should be explored using as many criteria as possible.
I discuss below different criteria that can be used to explore the quality of the spectra and to assign scores to the spectra.

\subsection{CCD}

Some of the spectra, 47/179, were not obtained with a CCD but with detectors whose measurements might be more uncertain because of the effects of non-linearity \citep[see, e.g.,][]{PT87}. On this account, the spectra are assigned a score of 10 when they have been observed with a CCD spectra and a score of 0 otherwise. The validity of this approach is examined below.

\subsection{Electron temperature}

The electron temperature is one of the critical parameters needed to determine nebular abundances. Besides, we know that nebulae can have different temperatures in regions that differ in their degree of ionization. The best estimates of the electron temperature in the high- and low-ionization regions are $T_{\rm{e}}$[\ion{O}{iii}] and $T_{\rm{e}}$[\ion{N}{ii}]. All the sample spectra used here have these two values of electron temperature and 20 points are assigned to these spectra because of this. Other spectra with just one of these temperatures might be assigned a score of 15, whereas spectra with no estimate of $T_{\rm{e}}$ would receive a score of 0 in this item.

\subsection{Electron density}

A good measurement of the electron density is also critical in the determination of nebular chemical abundances, especially when the density is larger than a few hundred particles per cm$^{-3}$, as in many of the observed PNe. The most useful diagnostics are [\ion{S}{ii}]~$\lambda6716/\lambda6731$, [\ion{O}{ii}]~$\lambda3726/\lambda3729$, [\ion{Cl}{iii}]~$\lambda5518/\lambda5538$, and [\ion{Ar}{iv}]~$\lambda4711/\lambda4740$. These diagnostics sample somewhat different ionization regions in the nebula. Thus, when several of them are measured, they can be used to obtain a better estimate of the density, and when measured with low uncertainty, they can also provide information on the density structure of the nebula. Besides, the number of density diagnostics that can be measured will depend not only on the physical conditions and degree of ionization of the PN, but also on the signal to noise ratio of the observed spectrum, so this number can be considered to provide some indication of the spectrum quality. Scores of 9, 11, 13, or 15 are assigned to the spectra when they contain one, two, three, or four working diagnostics. A spectrum with no working diagnostic would get an score of 0 in this item.

\subsection{[O~III] and [N~II] line ratios}

The [\ion{O}{iii}]~$\lambda5007/\lambda4959$ and [\ion{N}{ii}]~$\lambda6584/\lambda6548$ line intensity ratios involve transitions that arise from the same energy level in each ion. Hence, these line ratios must have a fixed value, determined by the energy of the transitions and the transition probabilities. Since the lines in both pairs are very close in wavelength, the values measured in any spectra for these line ratios should not be affected by problems with the flux calibration and extinction correction or with atmospheric differential refraction. A comparison between their observed values and the theoretical ones will reflect how well the intensities of relatively strong emission lines can be measured in the spectra.

\begin{figure} 
\begin{center}
\includegraphics[width=0.5\textwidth, trim=0 30 -20 0, clip=yes]{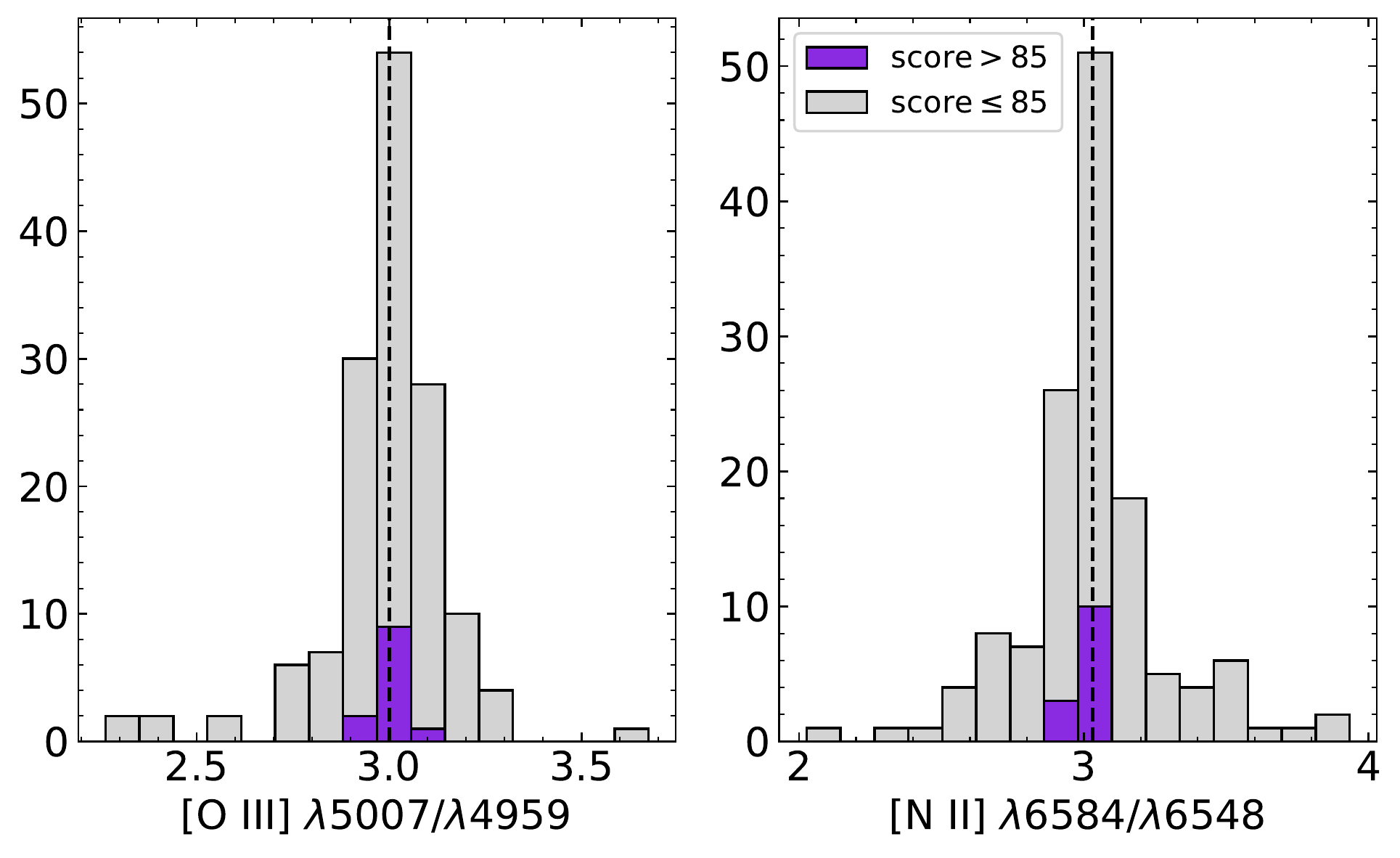}
\caption{Distributions of the values of the [\ion{O}{iii}]~$\lambda5007/\lambda4959$ and [\ion{N}{ii}]~$\lambda6584/\lambda6548$ line intensity ratios for the sample spectra. The dashed lines show the positions of the median values at 3.00 for [\ion{O}{iii}] and 3.03 for [\ion{N}{ii}]. The darker areas show the distributions for the spectra that achieve final scores larger than 85.} 
\label{figcNO}
\end{center}
\end{figure}

However, the theoretical predictions for these line ratios usually differ among themselves \citep[see, e.g.,][]{FFT09} and from the observations. With the atomic data used here (Table~\ref{tab:atdata}), the predicted values for the [\ion{O}{iii}] and [\ion{N}{ii}] line ratios are 2.98 and 2.94, respectively. These theoretical ratios are slightly lower than the ones obtained from the median values of the sample spectra: 3.00 for [\ion{O}{iii}] and 3.03 for [\ion{N}{ii}], and I decided to use these observational values as the reference values. Fig.~\ref{figcNO} shows the distributions of values of the [\ion{O}{iii}]~$\lambda5007/\lambda4959$ and [\ion{N}{ii}]~$\lambda6584/\lambda6548$ line ratios for the sample spectra, with the positions of the median values marked with dashed lines.

The scores based on each of these line ratios are assigned as follows. For each line ratio, a linear relation is used: $p_i=10-312.5\,|\log(r_i/r_{\rm{median}})|$, where $r_i$ is the observed line ratio and $r_{\rm{median}}$ its median. The values obtained for $p_i$ are rounded to the next integer. With $i=2$ for [\ion{N}{ii}] and $i=3$ for [\ion{O}{iii}], the values of $p_2$ and $p_3$ shown in Table~\ref{tab:scores} are obtained. The maximum score that can be obtained with this procedure is 10 for each line ratio; the minimum is 0. The minimum is assigned when the logarithmic differences from the median are larger or equal to 0.032~dex or when one of the lines is missing.

\subsection{He~I lines}
\label{hei}

There are many \ion{He}{i} lines across the nebular spectra that have strong to medium strengths, and most observers report the intensities of several of them. A comparison between the He$^+$ abundances derived from these lines can thus be used to assess the spectra quality.

Not all \ion{He}{i} lines are equally useful for this purpose, however. The intensities of some of them, like \ion{He}{i}~$\lambda7065$ and \ion{He}{i}~$\lambda3889$, depart from their predicted case B values because they are very sensitive to changes in the optical depth of transitions connecting the 2s~$^3$S and $^1$S energy levels with levels of higher energy.

In order to get a general idea of which lines follow closely the case B predictions for their intensities, I ran several simple models using the photoionization code {\sc cloudy} \citep[version C17.01,][]{Fer17}. I used the prescriptions described by these authors in their Section~3.1.1 for the numbers of resolved and collapsed levels required to get good estimates of the \ion{He}{i} line intensities, and changed the optical depths of each individual model by introducing microturbulence or by removing the outer ionized layers of the model. The models are spheres ionized by a blackbody with a temperature of $T_{\mathrm{eff}}=5\times10^4$ or $10^5$~K and a luminosity of $L=10^{37}$~erg s$^{-1}$. The hydrogen density of the models is constant and equal to $n_{\mathrm{H}}=10^{3}$, $10^{3.7}$, or $10^4$~cm$^{-3}$, the inner radius is $R_{\mathrm{in}}=10^{15}$~cm (for the models with $T_{\mathrm{eff}}=5\times10^4$~K) or $10^{16.5}$~cm (for $T_{\mathrm{eff}}=10^5$~K), and the models are either radiation-bounded, or matter-bounded with thickness in the range $\Delta{R}=10^{16.6}$--$10^{17}$~cm. Some of the models also include the effects of turbulence in the line widths, with velocities of either $v_{\mathrm{turb}}=15$ or 30~km s$^{-1}$. The models have the typical PN chemical abundances defined in {\sc cloudy} and the `ISM' dust grain properties that come by default with these abundances.

The different parameters of the models do not cover the range of PN typical values \citep[see, e.g.,][]{Mil16, DI14} and are not explored in a systematic way. However, the models are only used to explore the variations in the relative intensities of the \ion{He}{i} lines introduced by changes in the optical depth of the resonant \ion{He}{i} transitions. Hence, the results are expected to be insensitive to changes in the effective temperature of the ionizing radiation field or in its source (i.e., a blackbody versus a stellar atmosphere). Besides, the results can be compared with those obtained from the observations in order to check for their consistency.

The lines whose behaviour is explored with the models are \ion{He}{i}~$\lambda3614$, $\lambda3965$, $\lambda4026$, $\lambda4388$, $\lambda4438$, $\lambda4471$, $\lambda4713$, $\lambda4922$, $\lambda5016$, $\lambda5048$, $\lambda5876$, $\lambda6678$, $\lambda7065$, and $\lambda7281$. These lines sample different types of transitions: $2\,^1$S--$n\,^1$P$^{\mathrm{o}}$ ($\lambda3614$, $\lambda3965$, $\lambda5016$); $2\,^3$P$^{\mathrm{o}}$--$n\,^3$S ($\lambda4713$, $\lambda7065$); $2\,^1$P$^{\mathrm{o}}$--$n\,^1$S  ($\lambda4438$, $\lambda5048$, $\lambda7281$); $2\,^1$P$^{\mathrm{o}}$--$n\,^1$D ($\lambda4388$, $\lambda4922$,  $\lambda6678$); and $2\,^3$P$^{\mathrm{o}}$--$n\,^3$D ($\lambda4026$, $\lambda4471$, $\lambda5876$). The \ion{He}{i}~$\lambda3889$ line ($2\,^3$S--$3\,^3$P$^{\mathrm{o}}$) is not included because it is usually blended with the H8 line, but the optical depth of this transition, $\tau_{\lambda3889}$, whose value is provided by {\sc cloudy}, is used as a proxy for the optical depth of all \ion{He}{i} transitions. The models cover a relatively large range of optical depths, $\tau_{\lambda3889}=0.3$--$18$, which seems enough to reproduce the observed differences in the behaviour of the \ion{He}{i} lines.

\begin{figure} 
\begin{center}
\includegraphics[width=0.5\textwidth, trim=-20 20 -40 0, clip=yes]{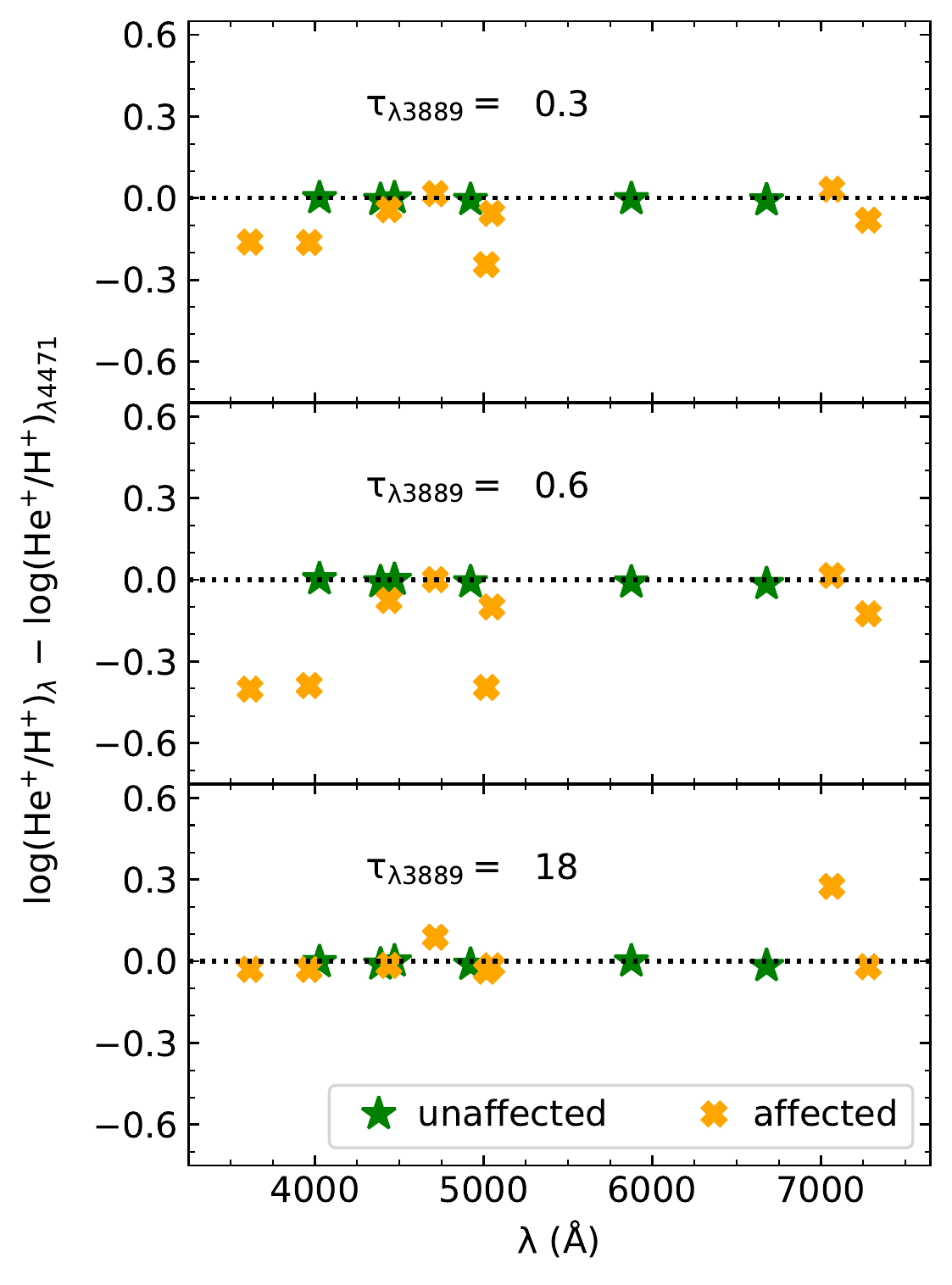}
\caption{Differences between the He$^+$ abundances implied by different \ion{He}{i} lines and those derived with \ion{He}{i}~$\lambda4471$ for three ionization models with different values of $\tau_{\lambda3889}$. The results corresponding to the \ion{He}{i}~$\lambda4026$, $\lambda4388$, $\lambda4471$, $\lambda4922$, $\lambda5876$, and $\lambda6678$ lines, which lead to very similar He$^+$ abundances, are shown with stars; the results for the lines that can be affected by changes in the optical depth, \ion{He}{i}~$\lambda3614$, $\lambda3965$, $\lambda4713$, $\lambda5016$, $\lambda5048$, $\lambda7065$, and $\lambda7281$, are represented with crosses.} 
\label{figtau}
\end{center}
\end{figure}

Fig.~\ref{figtau} shows the results for three models that sample the different behaviour of the \ion{He}{i} lines. The model with $\tau_{\lambda3889}=0.3$ has $T_{\mathrm{eff}}=5\times10^4$~K,  $n_{\mathrm{H}}=10^{3}$~cm$^{-3}$, $R_{\mathrm{in}}=10^{15}$~cm, $\Delta{R}=10^{17}$~cm, and $v_{\mathrm{turb}}=15$~km s$^{-1}$. The model with $\tau_{\lambda3889}=0.6$ has $T_{\mathrm{eff}}=10^5$~K,  $n_{\mathrm{H}}=10^{3.7}$~cm$^{-3}$, $R_{\mathrm{in}}=10^{16.5}$~cm, $\Delta{R}=10^{16.8}$~cm, and $v_{\mathrm{turb}}=15$~km s$^{-1}$. The third model shown in Fig.~\ref{figtau}, the one with $\tau_{\lambda3889}=18$, has $T_{\mathrm{eff}}=10^5$~K,  $n_{\mathrm{H}}=10^{4}$~cm$^{-3}$, $R_{\mathrm{in}}=10^{16.5}$~cm, $\Delta{R}$ unconstrained (i.e., radiation-bounded), and no turbulence. For each model, the He$^+$ abundances implied by the different lines are calculated with PyNeb using the average electron density of the model and the electron temperature in the O$^{++}$ region (the results are not sensitive to small changes in these parameters). Fig.~\ref{figtau} shows the differences between these abundances and those implied by the \ion{He}{i}~$\lambda4471$ line as a function of wavelength.

The results presented in Fig.~\ref{figtau} suggest that all the lines arising from the transitions $2\,^1$P$^{\mathrm{o}}$--$n\,^1$D and $2\,^3$P$^{\mathrm{o}}$--$n\,^3$D, whose results are represented with stars in Fig.~\ref{figtau}, are mostly insensitive to optical depth effects, with deviations from their case B values lower than 0.02~dex. Hence, the relative abundances implied by \ion{He}{i}~$\lambda4026$, $\lambda4388$, $\lambda4471$, $\lambda4922$, $\lambda5876$, and $\lambda6678$ can be used to assign a quality to the spectra.

Fig.~\ref{fighe} shows the differences between the He$^+$ abundances implied by the various \ion{He}{i} lines discussed above and the abundance implied by \ion{He}{i}~$\lambda4471$ for several of the sample spectra. The circles in Fig.~\ref{fighe} present the results for the \ion{He}{i} lines that are not expected to be affected by optical depth effects; the squares show the results for the other lines. Each panel in Fig.~\ref{fighe} is identified with the PN and the spectrum reference; the final score attained by each spectrum is also plotted. It can be seen in Fig.~\ref{fighe} that several spectra lead to a very good agreement in the abundances implied by the \ion{He}{i} lines that are expected to be insensitive to optical depth effects according to the models, indicating that there is a general consistency between models and observation in this issue. On the other hand, the spectra that show the largest differences usually show a dependence of these differences on wavelength, suggesting that the differences could be due to problems with the flux calibration.

Taking into account all of the above, I decided to assign scores to the spectra using the `unaffected' \ion{He}{i}~$\lambda4026$, $\lambda4388$, $\lambda4471$, $\lambda4922$, $\lambda5876$, and $\lambda6678$ lines. A score of 2 is assigned to those spectra that present the intensity of at least one \ion{He}{i} line from the six listed above. For spectra that include the measurement of several \ion{He}{i} lines, the He$^+$ abundance derived from each line is compared with the mean abundance implied by the brightest \ion{He}{i} lines, $\lambda4471$, $\lambda5876$, and $\lambda6678$. If the difference in the logarithmic abundances is lower than 0.04~dex, a score of 3 is added to the previous score; if the difference lies between 0.04 and 0.08~dex, a score of 2 is added; and if the difference is larger than 0.08~dex, a score of 1 is added (just because one additional line could be measured). This is done for each available line in the list of `unaffected' lines. With this procedure, the maximum score that can be achieved in this part is 20.

\subsection{H~I lines}
\label{hi}

Several \ion{H}{i} lines from the Balmer and Paschen series can be observed in the optical and near-infrared spectra of ionized nebulae. Most of them are outside the spectral range that is usually observed, and those that are generally measured, like H$\alpha$, H$\beta$, H$\gamma$, and H$\delta$, are used to determine the amount of extinction affecting the observed spectrum. This means that in most cases they cannot be used in the same way as the \ion{He}{i} lines to provide a quality assessment. However, if three or four of these Balmer lines are measured, the degree of agreement of their relative intensities with the predicted values provides some indication of the spectra quality.

Several spectra do include the measurements of Paschen lines and several other Balmer lines. These extra lines bracket the spectral range where the lines used in the abundance determination are located, and go beyond them in the blue and red spectral ends. The comparison of their extinction-corrected relative intensities with the theoretical ones can in principle be used to explore not only how well the spectrum allows the measurement of lines that have different intensities but also how well the flux calibration and extinction correction are working. There is one drawback, though: the Balmer and Paschen can depart from their predicted case~B values, especially when the quantum numbers $n$ of their upper levels are large \citep{Mes09}.

Fig.~\ref{fighbp} shows the departures of the observed, extinction-corrected intensities of the Balmer and Paschen lines relative to H$\beta$ from the values predicted by case~B for several of the sample spectra. The results are presented as a function of the principal quantum number $n$ of the upper level of the transition. In some cases, like those of the spectra for Cn~1-5, M~1-25, and PC~14 from \citet{GPMDR12} (GPMDR12 in Fig.~\ref{fighbp}) and the spectra for IC~418 from \citet{DASNA17} and \citet{SWBvH03} (DASNA17 and SWBvH03 in Fig.~\ref{fighbp}), the Balmer and Paschen lines show small departures from their case~B values and a dependence of these departures on $n$, which suggest that the deviations are real. Other cases are more difficult to interpret, like the different behaviour shown by the \ion{H}{i} lines in the three spectra of NGC~7009 from \citet{FL11}, \citet{HA95a} and \citet{HA95b} (see Fig.~\ref{fighbp}). In many cases, the departures of the \ion{H}{i} line ratios increase with $n$, an effect that might be real or due to the difficulties inherent in measuring the weaker lines arising from levels with higher values of $n$.

Taking all of the above into consideration, I decided to assign an score of 6 to those spectra that had independent measurements of at least two of the three lines H$\alpha$, H$\gamma$, and H$\delta$ (H$\beta$ is present in all the sample spectra) if all their intensities relative to H$\beta$ are within 0.05~dex of their expected values according to case~B. Besides, for those spectra with measurements of at least one Balmer and one Paschen line with upper levels in the range $n=9$ to 20, the ratios of their observed (extinction-corrected) to their predicted case~B intensities were used to obtain the median deviations of these Balmer and Paschen lines. These deviations were then compared (Paschen minus Balmer), and if they differed by 0.10 to 0.15~dex, 0.05 to 0.10~dex, or by less than 0.05~dex, the spectrum was assigned an extra score of 3, 6, or 9, respectively. This procedure is based on the assumption that Balmer and Paschen lines originating from levels with similar values of $n$ will show similar deviations from their case~B values. The validity of this assumption will need to be explored once a sufficient number of high-quality spectra have been identified. This could be done using the other quality criteria listed above.

The procedure described above results in scores in the range 0--15 based on the behaviour of the \ion{H}{i} lines.

\subsection{A comparison of the results obtained with the H~I and He~I lines}\label{o2hhe}

Since the \ion{He}{i} lines whose behaviour is compared here cover the wavelength range 4026--6678~\AA, they can be used to explore the trends introduced in this range by problems with the flux calibration or related effects. A comparison of these trends with those that can be inferred from the behaviour of the \ion{H}{i} Balmer and Paschen lines that lie to the blue and red of this wavelength range can shed light on the kind of observational uncertainties affecting the spectra.

\begin{figure} 
\begin{center}
\includegraphics[width=0.4\textwidth, trim=20 10 20 0, clip=yes]{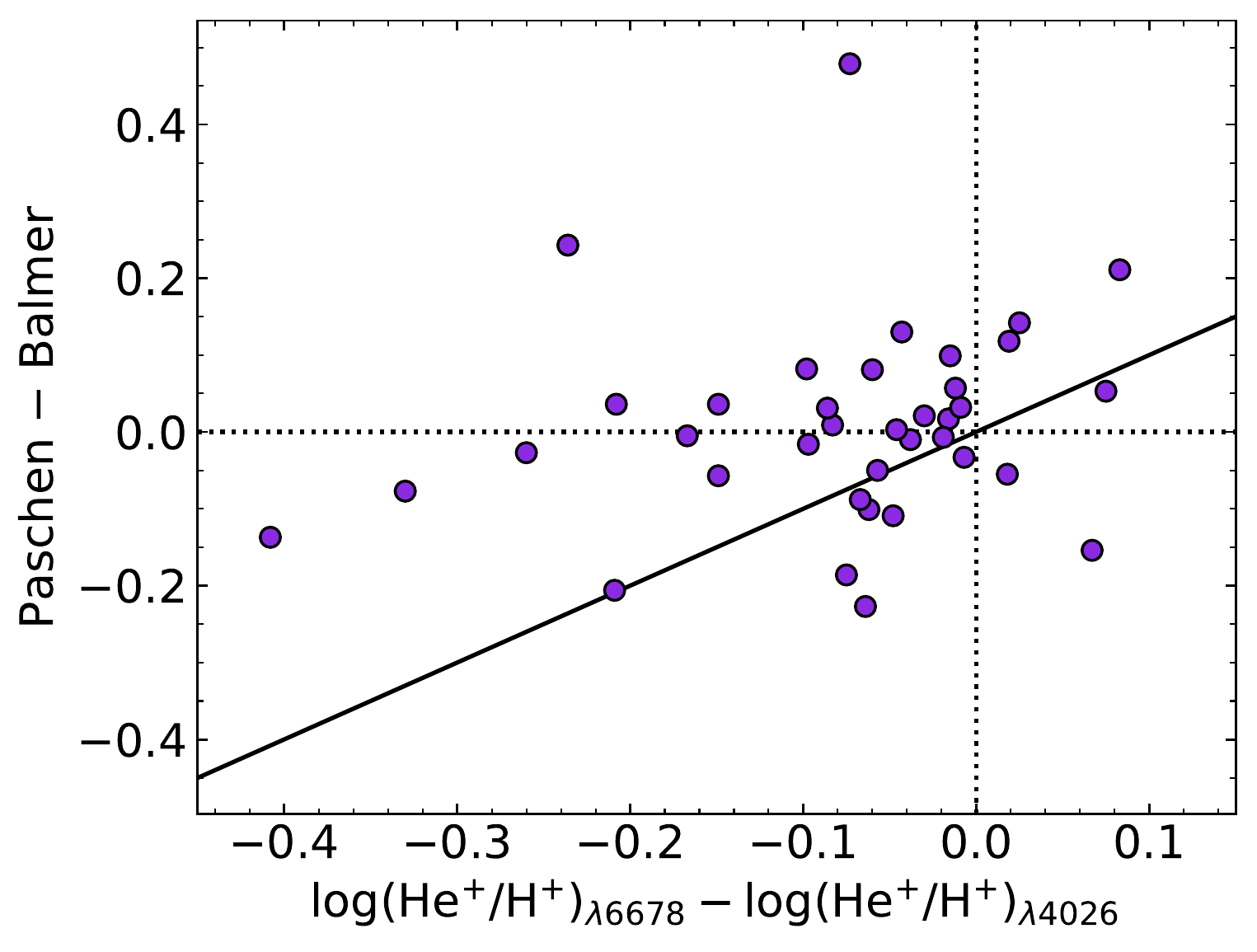}
\caption{Differences between the median deviations of the Paschen and Balmer lines from their expected values as a function of $\log(\mathrm{He}^+_{\lambda6678}/\mathrm{H}^+)- \log(\mathrm{He}^+_{\lambda4026}/\mathrm{H}^+)$. Dotted lines show zero differences; the continuous line represents equal deviations.} 
\label{figdd}
\end{center}
\end{figure}

Fig.~\ref{figdd} shows the difference between the median deviations of the Paschen and Balmer lines from their expected values, i.e., `Paschen $-$ Balmer' is the median value of $\log(I(\mathrm{Paschen})/I(\mathrm{H}\beta))-\log(I(\mathrm{Paschen})/I(\mathrm{H}\beta))_{\mathrm{case~B}}$ minus the median value of $\log(I(\mathrm{Balmer})/I(\mathrm{H}\beta))-\log(I(\mathrm{Balmer})/I(\mathrm{H}\beta))_{\mathrm{case~B}}$, where `Paschen' and `Balmer' stand for the available lines whose upper levels have quantum numbers in the range $n=9$ to 20. These differences are plotted as a function of $\log(\mathrm{He}^+_{\lambda6678}/\mathrm{H}^+)-\log(\mathrm{He}^+_{\lambda4026}/\mathrm{H}^+)$ for those spectra where all these values are available. The dotted lines in this figure show the expected values of zero for these quantities, although for the \ion{H}{i} lines, as discussed in Section~\ref{hi}, it is not clear whether zero deviations are to be expected. The continuous line in Fig.~\ref{figdd} shows the effects of equal deviations for \ion{He}{i} and \ion{H}{i}.

The results shown in Fig.~\ref{figdd} indicate that there is some correlation between the trends seen in the \ion{He}{i} and \ion{H}{i} lines, although with some dispersion, which is to be expected if the deviations are not linear with wavelength, as illustrated by the behaviour of the different \ion{He}{i} lines in several spectra (see Fig.~\ref{fighe}). On the other hand, whereas half of the spectra are redder than expected for `Paschen $-$ Balmer' and half are bluer, most of the spectra are bluer than expected according to the \ion{He}{i} lines. Since large changes in electron density or temperature do not change significantly the behaviour of the \ion{He}{i} and \ion{H}{i} lines, this could be a real systematic effect of the flux calibration. As an alternative explanation of the bluer than expected spectra implied by the \ion{He}{i} lines, one might consider the effects of \ion{He}{i} absorption lines in the central stars of the PNe. However, most of the sample PNe are extended objects, and this kind of effect is found in some spectra that do not include the central star in the observed area, like the spectrum of \citet{HA98} for NGC~2440 (see Fig.~\ref{fighe}).

\subsection{A comparison with the results obtained with the blue and red [O~II] lines}

Since the red and blue lines of [\ion{O}{ii}] are located at 3727 and 7325~\AA, a comparison of the O$^+$ abundances implied by these two sets of lines can provide complementary information on the effects of the flux calibration explored with the \ion{He}{i} and \ion{H}{i} lines. However, as discussed in Section~\ref{oii}, the relative abundances derived with the red and blue [\ion{O}{ii}] lines with the sample spectra, O$^+_{\mathrm{r}}$/O$^+_{\mathrm{b}}$, are larger than one in most cases, an effect that can be real or due to the choice of the density structure and of the atomic data. Nevertheless, a comparison of the trends followed by O$^+_{\mathrm{r}}$/O$^+_{\mathrm{b}}$ with those implied by the \ion{He}{i} and \ion{H}{i} lines might be valuable.

\begin{figure*} 
\begin{center}
\includegraphics[width=0.9\textwidth, trim=30 10 20 10, clip=yes]{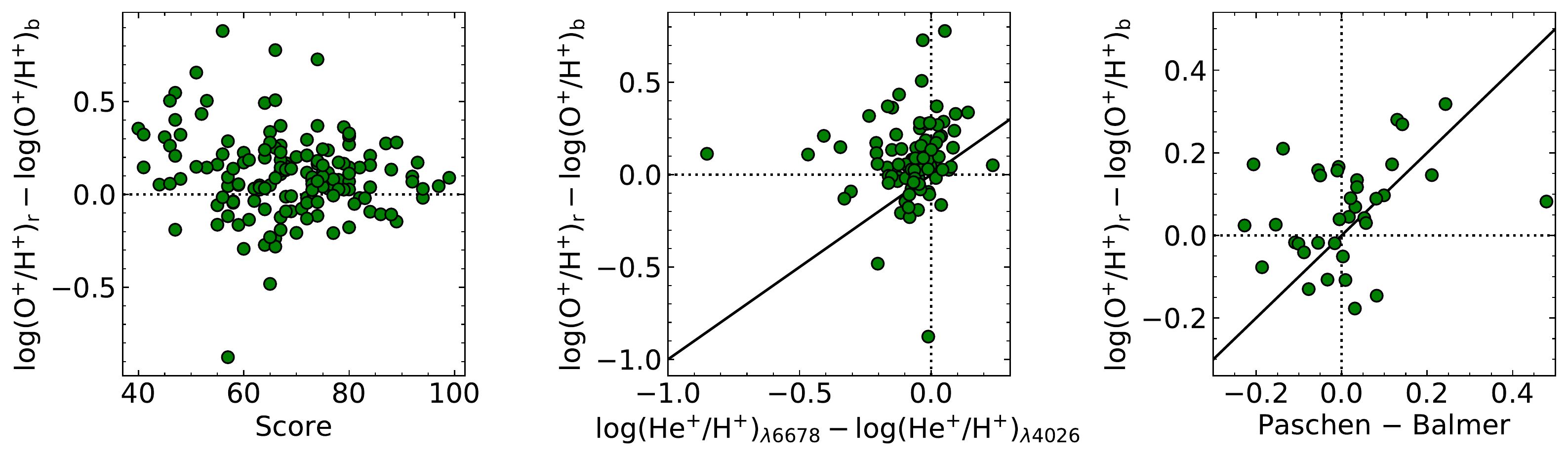}
\caption{Differences in the O$^+$ abundances implied by the red and blue [\ion{O}{ii}] lines (identified with the subscripts `r' and `b', respectively) plotted as a function of the final scores of the spectra (left panel), the differences in He$^+$ implied by \ion{He}{i}~$\lambda6678$ and $\lambda4026$ (middle panel), and the differences in the deviations of the Paschen and Balmer lines from their expected values (right panel). Dotted lines show zero deviations; continuous lines are for equal deviations. Filled/empty circles show the results of CCD/non-CCD spectra.} 
\label{figdoh}
\end{center}
\end{figure*}

The left panel of Fig.~\ref{figdoh} shows the values of O$^+_{\mathrm{r}}$/O$^+_{\mathrm{b}}$ as a function of the final scores of the spectra calculated following the procedure described above. Filled circles show the results obtained from CCD spectra, empty circles are for non-CCD spectra. The dispersion in O$^+_{\mathrm{r}}$/O$^+_{\mathrm{b}}$ can be seen to decrease for the spectra with higher scores, suggesting that most of the dispersion is introduced by observational errors.

The middle and right panels of Fig.~\ref{figdoh} show the O$^+_{\mathrm{r}}$/O$^+_{\mathrm{b}}$ abundance ratio as a function of $\log(\mathrm{He}^+_{\lambda6678}/\mathrm{H}^+)-\log(\mathrm{He}^+_{\lambda4026}/\mathrm{H}^+)$ and of `Paschen $-$ Balmer', with the dotted and continuous lines plotted at zero and at equal differences, respectively. It can be seen in these panels that the values of O$^+_{\mathrm{r}}$/O$^+_{\mathrm{b}}$ follow similar trends to those defined by the \ion{He}{i} and \ion{H}{i} lines, although there is dispersion probably arising from the effects of different sources of observational errors.

The results for the [\ion{O}{ii}], \ion{He}{i}, and \ion{H}{i} lines, if interpreted in terms of the relative intensities of the blue and red lines of each ion, indicate that the spectra are bluer than expected between 4026 and 6678~\AA\ (\ion{He}{i} lines), redder than expected between 3727 and 7235~\AA\ ([\ion{O}{ii}] lines), and about right between $\sim3700$ and $\sim8500$~\AA\ (\ion{H}{i} lines). Since the spectra have been obtained using different techniques and corrected for extinction with different extinction laws, it seems difficult to explain these colour variations using a single cause, such as the effect of biases introduced by the flux calibration or the extinction correction. On the other hand, both the flux calibration and the extinction correction could be introducing systematic errors in the line intensities. Besides, as discussed in Section~\ref{oii} above, atmospheric differential refraction and atomic data can also be responsible for the discrepancy between the results implied by the blue and red [\ion{O}{ii}] lines.

\subsection{Final scores and observational uncertainties}
\label{fsco}

The final score of each spectrum results from the addition of the scores obtained in each of the items described in this section. The maximum possible score for each spectrum is 100, and the spectra analysed here have scores in the range 40--99. Table~\ref{tab:scores} lists the partial and final scores for all the spectra analysed here. Columns~1 and 2 identify the PN and the reference for the spectrum; column~3 provides the score related to whether the spectra was obtained with a CCD ($p_0=10$) or not ($p_0=0$); column~4 gives the score related to the available temperature diagnostics ($p_1=20$ for the two diagnostics available in the spectra of this sample); columns~5 and 6 list the number of density diagnostics and the related score ($p_2=9$--15 for 1--4 working density diagnostics); columns~7--10 give the values of the [\ion{O}{iii}]~$\lambda5007/\lambda4959$ and [\ion{N}{ii}]~$\lambda6584/\lambda6548$ line ratios and their associated scores ($p_3$ and $p_4$, both in the range 0--10); columns~11 and 12 give the number of \ion{He}{i} lines unaffected by optical depth effects available in the spectra (out of the six listed above in Section~\ref{hei}) and the score assigned according to how consistent are the He$^+$ abundances implied by them ($p_5=2$--20, where 2 is the score when only one \ion{He}{i} line is available); columns~13--15 list the number of Balmer lines (out of H$\alpha$, H$\gamma$, and H$\delta$) whose intensities are listed in the spectra, the difference between the median deviations of the Paschen and Balmer lines from their expected values, and the score assigned the behaviour of all these lines ($p_6=0$--15, see Section~\ref{hi}); and column~16 provides the final score obtained from the addition of the previous ones ($p_{\mathrm{tot}}$, with a maximum value of 100).

An inspection of Table~\ref{tab:scores} shows that some spectra have low scores because they do not contain all the information used to assign scores. This can be due to their limited wavelength coverage or spectral resolution, to their low signal-to-noise not allowing the measurement of relatively weak lines, or to the presence of saturated features (like [\ion{O}{iii}]~$\lambda5007$). The case of the spectra presented by \citet{PSEK98} and \citet{PSM01} is particularly noteworthy in this respect, since these are the CCD spectra that show the lowest scores. As will be shown below, these spectra are not performing as badly as one might expect from their scores, but the low scores arise from an extreme lack of information, since the authors only present the intensities of the brightest lines that they require for their analysis.

\begin{figure} 
\begin{center}
\includegraphics[width=0.35\textwidth, trim=0 15 0 0, clip=yes]{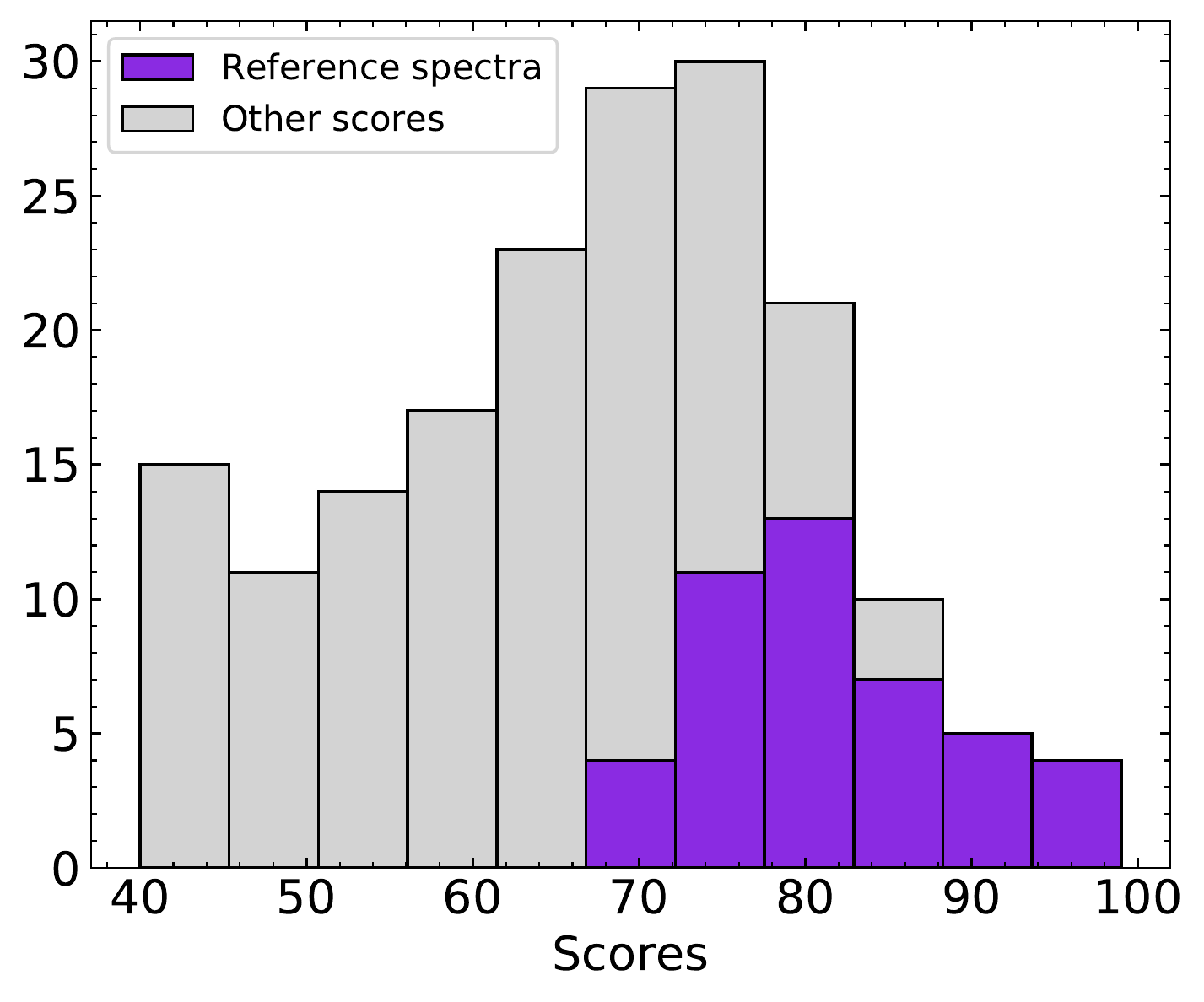}
\caption{Distribution of the final scores assigned to the spectra. The maximum scores attained by the spectra of each object, which are used to identify its reference spectra, are also identified.} 
\label{figsco}
\end{center}
\end{figure}

The distribution of final scores for the 179 spectra analyzed here is shown in Fig.~\ref{figsco}. The figure also shows the distribution of maximum scores obtained for the spectra of each object, which are in the range 68--99. In all cases but one, there is one spectrum that can be considered as the best available spectrum of the PN. The exception is for NGC~6302, where three of the four available spectra reach the maximum score for this object, $p_{\mathrm{tot}}=80$ \citep{KC06, RCK14, TBLDS03}. Three spectra with a final score of 80 might seem a pretty good result at first sight, suggesting that we can obtain good estimates of the final abundances in this PN by averaging the results of these three spectra, but a comparison of the final abundances implied by these spectra suggests otherwise. The maximum differences in the total abundances implied by these three spectra reach 0.11~dex for O/H and are in the range 0.16--0.25~dex for Cl/H, N/H, Ne/H, and S/H. As a matter of fact, the three spectra show extreme behaviour in some of the quality criteria explored above: the spectrum of \citet{RCK14} has $\log(\mathrm{O}^+_{\mathrm{r}}/\mathrm{O}^+_{\mathrm{b}})=0.32$, whereas the spectra of \citet{KC06} and \citet{TBLDS03} have $\log(\mathrm{He}^+_{\lambda6678}/\mathrm{He}^+_{\lambda4026})=-0.36$ and $-0.85$, respectively. Hence, no reference spectrum is defined or used for this PN.

There are other spectra that achieve the maximum score for their object, with $p_{\mathrm{tot}}=72$--80, and have similar problems, like the one for NGC~7662 of \citet{HA97b} with [\ion{N}{ii}]~$\lambda6584/\lambda6548=3.8$ and $\log(\mathrm{He}^+_{\lambda6678}/\mathrm{He}^+_{\lambda4026})=-0.33$; the one for NGC~6790 of \citet{KH01} with $\log(\mathrm{O}^+_{\mathrm{r}}/\mathrm{O}^+_{\mathrm{b}})=0.73$; and the spectra for NGC~6751,  Hu~1-2, and  M~2-23 of \citet{MKHS10}, \citet{SCT87}, and \citet{WL07} with $\log(\mathrm{O}^+_{\mathrm{r}}/\mathrm{O}^+_{\mathrm{b}})=0.33$--$0.37$. The discrepancies in $\log(\mathrm{O}^+_{\mathrm{r}}/\mathrm{O}^+_{\mathrm{b}})$ might be lower assuming different density structures or atomic data, but the problems presented by these spectra suggest that we should aim for spectra that have scores higher than 80 for each object of interest.

\begin{figure*} 
\begin{center}
\includegraphics[width=0.9\textwidth, trim=30 10 20 10, clip=yes]{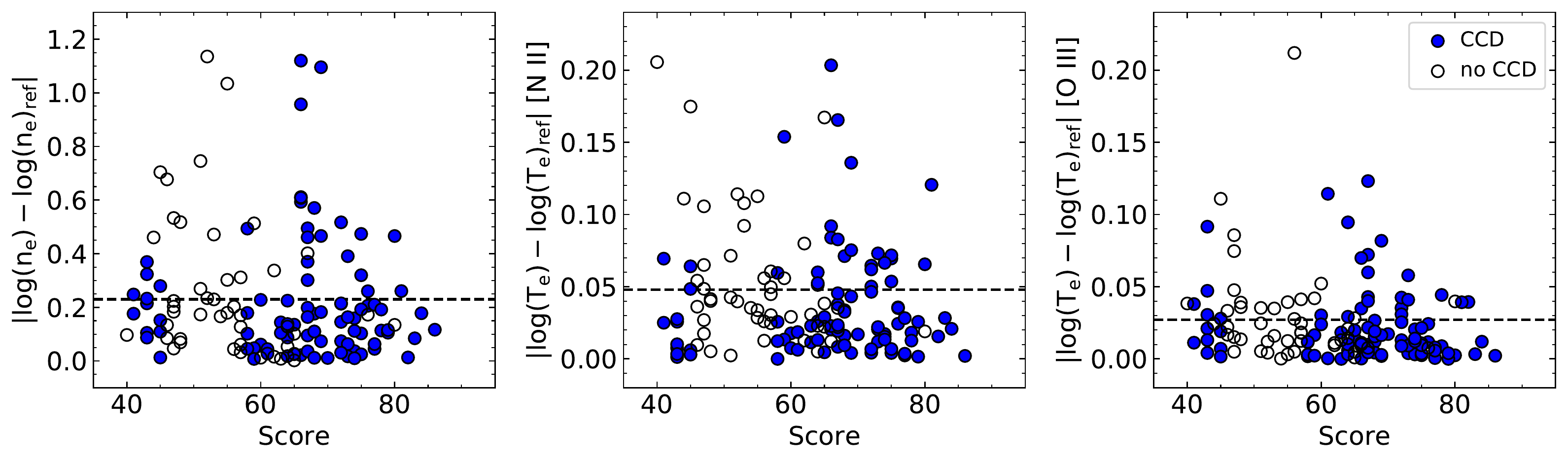}
\caption{Absolute differences between the values of $n_{\rm{e}}$, $T_{\rm{e}}$[\ion{N}{ii}], and $T_{\rm{e}}$[\ion{O}{iii}] derived with different spectra and those obtained with the reference spectrum for each object, plotted as a function of the scores assigned to the spectra. The dashed lines, located at 0.23~dex for $n_{\rm{e}}$, 0.05~dex for $T_{\rm{e}}$[\ion{N}{ii}], and 0.03~dex for $T_{\rm{e}}$[\ion{O}{iii}], are one-$\sigma$ uncertainties.} 
\label{figdnt}
\end{center}
\end{figure*}

The spectrum with the highest score for each object can now be used as a reference, and the differences between the results obtained with the available spectra and with the reference spectra will provide estimates of the uncertainties introduced by observational problems. Fig.~\ref{figdnt} shows the absolute values of the differences in the adopted densities, $T_{\rm{e}}$[\ion{N}{ii}], and $T_{\rm{e}}$[\ion{O}{iii}] as a function of the final scores. Some of these differences might be due to the fact that the different spectra can be sampling different regions of the PNe. However, since the magnitude of the differences decreases for higher scores, most of them seem to be attributable to observational errors. The dashed lines in Fig.~\ref{figdnt} are the upper limits that contain 68 per cent of the differences, and can thus be interpreted as one-$\sigma$ errors: 0.23~dex for $n_{\rm{e}}$, 0.05~dex for $T_{\rm{e}}$[\ion{N}{ii}], and 0.03~dex for $T_{\rm{e}}$[\ion{O}{iii}].

\begin{figure*} 
\begin{center}
\includegraphics[width=0.9\textwidth, trim=30 10 20 10, clip=yes]{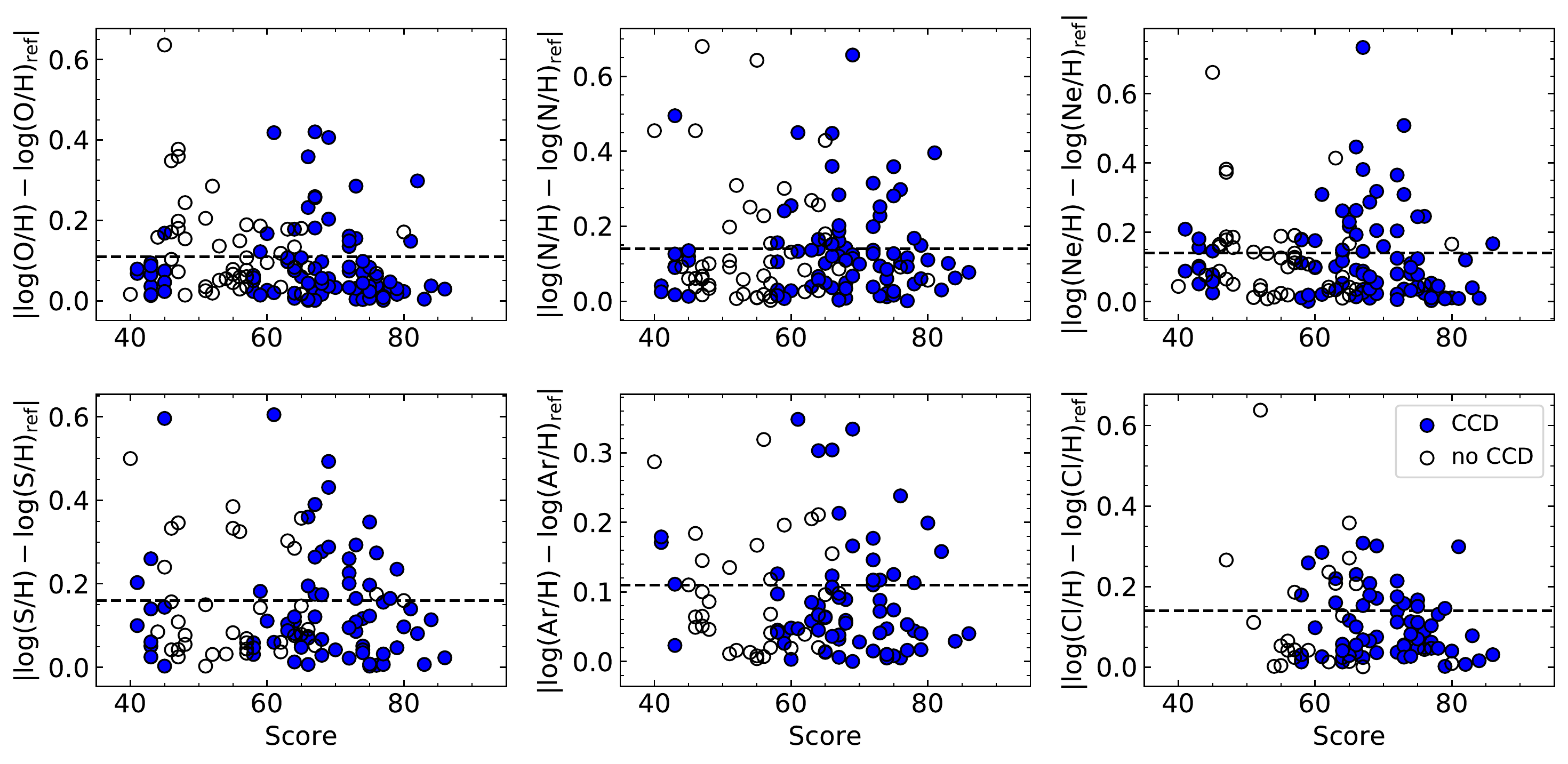}
\caption{Absolute values of the differences in chemical abundances implied by different spectra relative to the abundances implied by the reference spectrum as a function of the final scores of the spectra. Dashed lines show the one-$\sigma$ uncertainties at 0.11~dex for O/H, 0.14~dex for N/H, 0.14~dex for Ne/H, 0.16~dex for S/H, 0.11~dex for Ar/H, and 0.14~dex for Cl/H.} 
\label{figdab}
\end{center}
\end{figure*}

Fig.~\ref{figdab} shows the absolute values of the differences in the chemical abundances relative to the reference spectrum of each PN. These abundances have all been derived using the O$^+$ abundances implied by the blue [\ion{O}{ii}] lines. The differences decrease for higher scores, but not in a completely consistent way: it seems that a larger sample will be needed to relate a score with its observational uncertainty. The one-$\sigma$ observational uncertainties are equal to 0.11~dex for O/H, 0.14~dex for N/H, 0.14~dex for Ne/H, 0.16~dex for S/H, 0.11~dex for Ar/H, and 0.14~dex for Cl/H. These uncertainties are plotted with dashed lines in the figure. The uncertainties are clearly non-negligible and higher than those reported by many authors. Besides, several spectra introduce differences that reach or surpass factors of 2--4 in some of the abundance ratios.

In the case of helium (whose results are not plotted in Fig.~\ref{figdab}), the maximum difference in He/H~$\simeq$~He$^+$/H$^+ +$~He$^{++}$/H$^+$ is lower than 0.18~dex, and the one-$\sigma$ difference is equal to 0.05~dex. Part of these differences can be due to real variations in the ionization state of helium for the different areas of each PN sampled by different observations. Because of that, 0.05~dex should be considered an upper limit to the observational uncertainty in He/H. Something similar could be happening with the other elements if the ionization correction factors introduce biases that depend on the degree of ionization. However, since the abundance differences of the other elements decrease for spectra with higher scores, as shown in Fig.~\ref{figdab}, this is likely to be a minor effect.

A similar analysis can be performed for the abundances relative to oxygen, which leads to the following observational one-$\sigma$ uncertainties: 0.17~dex for N/O, 0.09~dex for Ne/O, 0.18~dex for S/O, 0.15~dex for Ar/O, and 0.14~dex for Cl/O.

The behaviour of the spectra which were not observed with a CCD, which are shown as open circles in Figs.~\ref{figdnt} and \ref{figdab}, seems to be broadly consistent with the scores assigned to these spectra (10 points lower than those given to CCD spectra), although many of the non-CCD spectra show low differences with their reference spectra, suggesting that their measurements can be quite reliable.

\subsection{The best spectra}
\label{best}

There are seven spectra that have final scores higher than 90: the one observed by \citet{SWBvH03} for IC~418 and the spectra of \citet{GPMDR12} for Cn~1-5, He~2-86, M~1-25, NGC~6369, PC~14, and Pe~1-1. All of these are deep, high-spectral resolution spectra that cover a wide wavelength range, and were obtained with echelle spectrographs in 4-m and 6.5-m telescopes. Thus, it is not surprising that these spectra allow the measurement of all the lines required to achieve a high score.

However, there are other spectra that were observed in a similar way and do not get scores higher than 90. As a case in point, the sample includes three other spectra of \citet{GPMDR12} that have final scores lower than 90: those for Hb~4, M~1-32 and M~3-15. The spectra for Hb~4 (score of 89) and M~1-32 (score of 80) do not include measurements of all the lines required to achieve the highest scores, but the M~3-15 spectrum, with a score of 80, does. In this case, the relatively low score is mainly due to problems with the relative intensities of the \ion{H}{I} lines at $\lambda<4400$~\AA. Below this wavelength, the \ion{He}{I} and \ion{H}{I} start to depart from their expected values. This only affects slightly the \ion{He}{I} lines at 4026 and 4388~\AA, but the bluest \ion{H}{I} lines are seriously affected and the spectrum gets a score of zero (out of 15) in the part based on the behaviour of the \ion{H}{I} lines. This suggest that there might be a problem with the flux calibration at the blue end of the spectrum. However, this spectrum was flux calibrated in the same way as the one observed by the same authors for Cn~1-5 (Garc\'ia-Rojas, private communication), which, with a score of 97, is one of the best spectra in the sample. Besides, both objects were observed at very low airmasses. This illustrates how easily observational errors of uncertain origin can creep into our results.

On the other hand, there are many spectra that do not include measurements of all the lines required to obtain a high score in the procedure outlined above, but that are good enough to derive reliable chemical abundances. In those cases where the reference spectrum of each object provides reliable chemical abundances, these spectra can be identified by comparing their abundances to those derived from the reference spectrum. For each object, I have compared the abundances of O/H, N/H, and S/H, derived with either the blue or red [\ion{O}{II}] lines, with those obtained with the reference spectrum. The O/H abundance ratio is chosen because it is generally used as a measure of the metallicity of PNe; N/H and S/H because these ratios are very sensitive to observational errors, as shown above, and can be estimated in most objects. A total of 23 spectra lead to O/H, N/H, and S/H abundance ratios that are within $\pm0.1$~dex of those derived using the blue [\ion{O}{II}] lines in the corresponding reference spectrum. These spectra are identified with a `$\dagger$' in Tables~\ref{tab:neTe} to \ref{tab:scores}. In the case of the red [\ion{O}{II}] lines, there are 17 spectra whose abundance ratios are within $\pm0.1$~dex of the reference ones. These spectra are identified with a `$\ddagger$' in Tables~\ref{tab:neTe} to \ref{tab:scores}. There are spectra in common in the two samples, with 11 spectra that show agreement with the reference abundances derived using either the blue or red [\ion{O}{II}] lines. The scores of all these spectra span the full range covered by non-reference spectra, reaching values as low as 43. Note that the 20 PNe where two or more spectra lead to abundances very similar to those of the reference spectrum can be considered the ones where the derived chemical abundances are more reliable.

\section{Summary and conclusions}

A set of 179 optical spectra of 42 Galactic PNe compiled from the literature has been used to explore the effects of observational uncertainties on the physical conditions and chemical abundances derived for these objects.

The results based on the [\ion{O}{ii}] and [\ion{N}{ii}] lines have been corrected by the effects of recombination on the line intensities by including the effective recombination coefficients in the statistical equilibrium equations solved for these ions. Previous approaches to this problem considered the effects of recombination and collisional excitations separately and do not lead to the same results. The recombination corrections start to be important at O$^{++}$/O$^+>30$ and affect mainly the final abundances of N, S, and Cl (see Figs.~\ref{figrec} and \ref{figrec2}).

The O$^+$ abundances derived with the red [\ion{O}{ii}] lines at $\lambda\sim7325$~\AA\ are found to be higher than those obtained with the blue lines at $\lambda\sim3727$~\AA. The most likely explanation of this result is that it is related to the density structure of these objects or to the choice of atomic data involved in the calculation of this parameter. However, the relative O$^+$ abundances implied by the red and blue [\ion{O}{ii}] lines show a large dispersion that cannot be explained with these considerations (see Fig.~\ref{figbr1}). The dispersion is larger for spectra of poor quality, where it can reach discrepancies much larger than a factor of 2 (see the left panel of Fig.~\ref{figdoh}). This suggests that the relative intensities of lines that are not close in wavelength can be highly uncertain. Another implication is that the total abundances of elements where the ionization correction factor is based on the O ions can be very different when derived with either the blue or the red lines of [\ion{O}{ii}]. This is especially noticeable in the N abundances, where the differences are larger than 0.1~dex in most objects, and can be close to one order of magnitude in some cases (see Fig.~\ref{figbr2}).

The relative intensities of bright lines that are close in wavelength can also have important uncertainties, as illustrated by spread in the values measured for the [\ion{O}{iii}]~$\lambda5007/\lambda4959$ and [\ion{N}{ii}]~$\lambda6584/\lambda6548$ line ratios shown in Fig.~\ref{figcNO}.

The He$^+$ abundances implied by lines located at different wavelengths can also be used to explore the uncertainties introduced by the flux calibration or related effects (see Fig.~\ref{fighe}). The He$^+$ abundances derived with the \ion{He}{i}~$\lambda4026$ and $\lambda6678$ lines suggest that most of the spectra are bluer that expected (see Fig.~\ref{figdoh}).

The quality of the spectra has been assessed using several criteria: whether the spectrum was obtained with a CCD or with a detector that could have non-linearity problems; the number of available electron density and temperature diagnostics; the departures of the [\ion{O}{iii}]~$\lambda5007/\lambda4959$ and [\ion{N}{ii}]~$\lambda6584/\lambda6548$ intensity ratios from their expected values; and the behaviour of the \ion{He}{i} and \ion{H}{i} lines. These criteria have been used to assign a score to each spectrum that should provide an estimate of its quality.

The assigned scores are in the range 40--99, and the scores for the reference spectra (the one with the highest score for each PN) are in the range 68--99. The differences in abundances found for spectra having different scores suggest that reference spectra should ideally have scores higher than 80.

The spectrum with the highest score for each object has been used as a reference for the results, and the observational uncertainties affecting the total abundances have been estimated from the differences between the results obtained with different spectra and those implied by the reference spectrum. The derived one-$\sigma$ observational uncertainties are the following: 0.11~dex for O/H, 0.14~dex for N/H, 0.14~dex for Ne/H, 0.16~dex for S/H, 0.11~dex for Ar/H, and 0.14~dex for Cl/H. In the case of He/H, the one-$\sigma$ observational uncertainty should be lower than 0.05~dex.

These PNe can be considered the most easily observed PNe, since each of them has at least three available spectra with a minimum quality. The observational uncertainties affecting other objects might be larger.

In 20 of the 42 PNe in the sample, there is at least one additional spectrum that implies chemical abundances that differ in less than 0.1~dex from those derived with the reference spectrum. These 20 PNe are the ones with the more reliable chemical abundances.

Finally, many of the results found here should apply to spectra obtained in a similar way for objects other than PNe. In particular, the spectra of \ion{H}{ii} regions should be similarly affected.

\section*{Acknowledgements}
I thank R.~Manso Sainz, L.~Juan de Dios, G.~Dom\'inguez-Guzm\'an, and K.~Z.~Arellano C\'ordova for useful tips and discussions. I also thank the referee, B.~Melekh, for comments that improved this paper. I acknowledge support from Mexican CONACYT grant CB-2014-240562.



\bibliographystyle{mnras}
\bibliography{refs}



\appendix

\section{The behaviour of the He~I and H~I lines in some spectra}

\begin{figure*} 
\begin{center}
\includegraphics[width=0.9\textwidth, trim=30 10 20 10, clip=yes]{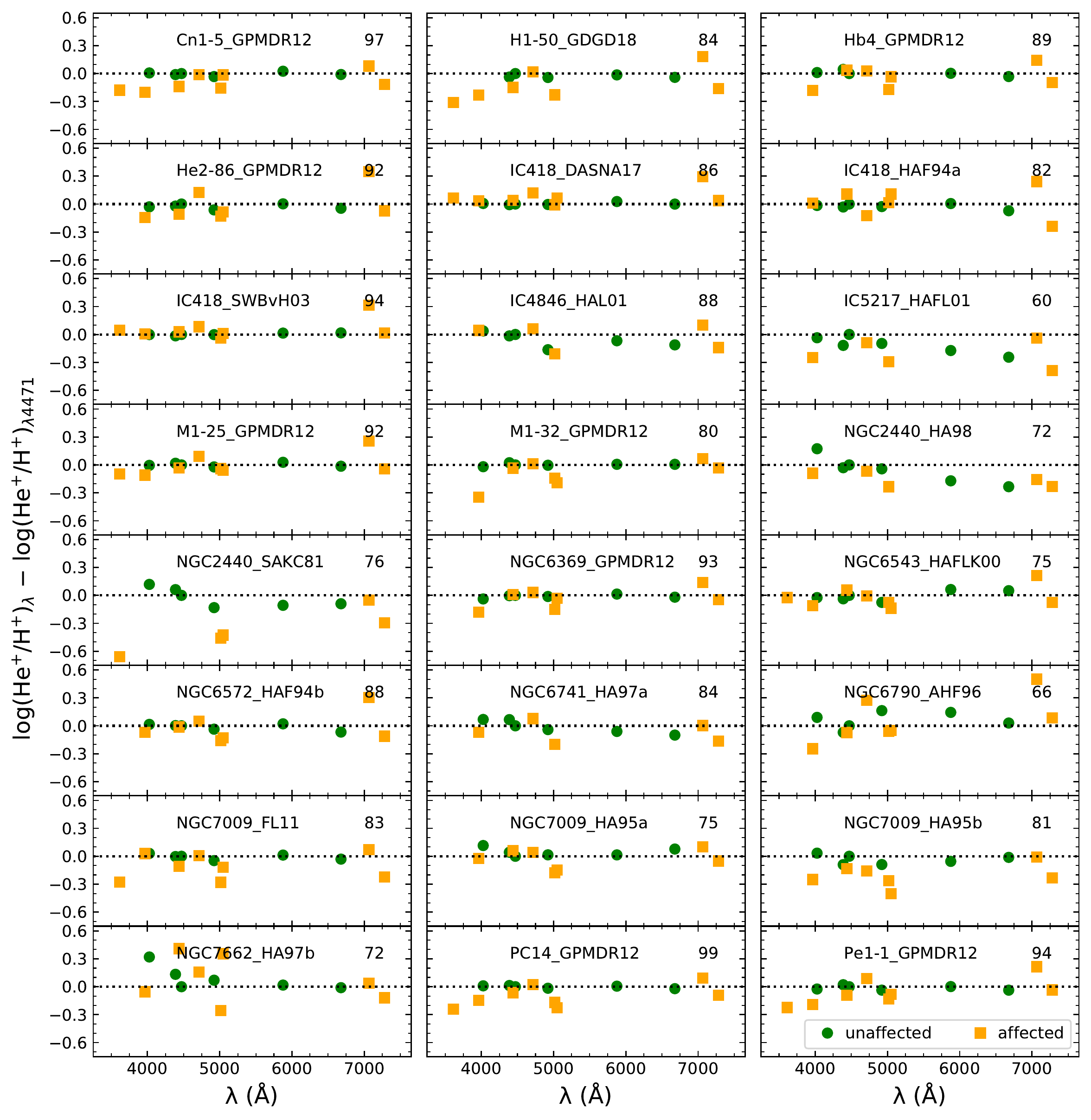}
\caption{Differences between the He$^+$ abundances obtained with different \ion{He}{i} lines and those derived with \ion{He}{i}~$\lambda4471$ as a function of wavelength for several sample spectra. Squares show the results for lines that are affected by optical depth effects; circles are for lines expected to be unaffected by these effects. The identification and score of each spectrum are shown in the panels (the references are provided in Table~\ref{tab:scores}).} 
\label{fighe}
\end{center}
\end{figure*}

\clearpage

\begin{figure*} 
\begin{center}
\includegraphics[width=0.9\textwidth, trim=30 10 20 10, clip=yes]{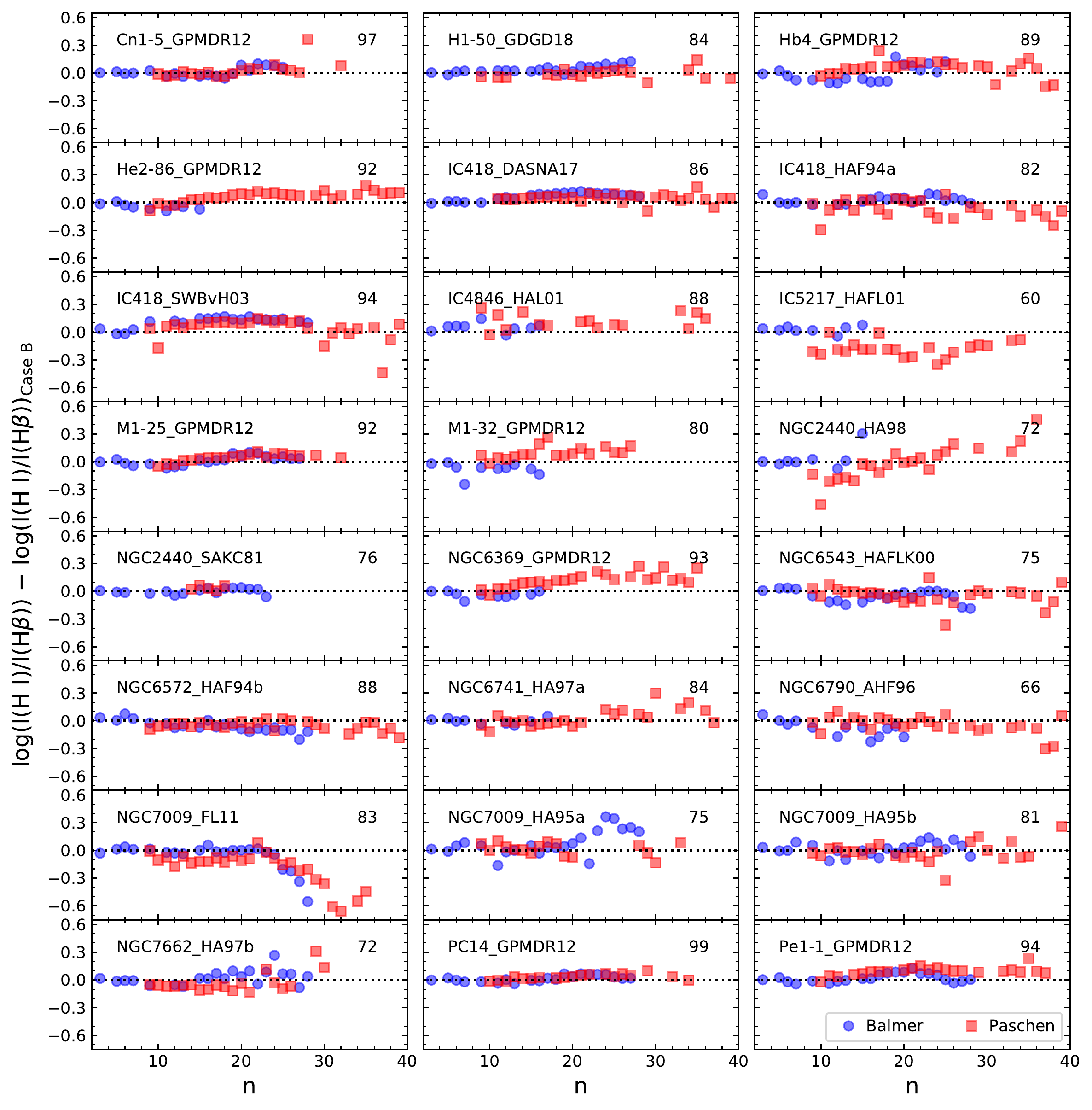}
\caption{Differences between the extinction-corrected intensities of \ion{H}{i} lines relative to H$\beta$ and their case~B values as a function of the principal quantum number $n$ of the upper levels of these lines for several sample spectra. Circles show the results obtained for the Balmer lines; squares are for the Paschen lines. The identification and score of each spectrum are shown in the panels (the references are provided in Table~\ref{tab:scores}).} 
\label{fighbp}
\end{center}
\end{figure*}

\clearpage

\section{Physical conditions, nebular abundances, and final scores}

\begin{table*}
\centering
\caption{Physical conditions of the sample PNe. The last column provides the final score of each spectrum. The spectra used as reference for each object are marked with asterisks. The references are identified in Table~\ref{tab:scores}.}
\label{tab:neTe}
\begin{tabular}{llrrrrrrrr}
\hline
PN & Ref. & $n_{\rm{e}}$[\ion{S}{ii}] & $n_{\rm{e}}$[\ion{O}{ii}] & $n_{\rm{e}}$[\ion{Cl}{iii}] & $n_{\rm{e}}$[\ion{Ar}{iv}] & $n_{\rm{e}}$ & $T_{\rm{e}}$[\ion{N}{ii}] & $T_{\rm{e}}$[\ion{O}{iii}] & $p_{\rm{tot}}$ \\
 & & (cm$^{-3}$) & (cm$^{-3}$) & (cm$^{-3}$) & (cm$^{-3}$) & (cm$^{-3}$) & (K) & (K) & \\
\hline
BoBn 1  & KHM03        & 8986  & --    & --    & --    & 8986  & 11932 & 12749 & 60 \\
        & KZGP08$^{*}$ & 4435  & --    & 13738 & --    & 7806  & 11456 & 13667 & 89 \\
        & OTHI10       & 4245  & 2916  & --    & 2573  & 3175  & 12008 & 13544 & 73 \\
Cn 1-5  & BK13	       & 3639  & --    & 3389  & --    & 3512  & 8201  & 9003  & 84  \\
        & EBW04        & 5977  & 5812  & --    & --    & 5905  & 8342  & 8433  & 59 \\
	& GPMDR12$^{*}$& 3773  & 5286  & 3889  & 10087 & 5291  & 8604  & 8759  & 97 \\
	& PSM01        & 3672  & 2836  & --    & --    & 3227  & 8755  & 10815 & 43  \\
        & WL07         & 1319  & --    & 3374  & 1333  & 1810  & 7399  & 8709  & 80 \\
DdDm1	& BC84$^{\dagger\ddagger}$& 3584  & --    & --    & --    & 3584  & 12521 & 11713 & 48 \\
        & CPT87$^{\dagger}$& 3901  & --    & --    & --    & 3901  & 11158 & 11744 & 57  \\
        & KH98$^{*}$   & 3504  & --    & 5043  & --    & 4203  & 12371 & 12085 & 74  \\
        & OHLIT09      & 3628  & 6592  & 5914  & 7668  & 5739  & 12502 & 11692 & 65  \\
H1-50	& EBW04	       & 2999  & --    & --    & --    & 2999  & 13269 & 11368 & 67  \\
        & GDGD18$^{*}$ & 6816  & 10959 & 13339 & 2455  & 7030  & 12165 & 11113 & 84  \\
        & WL07         & 5517  & 9659  & 10651 & 12220 & 9130  & 11666 & 10889 & 78 \\
H1-54	& EBW04	       & 13583 & --    & 19652 & --    & 16338 & 10737 & 8836  & 66 \\
        & GCSC09       & 9082  & --    & 18584 & --    & 12991 & 14076 & 9616  & 58 \\
        & WL07$^{*}$   & 9262  & --    & 17012 & 460   & 4169  & 13267 & 9576  & 87 \\
H4-1	& KHM03	       & 346   & --    & --    & --    & 346   & 11638 & 12830 & 68  \\
        & OT13$^{*}$   & 837   & 732   & 2249  & 1978  & 1287  & 10768 & 13187 & 84  \\
        & TP79         & 628   & --    & --    & --    & 628   & 11547 & 11992 & 57  \\
Hb4     & AK87$^{\dagger}$& 6368  & --    & 3483  & --    & 4709  & 10134 & 9655  & 55 \\
        & GKA07        & 5434  & --    & 7880  & --    & 6544  & 9875  & 9477  & 74 \\
        & GPMDR12$^{*}$& 5977  & 8535  & 6816  & 7384  & 7121  & 9491  & 9935  & 89 \\
        & MKHS10       & 5533  & --    & 6524  & --    & 6008  & 11012 & 9160  & 72 \\
        & PSM01        & 4075  & 2410  & --    & 5166  & 3742  & 10612 & 9508  & 45 \\
He2-86  & G14$^{\ddagger}$& 11785 & --    & 11011 & --    & 11392 & 10705 & 8615  & 76\\
        & GKA07        & 10334 & --    & 73362 & --    & 27534 & 8817  & 8502  & 64 \\
        & GPMDR12$^{*}$& 15632 & 14957 & 22244 & 35717 & 20731 & 10122 & 8387  & 92 \\
Hu1-2   & AC83         & 5115  & --    & 6524  & --    & 5777  & 11978 & 16847 & 56 \\
        & FGMR15       & 3611  & --    & 5104  & --    & 4293  & 12596 & 16581 & 69  \\
        & HPF04        & 4219  & 7360  & 4435  & 9944  & 6084  & 13502 & 20709 & 64  \\
        & LLLB04       & 3525  & 7108  & 4776  & 3457  & 4512  & 12963 & 19121 & 67   \\
        & MKHS10       & 3493  & --    & 6244  & --    & 4670  & 13196 & 16824 & 68   \\
        & PT87         & 3525  & --    & --    & --    & 3525  & 13598 & 18782 & 60  \\
        & SCT87$^{*}$  & 4870  & --    & 7219  & 1355  & 3625  & 12716 & 16657 & 79   \\
Hu2-1   & AC83         & --    & --    & 3143  & --    & 3143  & 13446 & 9181 & 52 \\
        & F81          & 8485    & --    & --    & --    & 8485    & 15462   & 11529 & 45  \\
        & KHM03$^{*}$  & 45827   & --    & 40183   & --    & 42912   & 10340   & 8931 & 77    \\
        & WLB05        & --    & 14657   & --    & --    & 14671   & 11423   & 10782 & 69   \\
IC418   & AC83$^{\dagger\ddagger}$& 48939 & 1312 & 6292 & -- & 7416 & 8936 & 8406 & 57 \\
        & DASNA17$^{\dagger\ddagger}$& 11280   & 19175   & 13666   & --    & 14353  & 9504 & 8725 & 86    \\
        & HAF94a       & 5860    & 18701   & 13219   & --    & 11309   & 9804    & 9602 & 82    \\
        & SWBvH03$^{*}$& 15030   & 16012   & 12444   & 4841    & 10974   & 9459    & 8769 & 94    \\
IC2003  & AC83         & 6114 & -- & 4232 & -- & 5087 & 13172 & 11529 & 51 \\
        & B78          & 1613    & --   & --   & --   & 1613    & 16976   & 13223 & 53   \\
        & KB94         & 8485    & --   & 2242    & --   & 4362    & 13788   & 12063 & 47   \\
        & WLB05$^{*}$  & 3773    & 3959    & 1753    & 2130    & 2737    & 13244   & 12201 & 75   \\
IC4846  & AC79         & 3773  & 13375 & 10256 & -- & 8042 & 13218 & 9871 & 57 \\
        & B78          & 3124    & --   & --   & --   & 3124    & 9535    & 10070 & 53   \\
        & HAL01$^{*}$  & 4974    & 22088   & 9014    & 7340    & 9250    & 11789   & 10447 & 88   \\
        & WL07         & 5241    & --   & 5720    & 6368    & 5758    & 12799   & 9880 & 76    \\
IC5217  & AC79         & 8745 & 9520 & 5127 & -- & 7529 & 11467 & 11194 & 65 \\
        & B78          & 1272    & --   & --   & --   & 1272    & 10626   & 11358 & 51   \\
        & F81          & 2212    & --   & --   & 2089    & 2149    & 13781   & 12184 & 48   \\
        & HAFL01       & 3525    & 5105    & 3865    & 4349    & 4189    & 12312   & 10618 & 60   \\
        & KH01$^{*}$   & 6983    & --   & 7175    & --   & 7078    & 12526   & 11219 & 73   \\
        & PSM01        & 9109    & 12366   & --   & 3457    & 7292    & 12701   & 11038 & 45   \\
M1-25   & GCSC09$^{\dagger}$& 5475 & -- & 14917 & -- & 9037 & 8307 & 8004 & 58 \\
        & GKA07$^{\dagger\ddagger}$& 6868    & --    & 9137    & --    & 7921    & 8019    & 7850 & 74    \\
        & GPMDR12$^{*}$& 7904    & 13500   & 14042   & --    & 11426   & 8307    & 7794 & 92    \\
        & MKHC02       & 3972    & --    & 16808   & --    & 8171    & 8389    & 8595 &  72   \\
\hline
\end{tabular}
\end{table*}

\begin{table*}
\centering
\contcaption{Physical conditions of the sample PNe. The last column provides the final score of each spectrum. The spectra used as reference for each object are marked with asterisks. The references are identified in Table~\ref{tab:scores}.}
\begin{tabular}{llrrrrrrrr}
\hline
PN & Ref. & $n_{\rm{e}}$[\ion{S}{ii}] & $n_{\rm{e}}$[\ion{O}{ii}] & $n_{\rm{e}}$[\ion{Cl}{iii}] & $n_{\rm{e}}$[\ion{Ar}{iv}] & $n_{\rm{e}}$ & $T_{\rm{e}}$[\ion{N}{ii}] & $T_{\rm{e}}$[\ion{O}{iii}] & $p_{\rm{tot}}$ \\
 & & (cm$^{-3}$) & (cm$^{-3}$) & (cm$^{-3}$) & (cm$^{-3}$) & (cm$^{-3}$) & (K) & (K) & \\
\hline
M1-25 & PSM01$^{\dagger}$& 8865    & 9083    & --    & --    & 8973    & 7829    & 7868 &  43   \\
M1-29 & EBW04 & 3234 & -- & 2901 & -- & 3063 & 8182 & 10061 & 64\\
   & GCSC09$^{\dagger\ddagger}$& 2870    & --    & 6022    & --    & 4158    & 9471    & 10835 & 58\\
   & WL07$^{*}$&  2500    & --    & 5127    & 4100    & 3746    & 9206    & 10761 & 73\\
M1-32 & GKA07 & 7927 & -- & 12729 & -- & 10045 & 8603 & 10744 & 73\\
   & GPMDR12$^{*}$&  8485    & 9679    & 13895   & --    & 10449   & 9051    & 9404 & 80\\
   & PSM01   & 5660    & 3523    & --    & --    & 4467    & 9265    & 9871 & 43\\
M1-74 & AC83 & 16732 & -- & 31342 & -- & 22900 & 11883 & 9477 & 55\\
   & KH01$^{*}$&  2117    & --    & --    & --    & 2117    & 15401   & 9546 & 76\\
   & WLB05   & 13441   & 12710   & --    & 41416   & 19168   & 12694   & 9743 & 66\\
M2-23 & EBW04 & 6244 & -- & 4606 & -- & 5363 & 18809 & 11788 & 66\\
   & GCSC09  & 15751   & --    & 13339   & --    & 14495   & 22605   & 12107 & 58\\
   & WL07$^{*}$&  12074   & --    & 16456   & 52818   & 21894   & 19709   & 11782 & 74\\
M3-15 & MKHC02 & 3912 & -- & 19183 & -- & 8663 & 14560 & 8295 & 59\\
   & GKA07   & 4697    & --    & 13502   & --    & 7963    & 10665   & 10849 & 61\\
   & GPMDR12$^{*}$&  5720    & 14803   & 9389    & 7668    & 8817    & 10218   & 8338 & 80\\
   & PSM01   & 7362    & 31067   & --    & --    & 15064   & 10249   & 8592 & 43\\
Me2-2 & AK87 & 2026 & -- & 6673 & -- & 3677 & 11907 & 10357 & 66\\
   & B78     & 561     & --    & --    & --    & 561     & 13927   & 10786 & 56\\
   & MKHS10$^{*}$&   5609    & --    & 39640   & --    & 14911   & 12246   & 10120 & 70 \\
   & WLB05   & 773     & 23601   & --    & 22685   & 7448    & 12804   & 11949 & 67 \\
NGC40 & AC79 & 966 & -- & 1957 & -- & 1375 & 8209 & 16977 & 56 \\
   & CSPT83$^{\ddagger}$& 1793    & --    & 5467    & --    & 3131    & 7792    & 10608 & 67 \\
   & LLLB04$^{*}$&   1424    & 1386    & 970     & --    & 1242    & 8451    & 10423 & 77 \\
   & PSM01$^{\dagger}$& 1584    & 1456    & --    & --    & 1518    & 8385    & 9716 & 43\\
NGC650 & AC83 & 295 & -- & -- & -- & 295 & 9782 & 10638 & 51 \\
  &   KHM03$^{*}$&    198   &   --   &   --   &   --   &   198   &   10790   &   11539 & 68 \\
  &   PT87    &   266   &   --   &   --   &   --   &   266   &   10236   &   11927 & 64 \\
NGC2392 & B91 & 635 & -- & 628 & -- & 631 & 9588 & 13581 & 54\\
 &  DRMV09$^{\dagger\ddagger}$&  1984  &  --  &  970  &  --  &  1388  &  12253   &  14431   & 68\\
 &  HKB00$^{\ddagger}$&  664  &  --  &  --  &  --  &  664  &  10679   &  13575   & 63\\
 &  Z76  &  546  &  993  &  --  &  --  &  740  &  16696   &  14829   & 40\\
 &  ZFCH12$^{*}$&   678  &  --  &  1262  &  --  &  925  &  10401   &  13575   & 71\\
NGC2440 & DKSH15 & 3639 & -- & -- & -- & 3639 & 10105 & 15169 & 67\\
 &  HA98  &  2612  &  5692  &  3420  &  3942  &  3767  &  9728  &  14006   & 72\\
 &  KB94  &  6653  &  --  &  --  &  --  &  6653  &  8690  &  13643   & 44\\
 &  KC06$^{*}$&  1333  &  --  &  3984  &  --  &  2304  &  11220   &  14431   & 80\\
 &  KHM03   &  1544  &  --  &  33546   &  --  &  7197  &  10667   &  13074   & 67\\
 &  PT87$^{\dagger}$&  2429  &  --  &  10334   &  --  &  5010  &  10897   &  14802   & 62\\
 &  SAKC81  &  2777  &  --  &  4232  &  --  &  3428  &  10347   &  14182   & 76\\
NGC2867 & AKRO81 & 1496 & -- & 6663 & -- & 3157 & 9317 & 11250 & 62\\
 &  GKA07   &  2462  &  --  &  3942  &  --  &  3116  &  10626   &  11798   & 66\\
 &  GPP09$^{*}$&  2130  &  3139  &  4309  &  3972  &  3274  &  11199   &  11497   & 82\\
 &  KB94  &  2382  &  --  &  8085  &  --  &  4388  &  9740  &  11029   & 57\\
 &  MKHC02$^{\dagger}$&  1812  &  --  &  3639  &  --  &  2568  &  10556   &  11497   & 79\\
 &  PSEK98  &  2182  &  --  &  --  &  --  &  2182  &  11864   &  11798   & 41\\
NGC6210 & B78 & 2628 & -- & -- & -- & 2628 & 9708 & 9597 & 52\\
 &  BERD15$^{\dagger}$&  2440  &  15057   &  3924  &  4336  &  4998  &  9969  &  9490  & 77\\
 &  DRMV09  &  3167  &  --  &  3960  &  --  &  3541  &  11215   &  9912  & 63\\
 &  F81  &  2798  &  --  &  --  &  2588  &  2691  &  11897   &  9840  & 47\\
 &  KH98$^{\ddagger}$&  3825  &  --  &  4676  &  --  &  4229  &  10795   &  9511  & 61\\
 &  LLLB04$^{*}$&   3133  &  5716  &  3694  &  6197  &  4506  &  10639   &  9503  & 80\\
NGC6302 & KC06 & 1716 & -- & 8865 & -- & 3901 & 13402 & 17476 & 80\\
 &  MKHS10  &  2862  &  --  &  7809  &  --  &  4727  &  12677   &  17476   & 67\\
 &  RCK14   &  2130  &  --  &  3813  &  --  &  2850  &  13487   &  17588   & 80\\
 &  TBLDS03 &  8121  &  8868  &  25793   &  13120   &  12493   &  13635   &  18445   & 80\\
NGC6369 & AK87 & 5816 & -- & 12257 & -- & 8444 & 11994 & 11639 & 55\\
 &  GKA07   &  1456  &  --  &  --  &  --  &  1456  &  8857  &  8007  & 67\\
 &  GPMDR12$^{*}$&  3504  &  4152  &  4245  &  5043  &  4210  &  12963   &  10632   & 93\\
 &  MKHS10  &  2400  &  --  &  3472  &  --  &  2887  &  10714   &  9694  & 67\\
 &  PSM01   &  2879  &  2660  &  --  &  4530  &  3273  &  11184   &  9970  & 45\\
NGC6543 & AC79 & 2981 & -- & 6058 & -- & 4250 & 9239 & 8189 & 64\\
\hline
\end{tabular}
\end{table*}

\begin{table*}
\centering
\contcaption{Physical conditions of the sample PNe. The last column provides the final score of each spectrum. The spectra used as reference for each object are marked with asterisks. The references are identified in Table~\ref{tab:scores}.}
\begin{tabular}{llrrrrrrrr}
\hline
PN & Ref. & $n_{\rm{e}}$[\ion{S}{ii}] & $n_{\rm{e}}$[\ion{O}{ii}] & $n_{\rm{e}}$[\ion{Cl}{iii}] & $n_{\rm{e}}$[\ion{Ar}{iv}] & $n_{\rm{e}}$ & $T_{\rm{e}}$[\ion{N}{ii}] & $T_{\rm{e}}$[\ion{O}{iii}] & $p_{\rm{tot}}$ \\
 & & (cm$^{-3}$) & (cm$^{-3}$) & (cm$^{-3}$) & (cm$^{-3}$) & (cm$^{-3}$) & (K) & (K) & \\
\hline
NGC6543 &  HAFLK00 &  --  &  17124   &  5533  &  2422  &  6090  &  8262  &  8001  & 75\\
 & PSM01$^{\dagger}$&  5154  &  12609   &  --  &  4990  &  6820  &  9283  &  7833  & 45\\
 & WL04$^{*}$& 4907 & 6191 & 5643 & 3091 & 4808 & 9347 & 7804  & 80\\
NGC6565 & EBW04 & 1401 & -- & 2104 & -- & 1717 & 9588 & 10899 & 65\\
 &  MKHC02$^{*}$&   1179  &  --  &  2225  &  --  &  1620  &  10249   &  10414   & 82\\
 &  WL07  &  1428  &  --  &  1807  &  387  &  999  &  10306   &  10226   & 77\\
NGC6572 & F81 & 4323 & -- & -- & 11661 & 7100 & 13905 & 11992 & 47\\
 &  GKA07   &  17481   &  --  &  13853   &  --  &  15562   &  11858   &  10157   & 75\\
 &  HAF94b$^{*}$&   10118   &  71091   &  21003   &  23345   &  24230   &  11971   &  10098   & 88\\
 &  KH01  &  9360  &  --  &  14386   &  --  &  11604   &  14050   &  10185   & 75\\
 &  LLLB04  &  14671   &  28646   &  18810   &  15121   &  18577   &  11928   &  10190   & 79\\
NGC6720 & B80 & 613 & -- & -- & -- & 613 & 9558 & 9379 & 47\\
 &  F81  &  574  &  --  &  --  &  784  &  670  &  9946  &  11013   & 46\\
 &  GMC97$^{\dagger\ddagger}$ &  537  &  --  &  --  &  --  &  537  &  10574   &  10889   & 70\\
 &  HM77  &  456  &  --  &  --  &  --  &  456  &  9272  &  9567  & 48\\
 &  KH98$^{\dagger\ddagger}$&  506  &  --  &  803  &  --  &  637  &  10453   &  10691   & 73\\
 &  LLLB04$^{*}$&   431  &  481  &  574  &  763  &  551  &  10169   &  10466   & 78\\
NGC6741 & HA97a$^{*}$& 4776 & 8594 & 8233 & 5345 & 6520 & 10851 & 12509 & 84\\
 &  LLLB04$^{\dagger\ddagger}$&  3942  &  6443  &  5584  &  7121  &  5640  &  10762   &  12345   & 77\\
 &  MKHS10  &  4697  &  --  &  5353  &  2932  &  4193  &  11172   &  11296   & 78\\
NGC6751 & AC79 & 2879 & -- & 224 & -- & 802 & 8247 & 10217 & 63\\
 &  CMJK91  &  1999  &  --  &  2959  &  --  &  2432  &  8713  &  10418   & 75\\
 &  KB94  &  1112  &  --  &  --  &  --  &  1112  &  8144  &  9541  & 46\\
 &  MKHS10$^{*}$&   1416  &  --  &  471  &  --  &  817  &  8851  &  9916  & 80\\
 &  PSM01   &  2192  &  1336  &  --  &  --  &  1721  &  9431  &  11053  &  43\\
NGC6790 & AC79 & 9728 & -- & 6765 & -- & 8112 & 17285 & 11644 & 59\\
 &  AHF96   &  11195   &  --  &  63699   &  49497   &  32803   &  12308   &  10919   & 66\\
 &  F81  &  8121  &  --  &  --  &  17245   &  11834   &  17351   &  13848   & 46\\
 &  KH01$^{*}$&  --  &  --  &  2489  &  --  &  2489  &  19657   &  12825   & 74\\
 &  LLLB04$^{\ddagger}$&  24280   &  25108   &  24280   &  62934   &  30995   &  14378   &  12743   & 69\\
NGC6884 & AC79 & 5677 & 11068 & 5475 & -- & 7006 & 11719 & 10786 & 65\\
 &  HAF97   &  15328   &  --  &  10555   &  6475  &  10156   &  9410  &  9707  & 73\\
 &  KH01$^{*}$&  6962  &  --  &  6983  &  --  &  6972  &  11134   &  10667   & 75\\
 &  LLLB04$^{\ddagger}$&  5860  &  8847  &  6197  &  9750  &  7485  &  11308   &  10889   & 72\\
NGC7009 & CA79 & 4362 & 7582 & 4349 & 3441 & 4726 & 9997 & 9044 & 80\\
 &  F81  &  20657   &  --  &  --  &  1355  &  5290  &  13322   &  12074   & 47\\
 &  FL11  &  3200  &  5710  &  4008  &  4245  &  4212  &  11149   &  9837  & 83\\
 &  HA95a$^{\dagger}$&  --  &  4363  &  4990  &  2249  &  3671  &  8855  &  9699  & 75\\
 &  HA95b   &  4303  &  21951   &  5896  &  2901  &  6318  &  13787   &  10849   & 81\\
 &  KC06$^{*}$&  2879  &  --  &  4175  &  --  &  3467  &  10445   &  9912  & 85\\
 &  KH98  &  3167  &  --  &  3595  &  --  &  3374  &  9690  &  9563  & 68\\
NGC7662 & AC83 & 2256 & 3392 & 3514 & -- & 3015 & 9676 & 13446 & 65\\
 &  HA97b$^{*}$&  3045  &  5248  &  4571  &  2825  &  3807  &  14219   &  12600   & 72\\
 &  LLLB04  &  2364  &  3191  &  2199  &  2318  &  2502  &  11956   &  13086   & 69\\
PB6 & GKA07$^{\dagger}$& 2159 & -- & -- & -- & 2159 & 10959 & 15552 & 64\\
     & GPP09$^{*}$& 2104    & 2988    & --      & 1812    & 2251    & 12367   & 15439   & 74\\
     & MKHS10  & 1599    & --      & 1179    & --      & 1373    & 11019   & 14379   & 72\\
     & PSEK98  & 1272    & --      & --      & --      & 1272    & 10540   & 14146   & 41\\
PC14 & GKA07$^{\dagger\ddagger}$& 2538 & -- & 5695 & -- & 3802 & 9603 & 9029 & 74\\
    &  GPMDR12$^{*}$&  3091    &  4456    &  3514    &  4697    &  3892    &  9904    &  9285    & 99\\
    &  MKHC02$^{\dagger\ddagger}$&  1999    &  --      &  12691   &  --    &  5037    &  8498 &  8988 & 74\\
Pe1-1 & BK13 & 6123 & -- & 72591 & -- & 21082 & 10696 & 10190 & 64\\
   & G14     & 7068    & --    & 10978   & --    & 8808    & 12268   & 10556   & 72\\
   & GKA07$^{\dagger}$& 21290   & --    & 23772   & --    & 22497   & 11265   & 9528    & 68\\
   & GPMDR12$^{*}$&  14042   & 40061   & 30965   & 40244   & 28958   & 11024   & 9956    & 94\\
\hline
\end{tabular}
\\
\raggedright
$^\dagger$ O/H, N/H, and S/H are within 0.1~dex of the reference spectrum for calculations based on the blue [\ion{O}{ii}] lines.\\
$^\ddagger$ O/H, N/H, and S/H are within 0.1~dex of the reference spectrum for calculations based on the red [\ion{O}{ii}] lines.\\
NOTE. In those cases where line intensities were presented for more than one region in the PN, the position with the largest number of measured lines was selected.\\
\end{table*}

\begin{table*}
\centering
\caption{Ionic abundances derived from collisionally excited lines, where $X^{i+}=12+\log(X^{i+}/\mbox{H}^+)$. O$^+$(b) and O$^+$(r) are the abundances calculated using the blue and red [\ion{O}{ii}] lines. The last column provides the final score of each spectrum. The spectra used as reference for each object are marked with asterisks. The references are identified in Table~\ref{tab:scores}.}
\label{tab:ion}
\begin{tabular}{llrrrrrrrrrr}
\hline
PN & Ref. & O$^+$(b) & O$^+$(r) & O$^{++}$ & N$^+$ & Ne$^{++}$ & S$^+$ & S$^{++}$ & Ar$^{++}$ & Cl$^{++}$ & $p_{\rm{tot}}$ \\
\hline
BoBn 1  & KHM03   & 6.72 & 6.43 & 7.70 & 6.61 & 8.04 & 4.15 & --   & 4.03 & -- & 60\\
        & KZGP08$^{*}$  & 6.87 & 6.72 & 7.67 & 6.81 & 7.96 & 4.08 & 4.99 & 4.07 & 3.18 & 89\\
        & OTHI10  & 6.65 & 6.74 & 7.69 & 6.82 & 7.99 & 3.97 & 4.57 & 4.12 & 3.19 & 73\\
Cn 1-5  & BK13	  & 8.22 & 8.43 & 8.61 & 8.11 & 8.30 & 6.45 & 6.99 & 6.61 & 5.42 & 84\\
        & EBW04   & 8.28 & 8.33 & 8.81 & 8.07 & 8.37 & 6.50 & --   & 6.58 & -- & 59\\
   & GPMDR12$^{*}$& 8.16    & 8.20    & 8.69    & 8.07    & 8.37    & 6.51    & 7.08    & 6.62    & 5.42 & 97\\
   & PSM01   & 8.57    & --    & 8.31    & 8.02    & 8.05    & 6.35    & --    & --    & -- & 43\\
   & WL07    & 8.20    & 8.51    & 8.70    & 8.20    & 8.36    & 6.35    & 7.01    & 6.43    & 5.46 & 80\\
DdDm1 & BC84$^{\dagger\ddagger}$& 7.43 & 7.51 & 7.95 & 6.77 & 7.36 & 5.39 & 6.37 & 5.59 & -- & 48\\
   & CPT87$^{\dagger}$& 7.62    & 7.91    & 7.94    & 6.94    & 7.33    & 5.46    & 6.39    & 5.69    & -- & 57\\
   & KH98$^{*}$& 7.43    & 7.60    & 7.93    & 6.82    & 7.30    & 5.38    & 6.29    & 5.68    & 4.25 & 74\\
   & OHLIT09 & 7.58    & 7.10    & 7.96    & 6.81    & 7.46    & 5.44    & 6.25    & 5.71    & 4.37 & 65\\
H1-50 & EBW04 & 6.88 & 7.14 & 8.60 & 6.83 & 8.14 & 5.51 & 6.53 & 6.07 & -- & 67\\
   & GDGD18$^{*}$& 7.23    & 7.39    & 8.60    & 6.89    & 8.10    & 5.59    & --    & 6.08    & 4.66 & 84\\
   & WL07    & 7.32    & 7.40    & 8.63    & 6.99    & 8.12    & 5.70    & 6.62    & 6.04    & 4.80 & 78\\
H1-54 & EBW04 & 7.93 & 8.18 & 8.41 & 7.22 & 7.70 & 5.62 & 6.81 & 6.24 & 4.54 & 66\\
   & GCSC09  & 7.28    & 7.24    & 8.28    & 6.71    & 7.60    & 5.22    & 6.69    & 5.97    & 4.37 & 58\\
   & WL07$^{*}$& 7.27    & 7.54    & 8.26    & 6.82    & 7.56    & 4.97    & 6.67    & 5.85    & 4.36 & 87\\
H4-1 & KHM03 & 7.60 & 7.77 & 8.02 & 7.18 & 6.44 & 4.60 & 5.25 & 4.33 & -- & 68\\
    & OT13$^{*}$& 7.76    & 7.66    & 8.00    & 7.14    & 6.41    & 4.47    & 4.94    & 4.25    & 3.81 & 84\\
    & TP79    & 7.68    & 7.72    & 8.17    & 7.14    & 6.57    & 4.45    & --      & --      & -- & 57\\
Hb4 & AK87$^{\dagger}$& 7.25 & 7.41 & 8.68 & 7.15 & 8.14 & 5.69 & 6.83 & 6.33 & 5.05 & 55\\
     & GKA07   & 7.42    & 7.30    & 8.69    & 7.32    & 8.26    & 5.89    & 6.90    & 6.36    & 4.99 & 74\\
     & GPMDR12$^{*}$& 7.35    & 7.63    & 8.60    & 7.31    & 8.16    & 5.83    & 6.77    & 6.33    & 5.09 & 89\\
     & MKHS10  & 7.21    & 7.33    & 8.77    & 7.23    & 8.27    & 5.81    & 7.00    & 6.44    & 4.91 & 72\\
     & PSM01   & 7.31    & --      & 8.71    & 7.31    & 8.05    & 5.85    & 6.92    & --      & -- & 45\\
He2-86 & G14$^{\ddagger}$& 7.41 & 7.44 & 8.71 & 7.36 & 8.34 & 5.71 & 7.10 & 6.65 & 4.95 & 76\\
  & GKA07   & 7.78    & 7.50    & 8.73    & 7.64    & 8.11    & 6.15    & 7.06    & 6.52    & 5.12 & 64\\
  & GPMDR12$^{*}$& 7.37    & 7.46    & 8.75    & 7.38    & 8.33    & 5.88    & 7.08    & 6.41    & 5.02 & 92\\
Hu1-2 & AC83 & 7.30 & 7.52 & 7.80 & 7.41 & 7.34 & 5.70 & 6.06 & 5.54 & 4.60 & 56\\
   & FGMR15  & 7.26    & --    & 7.79    & 7.29    & 7.28    & 5.56    & 6.15    & --    & 4.59 & 69\\
   & HPF04   & 7.02    & --    & 7.56    & 7.27    & 7.00    & 5.64    & 5.79    & 5.42    & 4.54 & 64\\
   & LLLB04  & 7.15    & 7.26    & 7.66    & 7.21    & 7.20    & 5.55    & 5.94    & 5.42    & 4.49 & 67\\
   & MKHS10  & 7.07    & 7.23    & 7.80    & 7.26    & 7.25    & 5.53    & 6.10    & 5.54    & 4.56 & 68\\
   & PT87    & 7.05    & 7.28    & 7.66    & 7.22    & 7.18    & 5.48    & 6.01    & 5.48    & -- & 60\\
   & SCT87$^{*}$& 7.12    & 7.49    & 7.83    & 7.33    & 7.35    & 5.56    & --    & 5.53    & 4.55 & 79\\
Hu2-1 & AC83 & 7.19 & 7.63 & 8.30 & 6.79 & 7.59 & 4.18 & 6.15 & 5.96 & 3.85 & 52\\
   & F81     & 7.12    & 7.43    & 7.92    & 6.70    & 7.08    & 4.69    & 6.02    & 5.88    & -- & 45\\
   & KHM03$^{*}$& 8.24    & 8.04    & 8.39    & 7.25    & 7.59    & 5.46    & 6.38    & 6.09    & 4.62 & 77\\
   & WLB05   & 7.73    & 7.64    & 8.04    & 7.02    & 7.33    & 4.79    & 5.88    & 5.73    & 4.31 & 69\\
IC418 & AC83$^{\dagger\ddagger}$& 8.34 & 8.44 & 7.98 & 7.68 & 6.93 & 5.65 & 6.57 & 5.89 & 4.65 & 57\\
   & DASNA17$^{\dagger\ddagger}$& 8.41    & 8.31    & 8.03    & 7.61    & 7.03    & 5.88    & 6.60    & 5.96    & 4.72 & 86\\
   & HAF94a  & 8.14    & 8.28    & 7.55    & 7.70    & 6.52    & 5.96    & 6.45    & 5.79    & 4.68 & 82\\
   & SWBvH03$^{*}$& 8.34    & 8.32    & 8.09    & 7.64    & 6.88    & 5.79    & 6.60    & 6.02    & 4.70 & 94\\
IC2003 & AC83 & 6.81 & -- & 8.39 & 6.54 & 7.78 & 5.11 & 6.19 & 5.76 & 4.45 & 51\\
  & B78   & 6.30    & 6.45    & 8.20    & 6.07    & 7.64    & 4.55    & --   & --   & -- & 53\\
  & KB94    & 6.54    & --   & 8.29    & 6.18    & 7.72    & 4.77    & 5.91    & 5.48    & 4.65 & 47\\
  & WLB05$^{*}$& 6.75    & 6.86    & 8.29    & 6.40    & 7.73    & 4.97    & 6.14    & 5.71    & 4.51 & 75\\
IC4846 & AC79 & 6.89 & 6.78 & 8.62 & 6.14 & 7.95 & 5.02 & 6.45 & 5.88 & 4.44 & 57\\
  & B78   & 7.26    & 7.77    & 8.55    & 6.67    & 8.03    & 5.32    & --   & --   & -- & 53\\
  & HAL01$^{*}$& 7.01    & 7.15    & 8.51    & 6.42    & 8.14    & 5.29    & 6.42    & 5.96    & 4.68 & 88\\
  & WL07    & 6.65    & 6.89    & 8.58    & 6.29    & 7.90    & 5.11    & 6.59    & 5.95    & 4.53 & 76\\
IC5217 & AC79 & 7.02 & 7.08 & 8.46 & 6.58 & 7.97 & 5.25 & 6.36 & 5.84 & 4.66 & 65\\
  & B78   & 6.80    & 7.46    & 8.45    & 6.66    & 7.96    & 4.88    & --   & --   & -- & 51\\
  & F81   & 6.51    & 6.83    & 8.32    & 6.31    & 7.81    & 4.99    & 6.32    & 5.89    & -- & 48\\
  & HAFL01  & 6.64    & 6.81    & 8.63    & 6.48    & 8.14    & 5.07    & 6.41    & 5.87    & 4.43 & 60\\
  & KH01$^{*}$& 6.75    & 6.83    & 8.47    & 6.50    & 7.97    & 5.15    & 6.39    & 5.93    & 4.60 & 73\\
  & PSM01   & 6.65    & --   & 8.51    & 6.46    & 7.94    & 5.19    & --   & --   & -- & 45\\
M1-25 & GCSC09$^{\dagger}$& 8.17 & 8.12 & 8.57 & 7.92 & 7.48 & 6.27 & 7.15 & 6.60 & 5.26 & 58\\
   & GKA07$^{\dagger\ddagger}$& 8.26    & 8.44    & 8.62    & 7.94    & 7.46    & 6.23    & 7.17    & 6.61    & 5.23 & 74\\
   & GPMDR12$^{*}$& 8.25    & 8.32    & 8.63    & 7.91    & 7.61    & 6.31    & 7.21    & 6.65    & 5.27 & 92\\
   & MKHC02  & 8.30    & 8.28    & 8.47    & 7.90    & 7.40    & 6.21    & 6.99    & 6.50    & 5.24 & 72\\
\hline
\end{tabular}
\end{table*}

\begin{table*}
\centering
\contcaption{Ionic abundances derived from collisionally excited lines, where $X^{i+}=12+\log(X^{i+}/\mbox{H}^+)$. O$^+$(b) and O$^+$(r) are the abundances calculated using the blue and red [\ion{O}{ii}] lines. The last column provides the final score of each spectrum. The spectra used as reference for each object are marked with asterisks. The references are identified in Table~\ref{tab:scores}.}
\begin{tabular}{llrrrrrrrrrr}
\hline
PN & Ref. & O$^+$(b) & O$^+$(r) & O$^{++}$ & N$^+$ & Ne$^{++}$ & S$^+$ & S$^{++}$ & Ar$^{++}$ & Cl$^{++}$ & $p_{\rm{tot}}$ \\
\hline
M1-25 & PSM01$^{\dagger}$& 8.26 & -- & 8.64 & 8.00 & 7.31 & 6.53 & 7.11 & -- & -- & 43\\
M1-29 & EBW04 & 8.02 & 8.51 & 8.67 & 8.09 & 8.18 & 6.59 & 6.92 & 6.64 & 5.33 & 64\\
   & GCSC09$^{\dagger\ddagger}$& 7.85    & 7.81    & 8.55    & 7.89    & 8.12    & 6.40    & 6.80    & 6.43    & 5.30 & 58\\
   & WL07$^{*}$& 7.95    & 7.92    & 8.58    & 7.93    & 8.12    & 6.41    & 6.88    & 6.33    & 5.27 & 73\\
M1-32 & GKA07 & 8.56 & 8.66 & 8.12 & 8.26 & 7.18 & 6.54 & 6.77 & 6.35 & 5.18 & 73\\
   & GPMDR12$^{*}$& 8.20    & 8.47    & 8.27    & 8.15    & 7.31    & 6.52    & 7.02    & 6.53    & 5.26 & 80\\
   & PSM01   & 8.19    & --    & 8.10    & 8.14    & 6.80    & 6.33    & 6.75    & --    & -- & 43\\
M1-74 & AC83 & 7.40 & 7.35 & 8.68 & 6.77 & 8.21 & 5.64 & 6.84 & 6.46 & 4.71 & 55\\
   & KH01$^{*}$& 6.36    & --    & 8.63    & 6.45    & 8.15    & 4.95    & 6.92    & 6.43    & -- & 76\\
   & WLB05   & 7.07    & 6.83    & 8.61    & 6.70    & 8.12    & 5.54    & 6.81    & 6.34    & 4.73 & 66\\
M2-23 & EBW04 & 6.06 & 6.84 & 8.38 & 6.01 & 7.83 & 4.49 & 6.33 & 5.95 & 4.00 & 66\\
   & GCSC09  & 6.08    & 6.22    & 8.28    & 5.82    & 7.81    & 4.59    & --    & 5.78    & 3.88 & 58\\
   & WL07$^{*}$& 6.11    & 6.48    & 8.34    & 5.95    & 7.64    & 4.89    & 6.45    & 5.83    & 4.05 & 74\\
M3-15 & MKHC02 & 6.82 & 6.66 & 8.80 & 6.68 & 8.28 & 5.40 & 7.02 & 6.42 & 4.57 & 59\\
   & GKA07   & 7.23    & 7.09    & 8.35    & 6.83    & 7.80    & 5.52    & 6.42    & 6.11    & 4.75 & 61\\
   & GPMDR12$^{*}$& 7.05    & 7.19    & 8.79    & 6.68    & 8.18    & 5.45    & 6.92    & 6.41    & 4.89 & 80\\
   & PSM01   & 7.44    & --    & 8.76    & 7.00    & 7.99    & 5.81    & 6.85    & --    & -- & 43\\
Me2-2 & AK87 & 6.74 & 7.25 & 8.35 & 7.33 & 7.84 & 4.46 & 6.12 & 5.79 & 4.13 & 66\\
   & B78     & 6.44    & 7.32    & 8.29    & 7.18    & 7.79    & 3.85    & --    & --    & -- & 56\\
   & MKHS10$^{*}$& 7.01    & 7.22    & 8.43    & 7.38    & 7.86    & 4.92    & 6.24    & 5.96    & 4.38 & 70\\
   & WLB05   & 6.78    & 7.15    & 8.17    & 7.28    & 7.58    & 4.71    & 5.85    & 5.75    & 4.45 & 67\\
NGC40 & AC79 & 8.71 & 8.70 & 6.00 & 7.98 & -- & 6.10 & 5.23 & 5.17 & 4.91 & 56\\
   & CSPT83$^{\ddagger}$& 8.91    & 8.72    & 6.71    & 8.04    & --    & 6.25    & 5.95    & 5.44    & 4.88 & 67\\
   & LLLB04$^{*}$& 8.64    & 8.69    & 7.10    & 7.95    & 5.88    & 6.14    & 6.19    & 5.73    & 4.94 & 77\\
   & PSM01$^{\dagger}$   & 8.57    & --    & 7.29    & 7.92    & --    & 6.13    & 6.30    & 5.82    & -- & 43\\
NGC650 & AC83 & 8.40 & 8.55 & 8.52 & 8.22 & 8.27 & 6.52 & 6.80 & 6.20 & -- & 51\\
  & KHM03$^{*}$& 8.15    & 8.28    & 8.31    & 7.98    & 8.10    & 6.37    & 6.62    & 6.31    & 4.89 & 68\\
  & PT87    & 8.32    & 8.52    & 8.39    & 8.05    & 8.08    & 6.45    & 6.68    & 6.29    & -- & 64\\
NGC2392 & B91 & 7.92 & -- & 8.14 & 7.60 & 7.61 & 6.17 & 5.99 & 5.78 & 4.92 & 54\\
 & DRMV09$^{\dagger\ddagger}$& 7.43  & 7.42  & 8.08  & 7.06  & 7.64  & 5.54  & 6.30  & 5.73  & 4.68 & 68\\
 & HKB00$^{\ddagger}$& 7.57  & 7.60  & 8.06  & 7.29  & 7.62  & 5.85  & 6.34  & 5.79  & 4.67 & 63\\
 & Z76  & 6.79  & 7.15  & 8.16  & 6.75  & 7.55  & 5.32  & 6.51  & 5.79  & -- & 40\\
 & ZFCH12$^{*}$& 7.80  & 7.73  & 8.12  & 7.29  & 7.60  & 5.89  & 6.25  & 5.75  & 4.92 & 71\\
NGC2440 & DKSH15 & 7.78 & 7.97 & 8.13 & 8.07 & 7.49 & 5.77 & 5.78 & 5.89 & -- & 67\\
 & HA98  & 7.81  & 8.02  & 8.30  & 8.25  & 7.66  & 5.71  & 6.00  & 5.92  & 4.94 & 72\\
 & KB94  & 8.19  & 8.24  & 8.15  & 8.22  & 7.66  & 5.90  & 6.09  & 5.99  & -- & 44\\
 & KC06$^{*}$& 7.66  & --  & 8.19  & 7.95  & 7.59  & 5.69  & 6.19  & --  & 4.76 & 80\\
 & KHM03   & 7.91  & 7.79  & 8.29  & 8.12  & 7.74  & 6.04  & 6.43  & 6.10  & 5.15 & 67\\
 & PT87$^{\dagger}$& 7.74  & 7.78  & 8.13  & 7.97  & 7.55  & 5.80  & 6.17  & 5.98  & 4.96 & 62\\
 & SAKC81  & 7.84  & 7.95  & 8.26  & 8.14  & 7.67  & 5.71  & 6.01  & 6.00  & 4.85 & 76\\
NGC2867 & AKRO81 & 7.96 & 7.92 & 8.53 & 7.39 & 7.97 & 5.80 & 6.47 & 6.01 & 4.93 & 62\\
 & GKA07   & 7.59  & 7.31  & 8.44  & 6.93  & 7.90  & 5.50  & 6.38  & 5.87  & 4.79 & 66\\
 & GPP09$^{*}$& 7.58  & --  & 8.46  & 7.05  & 7.94  & 5.70  & 6.45  & 5.99  & 4.87 & 82\\
 & KB94  & 7.89  & 7.01  & 8.64  & 7.27  & 8.07  & 5.84  & 6.53  & 6.06  & 4.92 & 57\\
 & MKHC02$^{\dagger}$& 7.60  & 7.77  & 8.48  & 7.11  & 7.92  & 5.64  & 6.40  & 5.98  & 4.87 & 79\\
 & PSEK98  & 7.47  & 7.62  & 8.42  & 7.04  & 7.87  & 5.53  & 6.36  & 5.83  & -- & 41\\
NGC6210 & B78 & 7.32 & -- & 8.63 & 6.66 & 8.15 & 5.40 & -- & -- & -- & 52\\
 & BERD15$^{\dagger}$& 7.19  & --  & 8.64  & 6.52  & 8.16  & 5.40  & 6.54  & --  & 4.80 & 77\\
 & DRMV09  & 7.12  & 7.17  & 8.55  & 6.50  & 8.08  & 5.31  & 6.46  & 5.98  & 4.55 & 63\\
 & F81  & 6.89  & 7.29  & 8.48  & 6.40  & 8.01  & 5.18  & 6.57  & 6.10  & -- & 47\\
 & KH98$^{\ddagger}$& 7.26  & 7.45  & 8.68  & 6.69  & 8.18  & 5.45  & 6.61  & 6.09  & 4.74 & 61\\
 & LLLB04$^{*}$& 7.22  & 7.29  & 8.66  & 6.53  & 8.16  & 5.39  & 6.55  & 6.04  & 4.70 & 80\\
NGC6302 & KC06 & 7.34 & -- & 7.98 & 7.83 & 7.54 & 5.90 & 6.25 & -- & 4.65 & 80\\
 & MKHS10  & 7.40  & 7.53  & 8.00  & 7.86  & 7.56  & 6.14  & 6.31  & 5.95  & 4.82 & 67\\
 & RCK14   & 7.19  & 7.51  & 7.99  & 7.89  & 7.52  & 6.11  & 6.29  & 5.97  & 4.83 & 80\\
 & TBLDS03 & 7.14  & 7.25  & 7.93  & 7.76  & 7.52  & 6.13  & 6.25  & 5.83  & 4.63 & 80\\
NGC6369 & AK87 & 7.27 & 7.11 & 8.45 & 6.95 & 7.80 & 5.56 & 6.24 & 6.03 & 4.80 & 55\\
 & GKA07   & 8.03  & 8.25  & 8.89  & 7.43  & 8.50  & 5.64  & 7.13  & 6.34  & -- & 67\\
 & GPMDR12$^{*}$& 6.93  & 7.10  & 8.52  & 6.57  & 7.98  & 5.30  & 6.54  & 6.17  & 4.66 & 93\\
 & MKHS10  & 7.40  & 7.55  & 8.68  & 7.06  & 8.11  & 5.53  & 6.73  & 6.23  & 4.88 & 67\\
 & PSM01   & 7.54  & --  & 8.65  & 7.14  & 8.08  & 5.58  & 7.25  & --  & -- & 45\\
NGC6543 & AC79 & 7.23 & 7.15 & 8.70 & 6.63 & 8.24 & 5.12 & 6.69 & 6.32 & 4.82 & 64\\
\hline
\end{tabular}
\end{table*}

\begin{table*}
\centering
\contcaption{Ionic abundances derived from collisionally excited lines, where $X^{i+}=12+\log(X^{i+}/\mbox{H}^+)$. O$^+$(b) and O$^+$(r) are the abundances calculated using the blue and red [\ion{O}{ii}] lines. The last column provides the final score of each spectrum. The spectra used as reference for each object are marked with asterisks. The references are identified in Table~\ref{tab:scores}.}
\begin{tabular}{llrrrrrrrrrr}
\hline
PN & Ref. & O$^+$(b) & O$^+$(r) & O$^{++}$ & N$^+$ & Ne$^{++}$ & S$^+$ & S$^{++}$ & Ar$^{++}$ & Cl$^{++}$ & $p_{\rm{tot}}$ \\
\hline
NGC6543 & HAFLK00 & 7.20 & 7.24 & 8.75 & 6.79 & 8.29 & 5.32 & 6.83 & 6.54 & 5.10 & 75\\
 & PSM01$^{\dagger}$& 7.06  & --  & 8.81  & 6.60  & 8.39  & 5.21  & 6.92  & --  & -- & 45\\
 & WL04$^{*}$& 7.31  & 7.34  & 8.78  & 6.88  & 8.31  & 5.50  & 6.97  & 6.53  & 4.95 & 80\\
NGC6565 & EBW04 & 8.17 & 8.50 & 8.41 & 8.03 & 7.96 & 6.59 & 6.72 & 6.41 & 5.10 & 65\\
 & MKHC02$^{*}$& 8.20  & 8.18  & 8.55  & 7.91  & 8.19  & 6.45  & 6.70  & 6.33  & 5.12 & 82\\
 & WL07  & 8.08  & 8.17  & 8.59  & 7.91  & 8.18  & 6.40  & 6.89  & 6.29  & 5.04 & 77\\
NGC6572 & F81 & 6.68 & 7.23 & 8.21 & 6.70 & 7.72 & 4.87 & 6.18 & 6.06 & -- & 47\\
 & GKA07   & 7.11  & 7.35  & 8.60  & 6.93  & 8.06  & 5.19  & 6.36  & 6.19  & 4.50 & 75\\
 & HAF94b$^{*}$& 7.25  & 7.14  & 8.58  & 6.79  & 8.10  & 5.19  & 6.18  & 6.13  & 4.49 & 88\\
 & KH01  & 6.88  & --  & 8.59  & 6.79  & 8.06  & 5.04  & 6.47  & 6.23  & 4.47 & 75\\
 & LLLB04  & 7.41  & 7.44  & 8.60  & 7.08  & 8.10  & 5.40  & 6.46  & 6.19  & 4.66 & 79\\
NGC6720 & B80 & 8.49 & 8.30 & 8.72 & 8.00 & 8.39 & 6.15 & -- & 6.45 & -- & 47\\
 & F81  & 8.11  & 8.37  & 8.43  & 7.91  & 8.03  & 6.20  & 6.71  & 6.38  & -- & 46\\
 & GMC97$^{\dagger\ddagger}$& 8.55  & 8.35  & 8.37  & 8.06  & 8.28  & 6.30  & 6.46  & 6.33  & -- & 70\\
 & HM77  & 8.61  & --  & 8.71  & 8.05  & 8.40  & 6.16  & --  & --  & -- & 48\\
 & KH98$^{\dagger\ddagger}$& 7.81  & 7.86  & 8.53  & 7.46  & 8.09  & 5.79  & 6.52  & 6.25  & 5.04 & 73\\
 & LLLB04$^{*}$& 8.27  & 8.29  & 8.50  & 7.91  & 8.19  & 6.15  & 6.53  & 6.33  & 5.00 & 78\\
NGC6741 & HA97a$^{*}$& 7.91 & 7.95 & 8.45 & 7.70 & 8.00 & 6.19 & 6.50 & 6.21 & 4.98 & 84\\
 & LLLB04$^{\dagger\ddagger}$& 8.05  & 8.04  & 8.43  & 7.75  & 7.98  & 6.24  & 6.52  & 6.21  & 4.98 & 77\\
 & MKHS10  & 7.83  & 8.00  & 8.57  & 7.73  & 7.99  & 6.18  & 6.71  & 6.32  & 4.94 & 78\\
NGC6751 & AC79 & 8.56 & 8.60 & 8.39 & 7.91 & 7.93 & 6.17 & 6.46 & 6.23 & 5.50 & 63\\
 & CMJK91  & 8.18  & --  & 8.53  & 7.77  & 8.05  & 6.20  & 6.81  & --  & 5.17 & 75\\
 & KB94  & 8.36  & 8.42  & 8.77  & 7.85  & 8.12  & 5.99  & 6.84  & 6.18  & -- & 46\\
 & MKHS10$^{*}$& 7.94  & 8.27  & 8.50  & 7.74  & 7.90  & 6.07  & 6.79  & 6.38  & 5.26 & 80\\
 & PSM01   & 8.03  & --  & 8.59  & 7.62  & 8.08  & 6.01  & 6.83  & 6.27  & -- & 43\\
NGC6790 & AC79 & 6.40 & 6.45 & 8.55 & 5.99 & 7.89 & 4.29 & 6.11 & 5.49 & 3.75 & 59\\
 & AHF96   & 7.10  & 7.19  & 8.71  & 6.46  & 8.04  & 5.19  & 6.21  & 5.74  & 4.15 & 66\\
 & F81  & 6.31  & 6.82  & 8.19  & 6.11  & 7.62  & 4.63  & 6.02  & 5.65  & -- & 46\\
 & KH01$^{*}$& 5.93  & 6.65  & 8.36  & 6.00  & 7.78  & 4.16  & 6.08  & 5.69  & 3.74 & 74\\
 & LLLB04$^{\ddagger}$& 6.93  & 6.92  & 8.38  & 6.31  & 7.84  & 5.02  & 6.00  & 5.55  & 4.12 & 69\\
NGC6884 & AC79 & 7.16 & 6.93 & 8.57 & 6.50 & 8.01 & 5.04 & 6.33 & 6.15 & 4.56 & 65\\
 & HAF97   & 7.11  & 7.13  & 8.88  & 6.82  & 8.36  & 5.40  & 6.49  & 6.19  & 4.86 & 73\\
 & KH01$^{*}$& 7.08  & 7.18  & 8.61  & 6.83  & 8.07  & 5.32  & 6.46  & 6.17  & 4.91 & 75\\
 & LLLB04$^{\ddagger}$& 7.12  & 7.07  & 8.57  & 6.76  & 8.09  & 5.26  & 6.38  & 6.06  & 4.75 & 72\\
NGC7009 & CA79 & 7.12 & 6.94 & 8.81 & 6.56 & 8.37 & 5.33 & 6.78 & 6.25 & 4.83 & 80\\
 & F81  & 6.38  & 6.59  & 8.25  & 5.73  & 7.79  & 4.56  & 6.19  & 6.01  & -- & 47\\
 & FL11  & 6.85  & 6.83  & 8.63  & 6.41  & 8.23  & 5.13  & 6.57  & 6.15  & 4.71 & 83\\
 & HA95a$^{\dagger}$& 7.52  & 7.68  & 8.65  & 7.19  & 8.28  & 5.84  & 6.73  & 6.28  & 5.10 & 75\\
 & HA95b   & 6.13  & 6.08  & 8.46  & 5.55  & 8.16  & 4.50  & 6.18  & 5.97  & 4.34 & 81\\
 & KC06$^{*}$& 6.94  & --  & 8.62  & 6.61  & 8.19  & 5.25  & 6.60  & --  & 4.81 & 85\\
 & KH98  & 6.36  & 6.50  & 8.69  & 6.09  & 8.13  & 4.84  & 6.53  & 6.12  & 4.88 & 68\\
NGC7662 & AC83 & 6.77 & 7.05 & 8.19 & 6.07 & 7.61 & 4.65 & 6.01 & 5.66 & 4.78 & 65\\
 & HA97b$^{*}$& 5.95  & 5.82  & 8.30  & 5.24  & 7.70  & 4.20  & 5.95  & 5.49  & 4.22 & 72\\
 & LLLB04  & 6.47  & 6.61  & 8.27  & 5.89  & 7.68  & 4.69  & 6.09  & 5.68  & 4.55 & 69\\
PB6 & GKA07$^{\dagger}$& 7.28 & 7.52 & 7.97 & 7.46 & 7.56 & 5.43 & 6.25 & 5.70 & -- & 64\\
     & GPP09$^{*}$& 7.20    & --      & 8.03    & 7.31    & 7.55    & 5.39    & 6.24    & 5.79    & -- & 74\\
     & MKHS10  & 7.40    & 7.61    & 8.11    & 7.64    & 7.63    & 5.62    & 6.23    & 5.85    & 4.84 & 72\\
     & PSEK98  & 7.55    & 7.88    & 8.13    & 7.78    & 7.68    & 5.74    & 6.29    & 5.88    & -- & 41\\
PC14 & GKA07$^{\dagger\ddagger}$& 7.48 & 7.55 & 8.78 & 6.94 & 8.25 & 5.78 & 6.86 & 6.28 & 4.97 & 74\\
    & GPMDR12$^{*}$& 7.38    & 7.47    & 8.74    & 6.82    & 8.27    & 5.67    & 6.81    & 6.27    & 4.98 & 99\\
    & MKHC02$^{\dagger\ddagger}$& 7.90    & 7.86    & 8.80    & 7.16    & 8.36    & 6.03    & 6.87    & 6.32    & 5.16 & 74\\
Pe1-1 & BK13 & 7.95 & 7.98 & 8.50 & 7.41 & 7.92 & 6.00 & 6.56 & 6.30 & 4.85 & 64\\
   & G14     & 7.42    & 7.71    & 8.44    & 7.19    & 7.92    & 5.54    & 6.63    & 6.25    & 4.69 & 72\\
   & GKA07$^{\dagger}$& 7.80    & 7.71    & 8.63    & 7.27    & 8.15    & 5.75    & 6.68    & 6.24    & 4.66 & 68\\
   & GPMDR12$^{*}$& 7.82    & 7.85    & 8.56    & 7.32    & 8.02    & 5.91    & 6.66    & 6.31    & 4.86 & 94\\
\hline
\end{tabular}
\\
\raggedright
$^\dagger$ O/H, N/H, and S/H are within 0.1~dex of the reference spectrum for calculations based on the blue [\ion{O}{ii}] lines.\\
$^\ddagger$ O/H, N/H, and S/H are within 0.1~dex of the reference spectrum for calculations based on the red [\ion{O}{ii}] lines.\\
\end{table*}

\begin{table*}
\centering
\caption{Ionic and total abundances of helium, where $X^{i+}=12+\log(\mbox{He}^{i+}/\mbox{H}^+)$ and $\mbox{He}=12+\log(\mbox{He}/\mbox{H})$. The last column provides the final score of each spectrum. The spectra used as reference for each object are marked with asterisks. The references are identified in Table~\ref{tab:scores}.}
\label{tab:He}
\begin{tabular}{llrrrrrrrrrr}
\hline
PN & Ref. & He$^+$ & He$^+$ & He$^+$ & He$^+$ & He$^+$ & He$^+$ & He$^+$ & He$^{++}$ & He & $p_{\rm{tot}}$\\
 & & $\lambda4026$ & $\lambda4388$ & $\lambda4922$ & $\lambda4471$ & $\lambda5876$ & $\lambda6678$ & & & & \\
\hline
BoBn 1  & KHM03   & --	  & --	  & 10.67 & 10.83 & 10.81 & 10.78 & 10.81 & 10.24 & 10.91 & 60\\
        & KZGP08$^{*}$  & 10.99 & 10.92 & 10.88 & 10.92 & 10.86 & 10.89 & 10.89 & 10.18 & 10.97 & 89\\
        & OTHI10  & 11.01 & 10.94 & 10.96 & 10.96 & 11.04 & 10.93 & 10.98 & 10.33 & 11.06 & 73\\
Cn 1-5  & BK13	  & 11.10 & 11.16 & 11.02 & 11.12 & 11.13 & 11.14 & 11.13 & --	  & 11.13 & 84\\
        & EBW04   & 11.06 & --	  & --	  & 11.18 & 11.14 & 11.30 & 11.20 & --	  & 11.20 & 59\\
	& GPMDR12$^{*}$& 11.16	& 11.14	& 11.12	& 11.15	& 11.17	& 11.14	& 11.15	& --	& 11.15 & 97\\
	& PSM01	& --	& --	& --	& --	& 11.05	& --	& 11.05	& --	& 11.05 & 43\\
	& WL07	& --	& 11.11	& 11.13	& 11.12	& 11.11	& 11.09	& 11.11	& --	& 11.11 & 80\\
DdDm1	& BC84$^{\dagger\ddagger}$& --	& --	& --	& 10.89	& 10.96	& 10.95	& 10.93	& --	& 10.93 & 48\\
	& CPT87$^{\dagger}$& 10.88	& 10.87	& --	& 10.95	& 10.95	& 10.93	& 10.95	& --	& 10.95 & 57\\
	& KH98$^{*}$& 10.93	& --	& --	& 10.97	& 10.97	& 10.95	& 10.96	& --	& 10.96 & 74\\
	& OHLIT09 & 11.06	& 10.94	& 10.93	& 10.98	& 10.92	& 10.86	& 10.92	& --	& 10.92 & 65\\
H1-50	& EBW04	& --	& --	& --	& --	& 10.91	& 11.04	& 10.98	& 9.98	& 11.02 & 67\\
	& GDGD18$^{*}$& --	& 10.97	& 10.96	& 11.00	& 10.99	& 10.96	& 10.99	& 10.03	& 11.03 & 84\\
	& WL07	& 11.02	& 10.95	& 10.96	& 10.99	& 11.01	& 10.96	& 10.99	& 10.04	& 11.03 & 78\\
H1-54	& EBW04	& 11.07	& --	& --	& 10.96	& 10.94	& 11.03	& 10.98	& --	& 10.98 & 66\\
	& GCSC09  & --	& --	& --	& 10.99	& 11.02	& 10.93	& 10.98	& --	& 10.98 & 58\\
	& WL07$^{*}$& 10.94	& 10.94	& 10.97	& 10.99	& 11.00	& 10.92	& 10.97	& 8.43	& 10.97 & 87\\
H4-1	& KHM03	& --	& --	& 10.95	& 11.03	& 11.04	& 11.01	& 11.03	& 9.90	& 11.06 & 68\\
	& OT13$^{*}$& 10.92	& 10.92	& 10.86	& 10.95	& 11.00	& 10.91	& 10.95	& 10.19	& 11.02 & 84\\
	& TP79	& 10.99	& --	& 10.99	& 10.98	& 10.94	& 10.90	& 10.94	& 9.91	& 10.98 & 57\\
Hb4	& AK87$^{\dagger}$& --	& --	& --	& 10.98	& 11.01	& 10.93	& 10.97	& 10.25	& 11.05 & 55\\
	& GKA07	& --	& --	& --	& 10.99	& 10.97	& 11.00	& 10.98	& 10.10	& 11.04 & 74\\
	& GPMDR12$^{*}$& 10.99	& 11.02	& --	& 10.98	& 10.98	& 10.94	& 10.97	& 10.31	& 11.05 & 89\\
	& MKHS10  & --	& 10.76	& --	& 11.04	& 11.07	& 10.97	& 11.03	& 10.07	& 11.07 & 72\\
	& PSM01	& --	& --	& --	& --	& 11.09	& --	& 11.09	& 10.03	& 11.12 & 45\\
He2-86	& G14$^{\ddagger}$& --	& --	& --	& 11.16	& 11.14	& 11.10	& 11.13	& --	& 11.13 & 76\\
	& GKA07	& --	& --	& --	& 11.12	& 11.05	& 11.06	& 11.08	& --	& 11.08 & 64\\
	& GPMDR12$^{*}$& 11.09	& 11.10	& 11.06	& 11.12	& 11.13	& 11.08	& 11.11	& --	& 11.11 & 92\\
Hu1-2	& AC83	& 10.84	& --	& 10.17	& 10.81	& 10.69	& 10.70	& 10.74	& 10.89	& 11.12 & 56\\
	& FGMR15  & 10.97	& 10.79	& 10.65	& 10.68	& 10.68	& 10.69	& 10.68	& 10.92	& 11.12 & 69\\
	& HPF04	& 10.80	& 10.64	& 10.70	& 10.62	& 10.62	& 10.55	& 10.59	& 10.83	& 11.03 & 64\\
	& LLLB04  & 11.00	& 10.74	& 10.66	& 10.72	& 10.58	& 10.53	& 10.61	& 10.94	& 11.11 & 67\\
	& MKHS10  & --	& --	& --	& 10.68	& 10.69	& 10.67	& 10.68	& 10.90	& 11.10 & 68\\
	& PT87	& --	& --	& --	& 10.72	& 10.67	& 10.70	& 10.69	& 10.92	& 11.12 & 60\\
	& SCT87$^{*}$& 10.89	& --	& --	& 10.85	& 10.72	& 10.75	& 10.77	& 10.90	& 11.14 & 79\\
Hu2-1	& AC83	& 10.95	& --	& --	& 10.94	& 10.94	& 10.83	& 10.90	& --	& 10.90 & 52\\
	& F81	& --	& --	& --	& 10.96	& 11.02	& 10.91	& 10.96	& 8.32	& 10.96 & 45\\
	& KHM03$^{*}$& --	& --	& 10.89	& 10.98	& 10.99	& 10.93	& 10.97	& 8.22	& 10.97 & 77\\
	& WLB05	& 11.09	& 11.03	& 10.94	& 11.02	& 10.86	& 10.79	& 10.89	& 8.50	& 10.89 & 69\\
IC418	& AC83$^{\dagger\ddagger}$& --	& --	& --	& 10.83	& 10.83	& 10.73	& 10.80	& --	& 10.80 & 57\\
	& DASNA17$^{\dagger\ddagger}$& 10.90	& 10.88	& 10.89	& 10.89	& 10.92	& 10.89	& 10.90	& 6.60	& 10.90 & 86\\
	& HAF94a  & 10.83	& 10.81	& 10.81	& 10.84	& 10.85	& 10.77	& 10.82	& --	& 10.82 & 82\\
	& SWBvH03$^{*}$& 10.94	& 10.93	& 10.94	& 10.95	& 10.96	& 10.96	& 10.96	& --	& 10.96 & 94\\
IC2003	& AC83	& 10.79	& --	& 10.77	& 10.77	& 10.76	& --	& 10.76	& 10.50	& 10.96 & 51\\
	& B78	& --	& --	& --	& 10.88	& 10.83	& 10.80	& 10.84	& 10.66	& 11.06 & 53\\
	& KB94	& 11.06	& --	& --	& 10.57	& 10.69	& 10.51	& 10.59	& 10.71	& 10.96 & 47\\
	& WLB05$^{*}$& 10.81	& --	& 10.64	& 10.76	& 10.71	& 10.80	& 10.76	& 10.70	& 11.03 & 75\\
IC4846	& AC79	& --	& 11.21	& --	& 10.94	& 10.93	& 10.87	& 10.91	& 8.63	& 10.92 & 57\\
	& B78	& --	& --	& --	& 10.84	& 10.98	& 10.92	& 10.91	& --	& 10.91 & 53\\
	& HAL01$^{*}$& 11.08	& 11.03	& 10.88	& 11.04	& 10.98	& 10.93	& 10.98	& 8.69	& 10.99 & 88\\
	& WL07	& 10.88	& 10.93	& 10.98	& 10.97	& 11.01	& 10.97	& 10.99	& 8.51	& 10.99 & 76\\
IC5217	& AC79	& 10.89	& 10.87	& 10.77	& 10.90	& 10.88	& 10.91	& 10.90	& 10.07	& 10.96 & 65\\
	& B78	& --	& --	& --	& 10.99	& 10.99	& 10.98	& 10.99	& 9.79	& 11.01 & 51\\
	& F81	& --	& --	& --	& 10.94	& 11.00	& 10.96	& 10.97	& 9.80	& 11.00 & 48\\
	& HAFL01	& 11.03	& 10.95	& 10.97	& 11.06	& 10.89	& 10.82	& 10.92	& 9.95	& 10.97 & 60\\
	& KH01$^{*}$& 10.98	& --	& 10.92	& 11.00	& 10.97	& 10.95	& 10.97	& 9.90	& 11.01 & 73\\
	& PSM01	& --	& --	& --	& --	& 11.00	& --	& 11.00	& 9.96	& 11.04 & 45\\
M1-25	& GCSC09$^{\dagger}$& --	& --	& --	& 11.13	& 11.16	& 11.10	& 11.13	& --	& 11.13 & 58\\
	& GKA07$^{\dagger\ddagger}$& --	& --	& --	& 11.12	& 11.10	& 11.10	& 11.11	& --	& 11.11 & 74\\
	& GPMDR12$^{*}$& 11.10	& 11.12	& 11.08	& 11.11	& 11.13	& 11.09	& 11.11	& 7.59	& 11.11 & 92\\
	& MKHC02	& 11.14	& --	& 11.14	& 11.10	& 11.15	& 11.09	& 11.11	& 9.75	& 11.13 & 72\\
\hline
\end{tabular}
\end{table*}

\begin{table*}
\centering
\contcaption{Ionic and total abundances of helium, where $X^{i+}=12+\log(\mbox{He}^{i+}/\mbox{H}^+)$ and $\mbox{He}=12+\log(\mbox{He}/\mbox{H})$. The last column provides the final score of each spectrum. The spectra used as reference for each object are marked with asterisks. The references are identified in Table~\ref{tab:scores}.}
\begin{tabular}{llrrrrrrrrrr}
\hline
PN & Ref. & He$^+$ & He$^+$ & He$^+$ & He$^+$ & He$^+$ & He$^+$ & He$^+$ & He$^{++}$ & He & $p_{\rm{tot}}$ \\
 & & $\lambda4026$ & $\lambda4388$ & $\lambda4922$ & $\lambda4471$ & $\lambda5876$ & $\lambda6678$ & & & & \\
\hline
M1-25	& PSM01$^{\dagger}$& --	& --	& --	& --	& 11.13	& --	& 11.13	& --	& 11.13 & 43\\
M1-29	& EBW04	& --	 & --	 & --	 & 10.97   & 10.99   & 11.03   & 11.00   & 10.45   & 11.11 & 64\\
	& GCSC09$^{\dagger\ddagger}$& --	 & --	 & --	 & 11.05   & 11.06   & 11.00   & 11.04   & 10.47   & 11.14 & 58\\
	& WL07$^{*}$& --	 & --	 & 11.04   & 11.02   & 11.05   & 11.02   & 11.03   & 10.52   & 11.15 & 73\\
M1-32	& GKA07	& --	 & --	 & --	 & 11.05   & 11.05   & 11.07   & 11.06   & --	 & 11.06 & 73\\
	& GPMDR12$^{*}$& 11.08   & 11.12   & 11.09   & 11.10   & 11.10   & 11.10   & 11.10   & 8.57	 & 11.10 & 80\\
	& PSM01	& --	 & --	 & --	 & --	 & 11.09   & --	 & 11.09   & --	 & 11.09 & 43\\
M1-74	& AC83	& 11.00   & --	 & --	 & 10.95   & 11.00   & 10.95   & 10.97   & --	 & 10.97 & 55\\
	& KH01$^{*}$& --	 & --	 & 10.98   & 11.02   & 11.07   & 11.00   & 11.03   & --	 & 11.03 & 76\\
	& WLB05	& --	 & --	 & 10.98   & 10.91   & 11.01   & 10.97   & 10.96   & 8.61	 & 10.97 & 66\\
M2-23	& EBW04	& 10.91   & --	 & --	 & 10.99   & 10.84   & 10.96   & 10.93   & --	 & 10.93 & 66\\
	& GCSC09	& --	 & --	 & --	 & 10.99   & 10.99   & 10.89   & 10.96   & --	 & 10.96 & 58\\
	& WL07$^{*}$& 10.95   & 10.92   & 10.87   & 10.99   & 11.04   & 10.98   & 11.00   & --	 & 11.00 & 74\\
M3-15	& MKHC02	& 11.00   & --	 & --	 & 11.03   & 11.09   & 11.04   & 11.05   & 9.53	 & 11.07 & 59\\
	& GKA07	& --	 & --	 & --	 & 10.99   & 10.97   & 11.01   & 10.99   & --	 & 10.99 & 61\\
	& GPMDR12$^{*}$& 10.96   & 10.99   & 11.04   & 11.06   & 11.08   & 11.04   & 11.06   & --	 & 11.06 & 80\\
	& PSM01	& --	 & --	 & --	 & --	 & 11.14   & --	 & 11.14   & --	 & 11.14 & 43\\
Me2-2	& AK87	& 11.10   & 11.11   & 10.97   & 11.14   & 11.10   & 11.06   & 11.10   & --	 & 11.10 & 66\\
	& B78	& --	 & --	 & --	 & 11.21   & 11.22   & 11.17   & 11.20   & --	 & 11.20 & 56\\
	& MKHS10$^{*}$& 11.08   & 11.12   & 11.12   & 11.15   & 11.20   & 11.11   & 11.15   & --	 & 11.15 & 70\\
	& WLB05	& 11.28   & 11.17   & --	 & 11.26   & 11.11   & 11.11   & 11.16   & 8.10	 & 11.16 & 67\\
NGC40	& AC79	& 10.74   & --	 & 11.06   & 10.68   & 10.60   & 10.61   & 10.63   & --	 & 10.63 & 56\\
	& CSPT83$^{\ddagger}$& 10.66   & 10.62   & 10.63   & 10.65   & 10.61   & 10.61   & 10.62   & --	 & 10.62 & 67\\
	& LLLB04$^{*}$& 10.80   & 10.81   & 10.79   & 10.80   & 10.81   & 10.79   & 10.80   & 7.56	 & 10.80 & 77\\
	& PSM01$^{\dagger}$& --	 & --	 & --	 & --	 & 10.85   & --	 & 10.85   & --	 & 10.85 & 43\\
NGC650	& AC83	& 11.23   & --	 & --	 & 10.96   & 10.96   & 10.89   & 10.94   & 10.26   & 11.02 & 51\\
	& KHM03$^{*}$& --	 & --	 & 10.88   & 10.86   & 10.90   & 10.91   & 10.89   & 10.37   & 11.00 & 68\\
	& PT87	& --	 & --	 & --	 & 10.94   & 10.98   & 10.95   & 10.96   & 10.52   & 11.09 & 64\\
NGC2392	& B91	& --	 & --	 & --	 & 10.84   & 10.86   & 10.89   & 10.86   & 10.34   & 10.97 & 54\\
	& DRMV09$^{\dagger\ddagger}$& --	 & --	 & --	 & 10.76   & 10.88   & 10.84   & 10.83   & 10.44   & 10.98 & 68\\
	& HKB00$^{\ddagger}$& 10.63   & --	 & 10.72   & 10.78   & 10.75   & 10.70   & 10.74   & 10.29   & 10.87 & 63\\
	& Z76	& --	 & --	 & --	 & --	 & 10.75   & --	 & 10.75   & 10.64   & 11.00 & 40\\
	& ZFCH12$^{*}$& 10.88   & 10.89   & 10.92   & 10.92   & 10.78   & 10.80   & 10.83   & 10.31   & 10.95 & 71\\
NGC2440	& DKSH15	& --	 & --	 & --	 & 10.85   & 10.74   & 10.77   & 10.79   & 10.86   & 11.13 & 67\\
	& HA98	& 10.99   & 10.78   & 10.77   & 10.81   & 10.64   & 10.58   & 10.68   & 10.83   & 11.06 & 72\\
	& KB94	& --	 & --	 & --	 & 10.77   & 10.56   & 11.05   & 10.79   & 10.84   & 11.12 & 44\\
	& KC06$^{*}$& 10.96   & 10.68   & 10.70   & 10.73   & 10.70   & 10.82   & 10.75   & 10.81   & 11.08 & 80\\
	& KHM03	& --	 & --	 & 10.60   & --	 & 10.68   & 10.80   & 10.74   & 10.64   & 10.99 & 67\\
	& PT87$^{\dagger}$& --	 & --	 & --	 & 10.65   & 10.60   & 10.70   & 10.65   & 10.83   & 11.05 & 62\\
	& SAKC81	& 10.91   & 10.85   & 10.66   & 10.79   & 10.68   & 10.70   & 10.72   & 10.72   & 11.02 & 76\\
NGC2867	& AKRO81	& 11.00   & 10.79   & 10.97   & 10.91   & 10.85   & 10.87   & 10.88   & 10.42   & 11.01 & 62\\
	& GKA07	& --	 & --	 & --	 & 10.89   & 10.82   & 10.72   & 10.81   & 10.47   & 10.97 & 66\\
	& GPP09$^{*}$& 11.18   & --	 & 10.88   & 10.92   & 10.93   & 10.92   & 10.92   & 10.46   & 11.05 & 82\\
	& KB94	& 10.94   & --	 & --	 & 10.90   & 10.94   & 10.93   & 10.93   & 10.43   & 11.05 & 57\\
	& MKHC02$^{\dagger}$& 10.94   & --	 & --	 & 10.86   & 10.91   & 10.92   & 10.90   & 10.44   & 11.03 & 79\\
	& PSEK98	& --	 & --	 & --	 & --	 & 10.93   & --	 & 10.93   & 10.32   & 11.03 & 41\\
NGC6210	& B78	& --	 & --	 & --	 & 10.97   & 11.01   & 10.74   & 10.91   & 9.22	 & 10.92 & 52\\
	& BERD15$^{\dagger}$& 11.01   & 11.07   & 10.98   & 11.00   & 10.99   & 10.98   & 10.99   & 9.16	 & 11.00 & 77\\
	& DRMV09	& 11.06   & --	 & --	 & 11.03   & 11.02   & 11.00   & 11.02   & 9.28	 & 11.03 & 63\\
	& F81	& --	 & --	 & --	 & 11.06   & 11.10   & 10.97   & 11.04   & 9.38	 & 11.05 & 47\\
	& KH98$^{\ddagger}$& 11.00   & --	 & --	 & 11.03   & 11.03   & 10.98   & 11.01   & 8.99	 & 11.02 & 61\\
	& LLLB04$^{*}$& 11.01   & 11.00   & 10.98   & 11.01   & 10.96   & 10.97   & 10.98   & 9.09	 & 10.99 & 80\\
NGC6302	& KC06	& 11.12   & 10.94   & 10.95   & 11.01   & 10.99   & 10.76   & 10.92   & 10.80   & 11.17 & 80\\
	& MKHS10	& --	 & --	 & --	 & 11.04   & 11.00   & 10.93   & 10.99   & 10.74   & 11.18 & 67\\
	& RCK14	& 11.23   & 11.19   & 11.09   & 11.09   & 11.08   & 10.99   & 11.05   & 10.74   & 11.23 & 80\\
	& TBLDS03	& 11.03   & 10.90   & 10.91   & 10.97   & 10.87   & 10.18   & 10.67   & 10.83   & 11.06 & 80\\
NGC6369	& AK87	& --	 & --	 & 11.03   & 11.12   & 10.96   & 10.91   & 11.00   & 9.31	 & 11.01 & 55\\
	& GKA07	& --	 & --	 & --	 & 10.95   & 10.91   & 10.97   & 10.94   & --	 & 10.94 & 67\\
	& GPMDR12$^{*}$& 10.98   & 11.01   & 11.00   & 11.01   & 11.03   & 11.00   & 11.01   & 8.60	 & 11.01 & 93\\
	& MKHS10	& --	 & --	 & --	 & 10.94   & 11.09   & 11.00   & 11.01   & 9.17	 & 11.02 & 67\\
	& PSM01	& --	 & --	 & --	 & --	 & 11.00   & --	 & 11.00   & 9.92	 & 11.04 & 45\\
NGC6543	& AC79	& 11.08   & 11.09   & 10.95   & 11.08   & 11.01   & 11.04   & 11.04   & --	 & 11.04 & 64\\
\hline
\end{tabular}
\end{table*}

\begin{table*}
\centering
\contcaption{Ionic and total abundances of helium, where $X^{i+}=12+\log(\mbox{He}^{i+}/\mbox{H}^+)$ and $\mbox{He}=12+\log(\mbox{He}/\mbox{H})$. The last column provides the final score of each spectrum. The spectra used as reference for each object are marked with asterisks. The references are identified in Table~\ref{tab:scores}.}
\begin{tabular}{llrrrrrrrrrr}
\hline
PN & Ref. & He$^+$ & He$^+$ & He$^+$ & He$^+$ & He$^+$ & He$^+$ & He$^+$ & He$^{++}$ & He & $p_{\rm{tot}}$ \\
 & & $\lambda4026$ & $\lambda4388$ & $\lambda4922$ & $\lambda4471$ & $\lambda5876$ & $\lambda6678$ & & & & \\
\hline
NGC6543	& HAFLK00	& 11.05   & 11.04   & 11.00   & 11.07   & 11.14   & 11.12   & 11.11   & 7.60	 & 11.11 & 75\\
	& PSM01	& --	 & --	 & --	 & --	 & 11.07   & --	 & 11.07   & --	 & 11.07 & 45\\
        & WL04$^{*}$& 11.01	& 11.05	& 11.05	& 11.07	& 11.07	& 11.05	& 11.06	& --	& 11.06 & 80\\
NGC6565	& EBW04	& 10.90	& --	& --	& --	& 10.95	& 11.04	& 11.00	& 10.13	& 11.05 & 65\\
	& MKHC02$^{*}$& 11.09	& --	& --	& 11.00	& 11.00	& 10.99	& 11.00	& 10.11	& 11.05 & 82\\
	& WL07	& 11.02	& 10.93	& 10.98	& 11.00	& 11.01	& 11.00	& 11.00	& 10.19	& 11.07 & 77\\
NGC6572	& F81	& --	& --	& --	& 11.05	& 11.05	& 10.93	& 11.01	& 9.13	& 11.02 & 47\\
	& GKA07	& --	& --	& --	& 11.04	& 11.04	& 11.02	& 11.04	& 8.92	& 11.04 & 75\\
	& HAF94b$^{*}$& 11.02	& 11.01	& 10.97	& 11.00	& 11.02	& 10.94	& 10.99	& 8.51	& 10.99 & 88\\
	& KH01	& 11.03	& --	& --	& 11.04	& 11.07	& 10.98	& 11.03	& 8.70	& 11.03 & 75\\
	& LLLB04	& 11.02	& 11.00	& 11.03	& 11.05	& 11.04	& 10.98	& 11.02	& 8.53	& 11.02 & 79\\
NGC6720	& B80	& --	& --	& --	& 10.99	& 10.95	& 10.99	& 10.98	& 9.88	& 11.01 & 47\\
	& F81	& --	& --	& --	& 11.00	& 11.02	& 11.15	& 11.06	& 10.34	& 11.13 & 46\\
	& GMC97$^{\dagger\ddagger}$& 11.17	& 11.11	& 11.15	& 11.11	& 11.10	& 11.06	& 11.09	& 9.52	& 11.10 & 70\\
	& HM77	& --	& --	& --	& 11.00	& 10.96	& 10.95	& 10.97	& 9.91	& 11.01 & 48\\
	& KH98$^{\dagger\ddagger}$& 10.99	& --	& --	& 10.91	& 10.86	& 10.89	& 10.89	& 10.56	& 11.05 & 73\\
	& LLLB04$^{*}$& 11.04	& 10.99	& 11.00	& 11.00	& 10.98	& 10.97	& 10.99	& 10.25	& 11.06 & 78\\
NGC6741	& HA97a$^{*}$& 10.98	& 10.98	& 10.87	& 10.91	& 10.85	& 10.81	& 10.86	& 10.59	& 11.05 & 84\\
	& LLLB04$^{\dagger\ddagger}$& 10.99	& 10.92	& 10.92	& 10.93	& 10.89	& 10.83	& 10.89	& 10.49	& 11.03 & 77\\
	& MKHS10	& --	& 10.68	& --	& 10.89	& 10.95	& 10.87	& 10.90	& 10.45	& 11.03 & 78\\
NGC6751	& AC79	& --	& --	& --	& 11.02	& 11.00	& 10.98	& 11.00	& --	& 11.00 & 63\\
	& CMJK91	& --	& --	& --	& 11.07	& 11.07	& 11.06	& 11.07	& --	& 11.07 & 75\\
	& KB94	& 11.00	& --	& --	& 11.04	& 10.94	& 10.79	& 10.93	& 10.18	& 11.00 & 46\\
	& MKHS10$^{*}$& 10.99	& 11.04	& --	& 11.06	& 11.12	& 11.08	& 11.09	& --	& 11.09 & 80\\
	& PSM01	& --	& --	& --	& --	& 11.09	& --	& 11.09	& --	& 11.09 & 43\\
NGC6790	& AC79	& 11.00	& 10.76	& 11.01	& 10.98	& 10.93	& 10.88	& 10.93	& 9.31	& 10.94 & 59\\
	& AHF96	& 11.03	& 10.87	& 11.10	& 10.94	& 11.08	& 10.97	& 10.99	& 9.42	& 11.01 & 66\\
	& F81	& --	& --	& --	& 11.08	& 11.07	& 10.90	& 11.02	& 9.44	& 11.03 & 46\\
	& KH01$^{*}$& 11.04	& --	& 11.05	& 11.05	& 11.10	& 11.01	& 11.06	& 9.50	& 11.07 & 74\\
	& LLLB04$^{\ddagger}$& 11.05	& 11.02	& 11.02	& 11.06	& 10.97	& 10.90	& 10.98	& 9.52	& 10.99 & 69\\
NGC6884	& AC79	& 10.97	& 10.66	& 10.74	& 10.97	& 10.94	& 10.89	& 10.94	& 10.34	& 11.03 & 65\\
	& HAF97	& 10.89	& 10.65	& 10.99	& 10.87	& 10.92	& 10.83	& 10.87	& 10.29	& 10.98 & 73\\
	& KH01$^{*}$& 10.89	& --	& --	& 10.95	& 10.98	& 10.92	& 10.95	& 10.22	& 11.03 & 75\\
	& LLLB04$^{\ddagger}$& 11.00	& 10.96	& 10.94	& 10.98	& 10.93	& 10.84	& 10.92	& 10.19	& 10.99 & 72\\
NGC7009	& CA79	& 10.94	& 10.98	& 10.76	& 10.96	& 10.95	& 10.86	& 10.92	& 9.76	& 10.95 & 80\\
	& F81	& --	& --	& --	& 10.93	& 11.00	& 10.90	& 10.94	& 10.32	& 11.04 & 47\\
	& FL11	& 11.00	& 10.97	& 10.92	& 10.97	& 10.98	& 10.94	& 10.96	& 10.10	& 11.02 & 83\\
	& HA95a$^{\dagger}$& 11.08	& 11.01	& 10.98	& 10.97	& 10.98	& 11.04	& 11.00	& 9.69	& 11.02 & 75\\
	& HA95b	& 11.02	& 10.90	& 10.90	& 10.99	& 10.94	& 10.98	& 10.97	& 10.37	& 11.07 & 81\\
	& KC06$^{*}$& 11.02	& 10.95	& 11.03	& 10.99	& 11.02	& 10.92	& 10.98	& 10.15	& 11.04 & 85\\
	& KH98	& 10.98	& --	& 11.06	& 11.01	& 11.04	& 10.98	& 11.01	& 9.95	& 11.05 & 68\\
NGC7662	& AC83	& 10.69	& --	& 10.54	& 10.72	& 10.66	& 10.68	& 10.69	& 10.77	& 11.03 & 65\\
	& HA97b$^{*}$& 10.83	& 10.64	& 10.58	& 10.51	& 10.53	& 10.50	& 10.51	& 10.75	& 10.95 & 72\\
	& LLLB04	& 10.88	& 10.75	& 10.75	& 10.77	& 10.78	& 10.77	& 10.77	& 10.58	& 10.99 & 69\\
PB6	& GKA07$^{\dagger}$& --	& --	& --	& --	& 10.42	& --	& 10.42	& 11.09	& 11.17 & 64\\
	& GPP09$^{*}$& 10.64	& --	& --	& 10.59	& 10.49	& 10.50	& 10.53	& 11.14	& 11.23 & 74\\
	& MKHS10	& --	& --	& --	& 10.67	& 10.70	& 10.85	& 10.74	& 11.07	& 11.24 & 72\\
	& PSEK98	& --	& --	& --	& --	& 10.75	& --	& 10.75	& 11.01	& 11.20 & 41\\
PC14	& GKA07$^{\dagger\ddagger}$& --	& --	& --	& 11.03	& 11.03	& 11.03	& 11.03	& 9.16	& 11.04 & 74\\
	& GPMDR12$^{*}$& 11.03	& 11.03	& 11.01	& 11.02	& 11.03	& 11.00	& 11.02	& 9.58	& 11.03 & 99\\
	& MKHC02$^{\dagger\ddagger}$& 11.07	& --	& --	& 11.03	& 11.05	& 11.00	& 11.03	& 9.45	& 11.04 & 74\\
Pe1-1	& BK13	& --	& --	& --	& 11.09	& 11.01	& 11.03	& 11.05	& --	& 11.05 & 64\\
	& G14	& --	& --	& --	& 11.05	& 11.04	& 11.00	& 11.03	& --	& 11.03 & 72\\
	& GKA07$^{\dagger}$& --	& --	& --	& 11.00	& 11.00	& 10.97	& 10.99	& --	& 10.99 & 68\\
	& GPMDR12$^{*}$& 11.00	& 11.04	& 10.98	& 11.02	& 11.02	& 10.98	& 11.01	& 7.68	& 11.01 & 94\\
\hline
\end{tabular}
\\
\raggedright
$^\dagger$ O/H, N/H, and S/H are within 0.1~dex of the reference spectrum for calculations based on the blue [\ion{O}{ii}] lines.\\
$^\ddagger$ O/H, N/H, and S/H are within 0.1~dex of the reference spectrum for calculations based on the red [\ion{O}{ii}] lines.\\
\end{table*}

\begin{table*}
\centering
\caption{Total abundances, where $X=12+\log(X/\mbox{H})$. $X$(b) and $X$(r) are the abundances calculated using the blue and red lines of [\ion{O}{ii}]. The last column provides the final score of each spectrum. The spectra used as reference for each object are marked with asterisks. The references are identified in Table~\ref{tab:scores}.}
\label{tab:XH}
\begin{tabular}{llrrrrrrrrrrrrr}
\hline
PN & Ref. & O(b) & N(b) & Ne(b) & S(b) & Ar(b) & Cl(b) & O(r) & N(r) & Ne(r) & S(r) & Ar(r) & Cl(r) & $p_{\rm{tot}}$ \\
\hline
BoBn 1  & KHM03   & 7.80 & 7.69 & 8.10 & --   & 4.25 & --   & 7.78 & 7.96 & 8.10 & --   & 4.29 & --  & 60\\
        & KZGP08$^{*}$  & 7.78 & 7.72 & 8.00 & 5.21 & 4.25 & 3.42 & 7.76 & 7.85 & 8.00 & 5.26 & 4.27 & 3.44 & 89\\
        & OTHI10  & 7.78 & 7.95 & 8.04 & 4.92 & 4.34 & 3.47 & 7.79 & 7.87 & 8.04 & 4.90 & 4.33 & 3.46 & 73\\
Cn 1-5  & BK13	  & 8.76 & 8.65 & 8.73 & 7.12 & 6.68 & 5.55 & 8.83 & 8.51 & 8.83 & 7.09 & 6.65 & 5.53 & 84\\
        & EBW04   & 8.92 & 8.72 & 8.74 & --   & 6.66 & --   & 8.93 & 8.68 & 8.76 & --   & 6.66 & --   & 59\\
	& GPMDR12$^{*}$& 8.80	& 8.71	& 8.74	& 7.24	& 6.71	& 5.57	& 8.81	& 8.68	& 8.75	& 7.22	& 6.70	& 5.56 & 97\\
	& PSM01	& 8.76	& 8.21	& 8.84	& --	& --	& --	& --	& --	& --	& --	& --	& --  &  43\\
	& WL07	& 8.82	& 8.82	& 8.75	& 7.14	& 6.51	& 5.60	& 8.92	& 8.61	& 8.89	& 7.09	& 6.46	& 5.58 & 80\\
DdDm1	& BC84$^{\dagger\ddagger}$& 8.07& 7.41& 7.73 & 6.47 & 5.67 & --	& 8.09	& 7.35	& 7.76	& 6.45	& 5.66 & --  & 48\\
	& CPT87$^{\dagger}$& 8.11 & 7.43 & 7.79	& 6.45	& 5.74	& --	& 8.23	& 7.26	& 7.94	& 6.42	& 5.70	& --  &  57\\
	& KH98$^{*}$& 8.05	& 7.44	& 7.68	& 6.39	& 5.76	& 4.39	& 8.10	& 7.32	& 7.75	& 6.36	& 5.73	& 4.37 & 74\\
	& OHLIT09 & 8.11	& 7.34	& 7.89	& 6.34	& 5.77	& 4.51	& 8.02	& 7.73	& 7.72	& 6.47	& 5.85	& 4.58 & 65\\
H1-50	& EBW04	& 8.63	& 8.58	& 8.17	& 6.96	& 6.32	& --	& 8.64	& 8.33	& 8.17	& 6.90	& 6.31	& --   & 67\\
	& GDGD18$^{*}$& 8.64	& 8.30	& 8.13	& --	& 6.31	& 4.99	& 8.65	& 8.15	& 8.13	& --	& 6.30	& 4.96 & 84\\
	& WL07	& 8.68	& 8.34	& 8.15	& 6.97	& 6.27	& 5.12	& 8.68	& 8.27	& 8.15	& 6.95	& 6.26	& 5.10 & 78\\
H1-54	& EBW04	& 8.53	& 7.82	& 8.09	& 6.88	& 6.31	& 4.68	& 8.61	& 7.65	& 8.20	& 6.83	& 6.27	& 4.66 & 66\\
	& GCSC09  & 8.32	& 7.75	& 7.82	& 6.90	& 6.14	& 4.60	& 8.32	& 7.79	& 7.81	& 6.91	& 6.14	& 4.61 & 58\\
	& WL07$^{*}$& 8.30	& 7.86	& 7.64	& 6.87	& 6.01	& 4.58	& 8.33	& 7.62	& 7.67	& 6.79	& 5.97	& 4.53 & 87\\
H4-1	& KHM03	& 8.18	& 7.76	& 6.46	& 5.38	& 4.41	& --	& 8.23	& 7.65	& 6.46	& 5.35	& 4.39	& --   & 68\\
	& OT13$^{*}$& 8.24	& 7.62	& 6.45	& 5.11	& 4.33	& 3.97	& 8.20	& 7.68	& 6.45	& 5.12	& 4.34	& 3.98 & 84\\
	& TP79	& 8.31	& 7.78	& 6.59	& --	& --	& --	& 8.32	& 7.74	& 6.59	& --	& --	& --   & 57\\
Hb4	& AK87$^{\dagger}$& 8.74 & 8.64	& 8.19	& 7.21	& 6.58	& 5.42	& 8.75	& 8.48	& 8.19	& 7.17	& 6.57	& 5.38 & 55\\
	& GKA07	& 8.74	& 8.64	& 8.29	& 7.24	& 6.59	& 5.31	& 8.74	& 8.75	& 8.29	& 7.27	& 6.60	& 5.33 & 74\\
	& GPMDR12$^{*}$& 8.67	& 8.63	& 8.21	& 7.13	& 6.58	& 5.42	& 8.69	& 8.37	& 8.21	& 7.05	& 6.54	& 5.36 & 89\\
	& MKHS10  & 8.81	& 8.83	& 8.29	& 7.39	& 6.69	& 5.28	& 8.81	& 8.71	& 8.29	& 7.36	& 6.68	& 5.25 & 72\\
	& PSM01	& 8.75	& 8.75	& 8.07	& 7.27	& --	& --	& --	& --	& --	& --	& --	& --   & 45\\
He2-86	& G14$^{\ddagger}$& 8.73 & 8.68	& 8.51	& 7.39	& 6.85	& 5.24	& 8.74	& 8.65	& 8.52	& 7.39	& 6.85	& 5.23 & 76\\
	& GKA07	& 8.78	& 8.64	& 8.35	& 7.29	& 6.68	& 5.34	& 8.76	& 8.89	& 8.29	& 7.37	& 6.71	& 5.40 & 64\\
	& GPMDR12$^{*}$& 8.77	& 8.79	& 8.49	& 7.40	& 6.61	& 5.33	& 8.78	& 8.69	& 8.50	& 7.38	& 6.61	& 5.31 & 92\\
Hu1-2	& AC83	& 8.19	& 8.30	& 7.64	& 6.53	& 5.90	& 5.02	& 8.25	& 8.15	& 7.65	& 6.49	& 5.86	& 5.00 & 56\\
	& FGMR15  & 8.22	& 8.24	& 7.64	& 6.62	& --	& 5.05	& --	& --	& --	& --	& --	& --  &  69\\
	& HPF04	& 7.98	& 8.23	& 7.36	& 6.39	& 5.83	& 5.00	& --	& --	& --	& --	& --	& --   & 64\\
	& LLLB04  & 8.14	& 8.21	& 7.65	& 6.50	& 5.87	& 5.00	& 8.17	& 8.13	& 7.66	& 6.48	& 5.85	& 4.99 & 67\\
	& MKHS10  & 8.18	& 8.36	& 7.58	& 6.61	& 5.96	& 5.04	& 8.21	& 8.24	& 7.59	& 6.57	& 5.93	& 5.02 & 68\\
	& PT87	& 8.07	& 8.24	& 7.52	& 6.51	& 5.89	& --	& 8.12	& 8.07	& 7.54	& 6.46	& 5.85	& --   & 60\\
	& SCT87$^{*}$& 8.16	& 8.37	& 7.62	& --	& 5.91	& 4.98	& 8.25	& 8.09	& 7.63	& --	& 5.84	& 4.93 & 79\\
Hu2-1	& AC83	& 8.34	& 7.93	& 7.79	& 6.38	& 6.13	& 4.10	& 8.39	& 7.55	& 7.91	& 6.25	& 6.07	& 4.02 & 52\\
	& F81	& 7.99	& 7.57	& 7.18	& 6.18	& 6.01	& --	& 8.04	& 7.31	& 7.23	& 6.09	& 5.96	& --   & 45\\
	& KHM03$^{*}$& 8.62	& 7.63	& 7.84	& 6.41	& 6.12	& 4.74	& 8.55	& 7.76	& 7.78	& 6.44	& 6.15	& 4.75 & 77\\
	& WLB05	& 8.22	& 7.51	& 7.52	& 5.92	& 5.78	& 4.43	& 8.19	& 7.58	& 7.50	& 5.94	& 5.80	& 4.44 & 69\\
IC418	& AC83$^{\dagger\ddagger}$& 8.50 & 7.84	& 7.79	& 6.61	& 5.95	& 4.78	& 8.57	& 7.81	& 7.86 & 6.62 & 5.97& 4.79 & 57\\
	& DASNA17$^{\dagger\ddagger}$& 8.56 & 7.76 & 7.50 & 6.67 & 6.03	& 4.86	& 8.49	& 7.79	& 7.44 & 6.67 & 6.01 & 4.85 & 86\\
	& HAF94a  & 8.24	& 7.80	& 7.54	& 6.56	& 5.91	& 4.83	& 8.36	& 7.78	& 7.66	& 6.57	& 5.96	& 4.85 & 82\\
	& SWBvH03$^{*}$& 8.54	& 7.83	& 7.66	& 6.65	& 6.07	& 4.83	& 8.52	& 7.84	& 7.65	& 6.65	& 6.07	& 4.82 & 94\\
IC2003	& AC83	& 8.52	& 8.25	& 7.90	& 6.68	& 6.10	& 4.92	& --	& --	& --	& --	& --	& --   & 51\\
	& B78	& 8.35	& 8.12	& 7.78	& --	& --	& --	& 8.35	& 7.97	& 7.78	& --	& --	& --   & 53\\
	& KB94	& 8.56	& 8.20	& 7.98	& 6.57	& 5.97	& 5.30	& --	& --	& --	& --	& --	& --   & 47\\
	& WLB05$^{*}$& 8.48	& 8.14	& 7.92	& 6.68	& 6.11	& 5.03	& 8.49	& 8.02	& 7.92	& 6.65	& 6.11	& 5.01 & 75\\
IC4846	& AC79	& 8.63	& 7.88	& 8.00	& 6.84	& 6.11	& 4.83	& 8.63	& 7.99	& 8.00	& 6.87	& 6.12	& 4.85 & 57\\
	& B78	& 8.57	& 7.98	& 8.20	& --	& --	& --	& 8.62	& 7.52	& 8.31	& --	& --	& --   & 53\\
	& HAL01$^{*}$& 8.52	& 7.93	& 8.19	& 6.77	& 6.18	& 5.01	& 8.53	& 7.80	& 8.19	& 6.74	& 6.17	& 4.98 & 88\\
	& WL07	& 8.58	& 8.22	& 7.94	& 7.04	& 6.18	& 4.96	& 8.59	& 7.99	& 7.95	& 6.97	& 6.18	& 4.91 & 76\\
IC5217	& AC79	& 8.51	& 8.06	& 8.01	& 6.73	& 6.09	& 5.01	& 8.51	& 8.02	& 8.01	& 6.72	& 6.08	& 5.00 & 65\\
	& B78	& 8.47	& 8.34	& 7.97	& --	& --	& --	& 8.50	& 7.71	& 7.98	& --	& --	& --   & 51\\
	& F81	& 8.34	& 8.14	& 7.83	& 6.75	& 6.14	& --	& 8.35	& 7.83	& 7.83	& 6.67	& 6.12	& --   & 48\\
	& HAFL01	& 8.66	& 8.50	& 8.16	& 6.92	& 6.13	& 4.90	& 8.67	& 8.33	& 8.16	& 6.86	& 6.13	& 4.86 & 60\\
	& KH01$^{*}$& 8.50	& 8.25	& 7.99	& 6.81	& 6.18	& 5.00	& 8.50	& 8.17	& 7.99	& 6.79	& 6.18	& 4.98 & 73\\
	& PSM01	& 8.54	& 8.35	& 7.96	& --	& --	& --	& --	& --	& --	& --	& --	& --   & 45\\
M1-25	& GCSC09	& 8.72	& 8.47	& 7.91	& 7.23	& 6.67	& 5.39	& 8.70	& 8.50	& 7.89	& 7.23	& 6.68	& 5.40 & 58\\
	& GKA07	& 8.77	& 8.46	& 7.90	& 7.23	& 6.67	& 5.37	& 8.84	& 8.34	& 7.99	& 7.21	& 6.64	& 5.35 & 74\\
	& GPMDR12$^{*}$& 8.78	& 8.44	& 7.79	& 7.28	& 6.72	& 5.41	& 8.80	& 8.39	& 7.81	& 7.27	& 6.70	& 5.40 & 92\\
	& MKHC02	& 8.70	& 8.31	& 7.43	& 7.06	& 6.54	& 5.37	& 8.70	& 8.32	& 7.43	& 7.06	& 6.54	& 5.37 & 72\\
\hline
\end{tabular}
\end{table*}

\begin{table*}
\centering
\contcaption{Total abundances, where $X=12+\log(X/\mbox{H})$. $X$(b) and $X$(r) are the abundances calculated using the blue and red lines of [\ion{O}{ii}]. The last column provides the final score of each spectrum. The spectra used as reference for each object are marked with asterisks. The references are identified in Table~\ref{tab:scores}.}
\begin{tabular}{llrrrrrrrrrrrrr}
\hline
PN & Ref. & O(b) & N(b) & Ne(b) & S(b) & Ar(b) & Cl(b) & O(r) & N(r) & Ne(r) & S(r) & Ar(r) & Cl(r) & $p_{\rm{tot}}$ \\
\hline
M1-25	& PSM01	& 8.79	& 8.53	& 7.74	& 7.23	& --	& --	& --	& --	& --	& --	& --	& --   & 43\\
M1-29	& EBW04	& 8.82	 & 8.89	 & 8.25	 & 7.24	 & 6.81	 & 5.56	 & 8.96	 & 8.53	 & 8.25	 & 7.14	 & 6.73	 & 5.51 & 64\\
	& GCSC09$^{\dagger\ddagger}$& 8.69 & 8.74 & 8.18 & 7.11	 & 6.61	 & 5.53	 & 8.69	 & 8.77 & 8.18 & 7.12 & 6.61 & 5.54 & 58\\
	& WL07$^{*}$& 8.74	 & 8.72	 & 8.19	 & 7.16	 & 6.51	 & 5.50	 & 8.74	 & 8.74	 & 8.19	 & 7.16	 & 6.51	 & 5.51 & 73\\
M1-32	& GKA07	& 8.70	 & 8.39	 & 8.09	 & 6.96	 & 6.43	 & 5.32	 & 8.77	 & 8.37	 & 8.16	 & 6.96	 & 6.46	 & 5.33 & 73\\
	& GPMDR12$^{*}$& 8.54	 & 8.48	 & 7.58	 & 7.13	 & 6.55	 & 5.37	 & 8.69	 & 8.36	 & 7.69	 & 7.13	 & 6.58	 & 5.38 & 80\\
	& PSM01	& 8.45	 & 8.39	 & 7.49	 & 6.87	 & --	 & --	 & --	 & --	 & --	 & --	 & --	 & --   & 43\\
M1-74	& AC83	& 8.71	 & 8.07	 & 8.38	 & 7.13	 & 6.66	 & 5.00	 & 8.70	 & 8.12	 & 8.38	 & 7.15	 & 6.67	 & 5.01 & 55\\
	& KH01$^{*}$& 8.63	 & 8.71	 & 8.26	 & 7.52	 & 6.67	 & --	 & --	 & --	 & --	 & --	 & --	 & --  &  76\\
	& WLB05	& 8.63	 & 8.26	 & 8.17	 & 7.16	 & 6.56	 & 5.08	 & 8.62	 & 8.49	 & 8.16	 & 7.22	 & 6.58	 & 5.14 & 66\\
M2-23	& EBW04	& 8.39	 & 8.33	 & 7.94	 & 6.96	 & 6.19	 & 4.52	 & 8.40	 & 7.56	 & 7.97	 & 6.66	 & 6.17	 & 4.34 & 66\\
	& GCSC09	& 8.29	 & 8.02	 & 7.93	 & --	 & 6.03	 & 4.38	 & 8.29	 & 7.89	 & 7.93	 & --	 & 6.02	 & 4.35 & 58\\
	& WL07$^{*}$& 8.34	 & 8.18	 & 7.75	 & 7.03	 & 6.07	 & 4.56	 & 8.35	 & 7.81	 & 7.76	 & 6.87	 & 6.06	 & 4.47 & 74\\
M3-15	& MKHC02	& 8.81	 & 8.67	 & 8.29	 & 7.49	 & 6.66	 & 5.02	 & 8.81	 & 8.84	 & 8.29	 & 7.56	 & 6.67	 & 5.06 & 59\\
	& GKA07	& 8.38	 & 7.98	 & 8.00	 & 6.70	 & 6.29	 & 5.00	 & 8.37	 & 8.11	 & 7.98	 & 6.74	 & 6.30	 & 5.03 & 61\\
	& GPMDR12$^{*}$& 8.80	 & 8.43	 & 8.31	 & 7.31	 & 6.64	 & 5.28	 & 8.80	 & 8.29	 & 8.32	 & 7.27	 & 6.63	 & 5.25 & 80\\
	& PSM01	& 8.78	 & 8.34	 & 8.15	 & 7.17	 & --	 & --	 & --	 & --	 & --	 & --	 & --	 & --   & 43\\
Me2-2	& AK87	& 8.36	 & 8.96	 & 7.98	 & 6.47	 & 6.02	 & 4.49	 & 8.38	 & 8.47	 & 8.04	 & 6.35	 & 5.97	 & 4.38 & 66\\
	& B78	& 8.29	 & 9.03	 & 7.91	 & --	 & --	 & --	 & 8.33	 & 8.19	 & 8.02	 & --	 & --	 & --   & 56\\
	& MKHS10$^{*}$& 8.44	 & 8.80	 & 8.01	 & 6.56	 & 6.17	 & 4.70	 & 8.45	 & 8.61	 & 8.04	 & 6.52	 & 6.15	 & 4.65 & 70\\
	& WLB05	& 8.19	 & 8.68	 & 7.63	 & 6.17	 & 5.96	 & 4.76	 & 8.21	 & 8.34	 & 7.65	 & 6.08	 & 5.92	 & 4.68 & 67\\
NGC40	& AC79	& 8.71	 & 7.98	 & --	 & 6.13	 & 5.90	 & 5.28	 & 8.70	 & 7.98	 & --	 & 6.13	 & 5.90	 & 5.28 & 56\\
	& CSPT83$^{\ddagger}$& 8.91 & 8.04 & --	 & 6.40	 & 6.12	 & 5.21	 & 8.72	 & 8.05	 & --	 & 6.40	 & 6.09	 & 5.19 & 67\\
	& LLLB04$^{*}$& 8.65	 & 7.96	 & 7.21	 & 6.45	 & 6.22	 & 5.21	 & 8.70	 & 7.96	 & 7.26	 & 6.45	 & 6.24	 & 5.22 & 77\\
	& PSM01$^{\dagger}$& 8.59 & 7.94 & --	 & 6.51	 & 6.20	 & --	 & --	 & --	 & --	 & --	 & --	 & --   & 43\\
NGC650	& AC83	& 8.81	 & 8.64	 & 8.32	 & 7.02	 & 6.28	 & --	 & 8.88	 & 8.56	 & 8.32	 & 7.01	 & 6.27	 & --   & 51\\
	& KHM03$^{*}$& 8.61	 & 8.44	 & 8.17	 & 6.87	 & 6.41	 & 5.07	 & 8.67	 & 8.37	 & 8.17	 & 6.86	 & 6.40	 & 5.07 & 68\\
	& PT87	& 8.74	 & 8.47	 & 8.16	 & 6.95	 & 6.39	 & --	 & 8.84	 & 8.37	 & 8.16	 & 6.94	 & 6.41	 & --   & 64\\
NGC2392	& B91	& 8.41	 & 8.10	 & 7.68	 & 6.45	 & 5.88	 & 5.11	 & --	 & --	 & --	 & --	 & --	 & --   & 54\\
	& DRMV09$^{\dagger\ddagger}$& 8.26 & 7.90 & 7.73 & 6.55	 & 5.93	 & 4.93	 & 8.26	 & 7.91 & 7.73 & 6.56 & 5.93 & 4.94 & 68\\
	& HKB00$^{\ddagger}$& 8.26 & 7.98 & 7.70 & 6.59	 & 5.95	 & 4.89	 & 8.27	 & 7.96	 & 7.70	 & 6.58	 & 5.95	 & 4.89 & 63\\
	& Z76	& 8.34	 & 8.30	 & 7.71	 & 6.99	 & 6.16	 & --	 & 8.37	 & 7.97	 & 7.71	 & 6.89	 & 6.12	 & --   & 40\\
	& ZFCH12$^{*}$& 8.36	 & 7.85	 & 7.66	 & 6.48	 & 5.87	 & 5.11	 & 8.34	 & 7.90	 & 7.66	 & 6.50	 & 5.88	 & 5.12 & 71\\
NGC2440	& DKSH15	& 8.53	 & 8.82	 & 7.75	 & 6.32	 & 6.18	 & --	 & 8.59	 & 8.70	 & 7.75	 & 6.30	 & 6.16	 & --   & 67\\
	& HA98	& 8.69	 & 9.13	 & 7.96	 & 6.49	 & 6.27	 & 5.35	 & 8.75	 & 8.98	 & 7.97	 & 6.45	 & 6.24	 & 5.33 & 72\\
	& KB94	& 8.69	 & 8.72	 & 7.91	 & 6.50	 & 6.23	 & --	 & 8.72	 & 8.70	 & 7.91	 & 6.50	 & 6.24	 & --   & 44\\
	& KC06$^{*}$& 8.53	 & 8.82	 & 7.83	 & 6.59	 & --	 & 5.14	 & --	 & --	 & --	 & --	 & --	 & --  &  80\\
	& KHM03	& 8.61	 & 8.81	 & 7.91	 & 6.76	 & 6.33	 & 5.45	 & 8.58	 & 8.90	 & 7.91	 & 6.79	 & 6.35	 & 5.46 & 67\\
	& PT87$^{\dagger}$& 8.56 & 8.79	 & 7.88	 & 6.63	 & 6.33	 & 5.37	 & 8.57	 & 8.77	 & 7.88	 & 6.62	 & 6.32	 & 5.37 & 62\\
	& SAKC81	& 8.60	 & 8.91	 & 7.88	 & 6.41	 & 6.27	 & 5.18	 & 8.63	 & 8.82	 & 7.88	 & 6.39	 & 6.25	 & 5.17 & 76\\
NGC2867	& AKRO81	& 8.71	 & 8.14	 & 8.05	 & 6.69	 & 6.18	 & 5.16	 & 8.70	 & 8.17	 & 8.05	 & 6.70	 & 6.19	 & 5.17 & 62\\
	& GKA07	& 8.60	 & 7.94	 & 8.00	 & 6.68	 & 6.11	 & 5.09	 & 8.57	 & 8.19	 & 8.00	 & 6.77	 & 6.15	 & 5.15 & 66\\
	& GPP09$^{*}$& 8.59	 & 8.06	 & 8.01	 & 6.75	 & 6.22	 & 5.15	 & --	 & --	 & --	 & --	 & --	 & --  &  82\\
	& KB94	& 8.78	 & 8.16	 & 8.14	 & 6.80	 & 6.26	 & 5.17	 & 8.72	 & 8.98	 & 8.14	 & 7.03	 & 6.36	 & 5.35 & 57\\
	& MKHC02$^{\dagger}$& 8.61 & 8.12 & 8.00 & 6.71	 & 6.20	 & 5.15	 & 8.63	 & 7.97	 & 8.00	 & 6.65	 & 6.17	 & 5.12 & 79\\
	& PSEK98	& 8.52	 & 8.10	 & 7.92	 & 6.65	 & 6.05	 & --	 & 8.54	 & 7.97	 & 7.92	 & 6.61	 & 6.02	 & --   & 41\\
NGC6210	& B78	& 8.65	 & 7.99	 & 8.18	 & --	 & --	 & --	 & --	 & --	 & --	 & --	 & --	 & --   & 52\\
	& BERD15	& 8.66	 & 7.99	 & 8.21	 & 6.88	 & --	 & 5.13	 & --	 & --	 & --	 & --	 & --	 & --   & 77\\
	& DRMV09	& 8.57	 & 7.95	 & 8.12	 & 6.80	 & 6.20	 & 4.87	 & 8.57	 & 7.90	 & 8.12	 & 6.79	 & 6.19	 & 4.86 & 63\\
	& F81	& 8.49	 & 8.00	 & 8.03	 & 6.93	 & 6.32	 & --	 & 8.51	 & 7.62	 & 8.04	 & 6.84	 & 6.29	 & --   & 47\\
	& KH98$^{\ddagger}$& 8.69 & 8.12 & 8.24	 & 6.94	 & 6.30	 & 5.05	 & 8.70	 & 7.94	 & 8.25	 & 6.90	 & 6.29	 & 5.01 & 61\\
	& LLLB04$^{*}$& 8.67	 & 7.99	 & 8.22	 & 6.88	 & 6.26	 & 5.03	 & 8.68	 & 7.92	 & 8.22	 & 6.87	 & 6.25	 & 5.01 & 80\\
NGC6302	& KC06	& 8.23	 & 8.72	 & 7.70	 & 6.65	 & --	 & 4.97	 & --	 & --	 & --	 & --	 & --	 & --   & 80\\
	& MKHS10	& 8.22	 & 8.68	 & 7.68	 & 6.73	 & 6.17	 & 5.10	 & 8.25	 & 8.58	 & 7.68	 & 6.70	 & 6.15	 & 5.09 & 67\\
	& RCK14	& 8.16	 & 8.87	 & 7.62	 & 6.75	 & 6.21	 & 5.13	 & 8.22	 & 8.61	 & 7.63	 & 6.66	 & 6.16	 & 5.08 & 80\\
	& TBLDS03	& 8.27	 & 8.90	 & 7.81	 & 6.90	 & 6.24	 & 5.09	 & 8.29	 & 8.80	 & 7.81	 & 6.86	 & 6.22	 & 5.07 & 80\\
NGC6369	& AK87	& 8.48	 & 8.16	 & 7.84	 & 6.57	 & 6.23	 & 5.07	 & 8.48	 & 8.31	 & 7.83	 & 6.61	 & 6.24	 & 5.11 & 55\\
	& GKA07	& 8.95	 & 8.36	 & 8.76	 & 7.29	 & 6.49	 & --	 & 8.98	 & 8.16	 & 8.83	 & 7.23	 & 6.45	 & --   & 67\\
	& GPMDR12$^{*}$& 8.53	 & 8.17	 & 8.03	 & 6.91	 & 6.39	 & 5.02	 & 8.53	 & 8.00	 & 8.03	 & 6.87	 & 6.38	 & 4.98 & 93\\
	& MKHS10	& 8.71	 & 8.37	 & 8.17	 & 7.03	 & 6.43	 & 5.17	 & 8.72	 & 8.23	 & 8.18	 & 6.99	 & 6.42	 & 5.14 & 67\\
	& PSM01	& 8.70	 & 8.30	 & 8.10	 & 7.50	 & --	 & --	 & --	 & --	 & --	 & --	 & --	 & --  &  45\\
NGC6543	& AC79	& 8.72	 & 8.11	 & 8.39	 & 7.02	 & 6.54	 & 5.15	 & 8.71	 & 8.19	 & 8.38	 & 7.03	 & 6.54	 & 5.17 & 64\\
\hline
\end{tabular}
\end{table*}

\begin{table*}
\centering
\contcaption{Total abundances, where $X=12+\log(X/\mbox{H})$. $X$(b) and $X$(r) are the abundances calculated using the blue and red lines of [\ion{O}{ii}]. The last column provides the final score of each spectrum. The spectra used as reference for each object are marked with asterisks. The references are identified in Table~\ref{tab:scores}.}
\begin{tabular}{llrrrrrrrrrrrrr}
\hline
PN & Ref. & O(b) & N(b) & Ne(b) & S(b) & Ar(b) & Cl(b) & O(r) & N(r) & Ne(r) & S(r) & Ar(r) & Cl(r) & $p_{\rm{tot}}$ \\
\hline
NGC6543	& HAFLK00	& 8.76	 & 8.36	 & 8.33	 & 7.18	 & 6.76	 & 5.44	 & 8.76	 & 8.32	 & 8.33	 & 7.17	 & 6.75	 & 5.43 & 75\\
	& PSM01	& 8.82	 & 8.36	 & 8.52	 & 7.30	 & --	 & --	 & --	 & --	 & --	 & --	 & --	 & --   & 45\\
        & WL04$^{*}$& 8.80	& 8.37	& 8.46	& 7.30	& 6.75	& 5.28	& 8.80	& 8.35	& 8.46	& 7.29	& 6.75	& 5.27 & 80\\
NGC6565	& EBW04	& 8.64	& 8.50	& 7.99	& 6.99	& 6.49	& 5.25	& 8.79	& 8.32	& 8.00	& 6.97	& 6.47	& 5.25 & 65\\
	& MKHC02$^{*}$& 8.74	& 8.45	& 8.22	& 6.94	& 6.42	& 5.28	& 8.74	& 8.46	& 8.22	& 6.94	& 6.42	& 5.28 & 82\\
	& WL07	& 8.74	& 8.57	& 8.21	& 7.10	& 6.41	& 5.22	& 8.76	& 8.50	& 8.21	& 7.08	& 6.39	& 5.21 & 77\\
NGC6572	& F81	& 8.23	& 8.24	& 7.77	& 6.53	& 6.29	& --	& 8.26	& 7.72	& 7.80	& 6.40	& 6.23	& -- & 47\\
	& GKA07	& 8.61	& 8.43	& 8.12	& 6.71	& 6.41	& 4.83	& 8.62	& 8.20	& 8.13	& 6.65	& 6.39	& 4.78 & 75\\
	& HAF94b$^{*}$& 8.60	& 8.15	& 8.15	& 6.51	& 6.34	& 4.79	& 8.60	& 8.25	& 8.15	& 6.53	& 6.35	& 4.81 & 88\\
	& KH01	& 8.60	& 8.51	& 8.11	& 6.86	& 6.46	& 4.85	& --	& --	& --	& --	& --	& -- & 75\\
	& LLLB04	& 8.63	& 8.30	& 8.16	& 6.74	& 6.38	& 4.93	& 8.63	& 8.28	& 8.16	& 6.74	& 6.37	& 4.93 & 79\\
NGC6720	& B80	& 8.94	& 8.45	& 8.41	& --	& 6.51	& --	& 8.88	& 8.58	& 8.41	& --	& 6.54	& -- & 47\\
	& F81	& 8.64	& 8.44	& 8.07	& 6.88	& 6.47	& --	& 8.74	& 8.29	& 8.07	& 6.85	& 6.44	& -- & 46\\
	& GMC97$^{\dagger\ddagger}$& 8.78 & 8.28 & 8.39	& 6.68	& 6.38	& --	& 8.66	& 8.38	& 8.36	& 6.68 & 6.35 & -- & 70\\
	& HM77	& 8.99	& 8.43	& 8.42	& --	& --	& --	& --	& --	& --	& --	& --	& -- & 48\\
	& KH98$^{\dagger\ddagger}$& 8.71 & 8.37	& 8.20	& 6.81	& 6.48	& 5.32	& 8.72	& 8.32	& 8.20 & 6.79 & 6.47 & 5.31 & 73\\
	& LLLB04$^{*}$& 8.74	& 8.38	& 8.23	& 6.72	& 6.41	& 5.16	& 8.75	& 8.37	& 8.23	& 6.72	& 6.40	& 5.16 & 78\\
NGC6741	& HA97a$^{*}$& 8.68	& 8.46	& 8.12	& 6.85	& 6.41	& 5.25	& 8.69	& 8.43	& 8.12	& 6.84	& 6.41	& 5.24 & 84\\
	& LLLB04$^{\dagger\ddagger}$& 8.67 & 8.37 & 8.07 & 6.82	& 6.36	& 5.20	& 8.67	& 8.37	& 8.07 & 6.82 & 6.36 & 5.20 & 77\\
	& MKHS10	& 8.72	& 8.63	& 8.07	& 7.01	& 6.53	& 5.20	& 8.76	& 8.49	& 8.07	& 6.96	& 6.50	& 5.17 & 78\\
NGC6751	& AC79	& 8.79	& 8.14	& 8.67	& 6.62	& 6.26	& 5.62	& 8.81	& 8.12	& 8.69	& 6.62	& 6.27	& 5.62 & 63\\
	& CMJK91	& 8.69	& 8.28	& 8.50	& 6.92	& --	& 5.30	& --	& --	& --	& --	& --	& -- & 75\\
	& KB94	& 8.96	& 8.45	& 8.16	& 6.97	& 6.29	& --	& 8.97	& 8.41	& 8.16	& 6.96	& 6.28	& -- & 46\\
	& MKHS10$^{*}$& 8.61	& 8.41	& 8.25	& 6.93	& 6.47	& 5.41	& 8.70	& 8.17	& 8.40	& 6.86	& 6.42	& 5.38 & 80\\
	& PSM01	& 8.70	& 8.28	& 8.43	& 6.95	& 6.36	& --	& --	& --	& --	& --	& --	& -- & 43\\
NGC6790	& AC79	& 8.56	& 8.15	& 7.90	& 6.65	& 5.74	& 4.25	& 8.56	& 8.09	& 7.90	& 6.62	& 5.74	& 4.23 & 59\\
	& AHF96	& 8.73	& 8.09	& 8.06	& 6.60	& 5.97	& 4.52	& 8.73	& 8.00	& 8.06	& 6.58	& 5.97	& 4.50 & 66\\
	& F81	& 8.20	& 7.99	& 7.63	& 6.46	& 5.89	& --	& 8.21	& 7.50	& 7.63	& 6.34	& 5.86	& --   & 46\\
	& KH01$^{*}$& 8.37	& 8.45	& 7.80	& 6.79	& 5.94	& 4.29	& 8.38	& 7.73	& 7.80	& 6.46	& 5.92	& 4.13 & 74\\
	& LLLB04$^{\ddagger}$& 8.41 & 7.79 & 7.85 & 6.36 & 5.77	& 4.46	& 8.41	& 7.80	& 7.85	& 6.36	& 5.77	& 4.46 & 69\\
NGC6884	& AC79	& 8.65	& 7.99	& 8.07	& 6.71	& 6.42	& 4.94	& 8.64	& 8.21	& 8.07	& 6.76	& 6.44	& 4.99 & 65\\
	& HAF97	& 8.95	& 8.67	& 8.42	& 6.96	& 6.48	& 5.32	& 8.95	& 8.64	& 8.42	& 6.96	& 6.48	& 5.31 & 73\\
	& KH01$^{*}$& 8.66	& 8.41	& 8.11	& 6.85	& 6.44	& 5.29	& 8.67	& 8.32	& 8.11	& 6.83	& 6.43	& 5.27 & 75\\
	& LLLB04$^{\ddagger}$& 8.63 & 8.28 & 8.13 & 6.76 & 6.32	& 5.12	& 8.63	& 8.32	& 8.13	& 6.77	& 6.32	& 5.13 & 72\\
NGC7009	& CA79	& 8.84	& 8.28	& 8.39	& 7.18	& 6.50	& 5.23	& 8.83	& 8.45	& 8.39	& 7.23	& 6.51	& 5.27 & 80\\
	& F81	& 8.31	& 7.65	& 7.85	& 6.67	& 6.30	& --	& 8.31	& 7.45	& 7.85	& 6.61	& 6.29	& -- & 47\\
	& FL11	& 8.67	& 8.23	& 8.26	& 7.01	& 6.41	& 5.14	& 8.67	& 8.25	& 8.26	& 7.01	& 6.41	& 5.14 & 83\\
	& HA95a$^{\dagger}$& 8.69 & 8.36 & 8.30	& 7.02	& 6.47	& 5.37	& 8.70	& 8.21	& 8.30	& 6.98	& 6.45	& 5.34 & 75\\
	& HA95b	& 8.52	& 7.94	& 8.21	& 6.88	& 6.27	& 4.92	& 8.52	& 7.99	& 8.21	& 6.91	& 6.27	& 4.93 & 81\\
	& KC06$^{*}$& 8.67	& 8.33	& 8.22	& 7.02	& --	& 5.22	& --	& --	& --	& --	& --	& -- & 85\\
	& KH98	& 8.71	& 8.44	& 8.15	& 7.19	& 6.39	& 5.43	& 8.71	& 8.31	& 8.15	& 7.11	& 6.38	& 5.40 & 68\\
NGC7662	& AC83	& 8.44	& 7.74	& 7.86	& 6.57	& 6.11	& 5.33	& 8.46	& 7.48	& 7.86	& 6.50	& 6.08	& 5.27 & 65\\
	& HA97b$^{*}$& 8.62	& 7.91	& 8.02	& 6.92	& 6.05	& 5.06	& 8.62	& 8.04	& 8.02	& 7.01	& 6.05	& 5.09 & 72\\
	& LLLB04	& 8.42	& 7.84	& 7.82	& 6.64	& 6.05	& 5.10	& 8.42	& 7.70	& 7.82	& 6.60	& 6.04	& 5.06 & 69\\
PB6	& GKA07$^{\dagger}$& 8.65 & 8.82 & 8.38	& 7.01	& 6.42	& --	& 8.70	& 8.64	& 8.46	& 6.95	& 6.38	& -- & 64\\
	& GPP09$^{*}$& 8.64	& 8.76	& 8.26	& 7.00	& 6.48	& --	& --	& --	& --	& --	& --	& -- & 74\\
	& MKHS10	& 8.56	& 8.80	& 8.06	& 6.79	& 6.34	& 5.38	& 8.60	& 8.63	& 8.08	& 6.74	& 6.30	& 5.35 & 72\\
	& PSEK98	& 8.56	& 8.78	& 8.05	& 6.79	& 6.30	& --	& 8.65	& 8.55	& 8.09	& 6.73	& 6.25	& -- & 41\\
PC14	& GKA07$^{\dagger\ddagger}$& 8.81 & 8.27 & 8.31	& 7.18	& 6.49	& 5.27	& 8.81	& 8.20 & 8.32 & 7.16 & 6.48 & 5.25 & 74\\
	& GPMDR12$^{*}$& 8.76	& 8.21	& 8.28	& 7.14	& 6.48	& 5.29	& 8.77	& 8.12	& 8.28	& 7.11	& 6.47	& 5.27 & 99\\
	& MKHC02$^{\dagger\ddagger}$& 8.86 & 8.12 & 8.38 & 7.10	& 6.47	& 5.38	& 8.86	& 8.16 & 8.38 & 7.11 & 6.48 & 5.38 & 74\\
Pe1-1	& BK13	& 8.61	& 8.07	& 8.28	& 6.72	& 6.39	& 5.00	& 8.62	& 8.04	& 8.29	& 6.71	& 6.38	& 4.99 & 64\\
	& G14	& 8.48	& 8.25	& 8.14	& 6.87	& 6.42	& 4.93	& 8.52	& 7.99	& 8.22	& 6.78	& 6.37	& 4.87 & 72\\
	& GKA07	& 8.69	& 8.16	& 8.42	& 6.87	& 6.38	& 4.86	& 8.68	& 8.24	& 8.39	& 6.90	& 6.39	& 4.88 & 68\\
	& GPMDR12$^{*}$& 8.63	& 8.13	& 8.13	& 6.84	& 6.43	& 5.04	& 8.63	& 8.10	& 8.14	& 6.83	& 6.43	& 5.03 & 94\\
\hline
\end{tabular}
\\
\raggedright
$^\dagger$ O/H, N/H, and S/H are within 0.1~dex of the reference spectrum for calculations based on the blue [\ion{O}{ii}] lines.\\
$^\ddagger$ O/H, N/H, and S/H are within 0.1~dex of the reference spectrum for calculations based on the red [\ion{O}{ii}] lines.\\
\end{table*}

\begin{table*}
\centering
\caption{Partial and final scores for the sample spectra. The reference spectra for each object are marked with asterisks.}
\label{tab:scores}
\begin{tabular}{llcccrrrrrcrcrrl}
\hline
PN & Ref.& $p_0$ & $p_1$ & $n_{\rm{e}}$ & $p_2$ & [\ion{N}{ii}] & $p_3$ & [\ion{O}{iii}] & $p_4$ & \ion{He}{i} & $p_5$ & $\alpha\gamma\delta$ & P$-$B & $p_6$ & $p_{\rm{tot}}$ \\
\hline
BoBn 1	& KHM03   & 10 & 20 & 1 & 9  & 2.84 & 1  & --   & 0  & 4 &14&2&--   &6 &60\\
        & KZGP08$^{*}$  & 10 & 20 & 2 & 11 & 3.00 & 9  & 2.97 & 9  & 6 &18&3&0.08 &12&89\\
        & OTHI10  & 10 & 20 & 3 & 13 & 3.24 & 1  & 2.87 & 4  & 6 &19&3&0.48 &6 &73\\
Cn 1-5	& BK13    & 10 & 20 & 2 & 11 & 3.08 & 8  & 3.01 & 9  & 6 &20&3&--   &6 &84\\
        & EBW04   & 10 & 20 & 2 & 11 & 2.72 & 0  & 3.67 & 0  & 4 &12&3&--   &6 &59\\
        & GPMDR12$^{*}$ & 10 & 20 & 4 & 15 & 3.07 & 8  & 3.02 & 9  & 6 &20&3&0.02 &15&97\\
        & PSM01   & 10 & 20 & 2 & 11 & --   & 0  & --	& 0  & 1 &2 &0&--   &0 &43\\
        & WL07    & 10 & 20 & 3 & 13 & 3.14 & 5  & 2.98 & 9  & 5 &17&3&--   &6 &80\\
DdDm 1	& BC84$^{\dagger\ddagger}$    & 0  & 20 & 1 & 9  & --   & 0  & 2.83 & 2  & 3 &11&3&--   &6 &48\\
        & CPT87$^{\dagger}$   & 0  & 20 & 1 & 9  & --   & 0  & 2.90 & 5  & 5 &17&3&--   &6 &57\\
        & KH98$^{*}$	  & 10 & 20 & 2 & 11 & 3.12 & 6  & 3.06 & 7  & 4 &14&2&--   &6 &74\\
        & OHLIT09 & 10 & 20 & 4 & 15 & 3.54 & 0  & 3.16 & 3  & 6 &17&3&--   &0 &65\\
H 1-50	& EBW04   & 10 & 20 & 1 & 9  & 3.03 & 10 & 3.12 & 5  & 2 &7 &3&--   &6 &67\\
        & GDGD18$^{*}$  & 10 & 20 & 4 & 15 & 3.09 & 7  & 2.85 & 3  & 5 &17&3&-0.06&12&84\\
        & WL07    & 10 & 20 & 4 & 15 & 2.97 & 7  & 2.79 & 0  & 6 &20&3&--   &6 &78\\
H 1-54	& EBW04   & 10 & 20 & 2 & 11 & 2.83 & 1  & 3.07 & 7  & 4 &11&3&--   &6 &66\\
        & GCSC09  & 10 & 20 & 2 & 11 & --   & 0  & --   & 0  & 3 &11&3&--   &6 &58\\
        & WL07$^{*}$    & 10 & 20 & 3 & 13 & 3.03 & 10 & 3.04 & 8  & 6 &20&3&--   &6 &87\\
H 4-1	& KHM03   & 10 & 20 & 1 & 9  & 3.42 & 0  & 3.03 & 9  & 4 &14&2&--   &6 &68\\
        & OT13$^{*}$    & 10 & 20 & 4 & 15 & 2.91 & 5  & 2.98 & 9  & 6 &19&3&--	&6 &84\\
        & TP79    & 0  & 20 & 1 & 9  & 3.16 & 4  & 3.16 & 3  & 5 &15&2&--	&6 &57\\
Hb 4	& AK87$^{\dagger}$    & 0  & 20 & 2 & 11 & --   & 0  & 2.93 & 7  & 3 &11&3&--   &6 &55\\
        & GKA07   & 10 & 20 & 2 & 11 & 2.97 & 7  & 3.01 & 9  & 3 &11&3&--   &6 &74\\
        & GPMDR12$^{*}$ & 10 & 20 & 4 & 15 & 3.01 & 9  & 3.00 & 10 & 5 &16&3&0.13 &9 &89\\
        & MKHS10  & 10 & 20 & 2 & 11 & 3.05 & 9  & 3.03 & 9  & 4 &13&3&--   &0 &72\\
        & PSM01   & 10 & 20 & 3 & 13 & --   & 0  & --   & 0  & 1 &2 &0&--   &0 &45\\
He 2-86	& G14$^{\ddagger}$	  & 10 & 20 & 2 & 11 & 3.00 & 9  & 3.02 & 9  & 3 &11&3&--   &6 &76\\
        & GKA07   & 10 & 20 & 2 & 11 & 3.14 & 5  & 2.96 & 8  & 3 &10&3&--   &0 &64\\
        & GPMDR12$^{*}$ & 10 & 20 & 4 & 15 & 3.07 & 8  & 3.06 & 7  & 6 &20&3&0.10 &12&92\\
Hu 1-2	& AC83    & 0  & 20 & 2 & 11 & --   & 0  & 2.89 & 5  & 5 &14&3&--   &6 &56\\
        & FGMR15  & 10 & 20 & 2 & 11 & 3.18 & 3  & 3.02 & 9  & 6 &16&3&--   &0 &69\\
        & HPF04   & 10 & 20 & 4 & 15 & 2.90 & 4  & 3.24 & 0  & 6 &15&3&--   &0 &64\\
        & LLLB04  & 10 & 20 & 4 & 15 & 3.01 & 9  & --   & 0  & 6 &13&3&--   &0 &67\\
        & MKHS10  & 10 & 20 & 2 & 11 & 3.08 & 8  & 2.97 & 8  & 3 &11&3&--   &0 &68\\
        & PT87    & 0  & 20 & 1 & 9  & 2.95 & 6  & 2.95 & 8  & 3 &11&2&--   &6 &60\\
        & SCT87$^{*}$   & 10 & 20 & 3 & 13 & 3.00 & 9  & 3.00 & 10 & 4 &11&3&--   &6 &79\\
Hu 2-1	& AC83    & 0  & 20 & 1 & 9  & --   & 0  & 2.90 & 5  & 4 &12&3&--   &6 &52\\
        & F81     & 0  & 20 & 1 & 9  & 2.70 & 0  & 2.73 & 0  & 3 &10&3&--   &6 &45\\
        & KHM03$^{*}$   & 10 & 20 & 2 & 11 & 3.01 & 9  & 3.06 & 7  & 4 &14&2&--   &6 &77\\
        & WLB05   & 10 & 20 & 1 & 9  & 2.99 & 8  & 2.97 & 9  & 6 &13&3&--   &0 &69\\
IC 418	& AC83$^{\dagger\ddagger}$    & 0  & 20 & 3 & 13 & --   & 0  & 3.06 & 7  & 3 &11&3&--   &6 &57\\
        & DASNA17$^{\dagger\ddagger}$ & 10 & 20 & 3 & 13 & 2.99 & 8  & --   & 0  & 6 &20&3&-0.03&15&86\\
        & HAF94a  & 10 & 20 & 3 & 13 & 2.90 & 4  & 2.91 & 6  & 6 &20&3&-0.05&9 &82\\
        & SWBvH03$^{*}$ & 10 & 20 & 4 & 15 & 3.04 & 9  & 2.96 & 8  & 6 &20&3&-0.06&12&94\\
IC 2003	& AC83    & 0  & 20 & 2 & 11 & --   & 0  & 3.23 & 0  & 4 &14&3&--   &6 &51\\
        & B78     & 0  & 20 & 1 & 9  & --   & 0  & 2.96 & 8  & 3 &10&3&--   &6 &53\\
        & KB94    & 0  & 20 & 2 & 11 & --   & 0  & --   & 0  & 4 &10&3&--   &6 &47\\
        & WLB05$^{*}$   & 10 & 20 & 4 & 15 & 2.60 & 0  & 3.02 & 9  & 5 &15&3&--   &6 &75\\
IC 4846	& AC79    & 0  & 20 & 3 & 13 & --   & 0  & 3.09 & 6  & 4 &12&3&--   &6 &57\\
        & B78     & 0  & 20 & 1 & 9  & --   & 0  & 2.95 & 8  & 3 &10&3&--   &6 &53\\
        & HAL01$^{*}$   & 10 & 20 & 4 & 15 & 3.01 & 9  & 2.97 & 9  & 6 &16&3&0.04 &9 &88\\
        & WL07    & 10 & 20 & 3 & 13 & 2.98 & 8  & 2.89 & 5  & 6 &20&3&--   &0 &76\\
IC 5217	& AC79    & 0  & 20 & 3 & 13 & 3.16 & 4  & 2.82 & 2  & 6 &20&3&--   &6 &65\\
        & B78     & 0  & 20 & 1 & 9  & --   & 0  & 2.90 & 5  & 3 &11&3&--   &6 &51\\
        & F81     & 0  & 20 & 2 & 11 & --   & 0  & 2.76 & 0  & 3 &11&3&--   &6 &48\\
        & HAFL01  & 10 & 20 & 4 & 15 & 3.48 & 0  & 2.38 & 0  & 6 &15&3&-0.21&0 &60\\
        & KH01$^{*}$    & 10 & 20 & 2 & 11 & 2.81 & 0  & 3.02 & 9  & 5 &17&3&--   &6 &73\\
        & PSM01   & 10 & 20 & 3 & 13 & --   & 0  & --   & 0  & 1 &2 &0&--   &0 &45\\
M1-25	& GCSC09$^{\dagger}$  & 10 & 20 & 2 & 11 & --   & 0  & --   & 0  & 3 &11&3&--   &6 &58\\
        & GKA07$^{\dagger\ddagger}$   & 10 & 20 & 2 & 11 & 3.06 & 9  & 3.06 & 7  & 3 &11&3&--   &6 &74\\
        & GPMDR12$^{*}$ & 10 & 20 & 3 & 13 & 2.89 & 4  & 3.00 & 10 & 6 &20&3&0.03 &15&92\\
        & MKHC02  & 10 & 20 & 2 & 11 & 2.74 & 0  & 3.12 & 5  & 5 &17&3&-0.11&9 &72\\
\hline
\end{tabular}
\end{table*}

\begin{table*}
\centering
\contcaption{Partial and final scores for the sample spectra. The reference spectra for each object are marked with asterisks.}
\begin{tabular}{llcccrrrrrcrcrrr}
\hline
PN & Ref.& $p_0$ & $p_1$ & $n_{\rm{e}}$ & $p_2$ & [\ion{N}{ii}] & $p_3$ & [\ion{O}{iii}] & $p_4$ & \ion{He}{i} & $p_5$ & $\alpha\gamma\delta$ & P$-$B & $p_6$ & $p_{\rm{tot}}$ \\
\hline
M1-25	& PSM01$^{\dagger}$   & 10 & 20 & 2 & 11 & --   & 0  & --   & 0  & 1 &2 &0&--   &0 &43\\
M 1-29  & EBW04   & 10 & 20 & 2 & 11 & 3.11 & 6  & 2.75 & 0   &3&11&3&--   &6 &64\\
        & GCSC09$^{\dagger\ddagger}$  & 10 & 20 & 2 & 11 & --   & 0  & --   & 0   &3&11&3&--   &6 &58\\
        & WL07$^{*}$    & 10 & 20 & 3 & 13 & 3.20 & 2  & 3.03 & 8   &4&14&3&--   &6 &73\\
M 1-32  & GKA07   & 10 & 20 & 2 & 11 & 3.11 & 6  & 2.98 & 9   &3&11&3&--   &6 &73\\
        & GPMDR12$^{*}$ & 10 & 20 & 3 & 13 & 3.07 & 8  & 3.08 & 6   &6&20&3&0.14 &3 &80\\
        & PSM01   & 10 & 20 & 2 & 11 & --   & 0  & --   & 0   &1&2 &0&--   &0 &43\\
M 1-74  & AC83	  & 0  & 20 & 2 & 11 & --   & 0  & 2.87 & 4   &4&14&3&--   &6 &55\\
        & KH01$^{*}$    & 10 & 20 & 1 & 9  & 3.02 & 10 & 3.04 & 8   &4&13&3&--   &6 &76\\
        & WLB05   & 10 & 20 & 3 & 13 & 2.61 & 0  & 3.00 & 10  &4&13&3&--   &0 &66\\
M 2-23  & EBW04   & 10 & 20 & 2 & 11 & 2.58 & 0  & 3.10 & 6   &4&13&3&--   &6 &66\\
        & GCSC09  & 10 & 20 & 2 & 11 & --   & 0  & --   & 0   &3&11&3&--   &6 &58\\
        & WL07$^{*}$    & 10 & 20 & 3 & 13 & 2.78 & 0  & 3.12 & 5   &6&20&3&--   &6 &74\\
M 3-15  & MKHC02  & 10 & 20 & 2 & 11 & 2.03 & 0  & 3.13 & 4   &4&14&3&--   &0 &59\\
        & GKA07   & 10 & 20 & 2 & 11 & 2.88 & 3  & 2.92 & 6   &3&11&3&--   &0 &61\\
        & GPMDR12$^{*}$ & 10 & 20 & 4 & 15 & 3.11 & 6  & 3.02 & 9   &6&20&3&0.21 &0 &80\\
        & PSM01   & 10 & 20 & 2 & 11 & --   & 0  & --   & 0   &1&2 &0&--   &0 &43\\
Me 2-2  & AK87	  & 0  & 20 & 2 & 11 & --   & 0  & 2.99 & 9   &6&20&3&--   &6 &66\\
        & B78     & 0  & 20 & 1 & 9  & --   & 0  & 3.00 & 10  &3&11&3&--   &6 &56\\
        & MKHS10$^{*}$  & 10 & 20 & 2 & 11 & 3.16 & 4  & 3.08 & 6	&6&19&3&--   &0 &70\\
        & WLB05   & 10 & 20 & 3 & 13 & 3.11 & 6  & 2.88 & 5	&5&13&3&--   &0 &67\\
NGC 40   & AC79   & 0  & 20 & 2 & 11 & 3.16 & 4  & 3.16 & 3   &5&12&3&--   &6 &56\\
         & CSPT83$^{\ddagger}$ & 0  & 20 & 2 & 11 & 3.16 & 4  & 3.09 & 6	&6&20&3&--   &6 &67\\
         & LLLB04$^{*}$ & 10 & 20 & 3 & 13 & 3.08 & 8  & --   & 0	&6&20&3&--   &6 &77\\
         & PSM01$^{\dagger}$  & 10 & 20 & 2 & 11 & --   & 0  & --   & 0	&1&2 &0&--   &0 &43\\
NGC 650  & AC83   & 0  & 20 & 1 & 9  & --   & 0  & 3.13 & 4   &4&12&3&--   &6 &51\\
         & KHM03$^{*}$  & 10 & 20 & 1 & 9  & 3.06 & 9  & 3.26 & 0	&4&14&2&--   &6 &68\\
         & PT87   & 0  & 20 & 1 & 9  & 3.02 & 10 & 2.95 & 8	&3&11&2&--   &6 &64\\
NGC 2392 & B91	  & 0  & 20 & 2 & 11 & --   & 0  & 3.09 & 6   &3&11&3&--   &6 &54\\
         & DRMV09$^{\dagger\ddagger}$ & 10 & 20 & 2 & 11 & 2.87 & 3  & 2.95 & 8   &3&10&3&--   &6 &68\\
         & HKB00$^{\ddagger}$  & 10 & 20 & 1 & 9  & 2.95 & 6  & 3.20 & 1   &5&17&3&-0.15&0 &63\\
         & Z76    & 0  & 20 & 2 & 11 & 3.24 & 1  & 3.09 & 6   &1&2 &3&--   &0 &40\\
         & ZFCH12$^{*}$ & 10 & 20 & 2 & 11 & 3.09 & 7  & 3.02 & 9   &6&14&3&-0.19&0 &71\\
NGC 2440 & DKSH15 & 10 & 20 & 1 & 9  & 3.05 & 9  & 2.98 & 9   &3&10&3&--   &0 &67\\
         & HA98   & 10 & 20 & 4 & 15 & 3.65 & 0  & 3.08 & 6   &6&12&3&-0.14&9 &72\\
         & KB94   & 0  & 20 & 1 & 9  & 2.95 & 6  & --   & 0   &3&9 &3&--   &0 &44\\
         & KC06$^{*}$   & 10 & 20 & 2 & 11 & 3.04 & 10 & 2.91 & 6   &6&17&3&--   &6 &80\\
         & KHM03  & 10 & 20 & 2 & 11 & 3.02 & 10 & 3.26 & 0   &3&10&2&--   &6 &67\\
         & PT87$^{\dagger}$   & 0  & 20 & 2 & 11 & 2.95 & 7  & 2.95 & 8   &3&10&2&--   &6 &62\\
         & SAKC81 & 0  & 20 & 2 & 11 & 3.09 & 7  & 2.97 & 8   &6&15&3&0.04 &15&76\\
NGC 2867 & AKRO81 & 0  & 20 & 2 & 11 & 3.88 & 0  & 3.03 & 9   &6&16&3&--   &6 &62\\
         & GKA07  & 10 & 20 & 2 & 11 & --   & 0  & 2.99 & 10  &3&9 &3&--   &6 &66\\
         & GPP09$^{*}$  & 10 & 20 & 4 & 15 & 2.86 & 2  & 3.11 & 5   &5&15&3&-0.03&15&82\\
         & KB94   & 0  & 20 & 2 & 11 & 2.94 & 6  & --   & 0   &4&14&2&--   &6 &57\\
         & MKHC02$^{\dagger}$ & 10 & 20 & 2 & 11 & 2.34 & 0  & 3.00 & 10  &4&13&3&-0.01&15&79\\
         & PSEK98 & 10 & 20 & 1 & 9  & --   & 0  & --   & 0   &1&2 &0&--   &0 &41\\
NGC 6210 & B78	  & 0  & 20 & 1 & 9  & --   & 0  & 2.97 & 9   &3&8 &3&--   &6 &52\\
         & BERD15$^{\dagger}$ & 10 & 20 & 4 & 15 & 3.31 & 0  & 3.07 & 7   &6&19&3&--   &6 &77\\
         & DRMV09 & 10 & 20 & 2 & 11 & 2.70 & 0  & 2.83 & 2   &4&14&3&--   &6 &63\\
         & F81    & 0  & 20 & 2 & 11 & 2.64 & 0  & 2.30 & 0   &3&10&3&--   &6 &47\\
         & KH98$^{\ddagger}$   & 10 & 20 & 2 & 11 & 3.49 & 0  & --   & 0   &4&14&2&--   &6 &61\\
         & LLLB04$^{*}$ & 10 & 20 & 4 & 15 & 3.04 & 9  & --   & 0   &6&20&3&--   &6 &80\\
NGC 6302 & KC06   & 10 & 20 & 2 & 11 & 3.08 & 8  & 2.99 & 10  &6&15&3&--   &6 &80\\
         & MKHS10 & 10 & 20 & 2 & 11 & 3.08 & 8  & 3.05 & 8   &3&10&3&--   &0 &67\\
         & RCK14  & 10 & 20 & 2 & 11 & 3.06 & 8  & 3.00 & 10  &6&15&3&0.24 &6 &80\\
         & TBLDS03& 10 & 20 & 4 & 15 & 3.00 & 9  & 3.01 & 10  &6&10&3&--   &6 &80\\
NGC 6369 & AK87   & 0  & 20 & 2 & 11 & --   & 0  & 2.92 & 6   &4&12&3&--   &6 &55\\
         & GKA07  & 10 & 20 & 1 & 9  & 2.97 & 7  & 3.01 & 10  &3&11&3&--   &0 &67\\
         & GPMDR12$^{*}$& 10 & 20 & 4 & 15 & 3.05 & 9  & 3.01 & 10  &6&20&3&0.12 &9 &93\\
         & MKHS10 & 10 & 20 & 2 & 11 & 3.07 & 8  & 2.97 & 9   &3&9 &3&--   &0 &67\\
         & PSM01  & 10 & 20 & 3 & 13 & --   & 0  & --   & 0   &1&2 &0&--   &0 &45\\
NGC 6543 & AC79   & 0  & 20 & 2 & 11 & --   & 0  & 2.95 & 8   &6&19&3&--   &6 &64\\
\hline
\end{tabular}
\end{table*}

\begin{table*}
\begin{minipage}{155mm}
\contcaption{Partial and final scores for the sample spectra. The reference spectra for each object are marked with asterisks.}
\begin{tabular}{llcccrrrrrcrcrrr}
\hline
PN & Ref.& $p_0$ & $p_1$ & $n_{\rm{e}}$ & $p_2$ & [\ion{N}{ii}] & $p_3$ & [\ion{O}{iii}] & $p_4$ & \ion{He}{i} & $p_5$ & $\alpha\gamma\delta$ & P$-$B & $p_6$ & $p_{\rm{tot}}$ \\
\hline
NGC 6543 & HAFLK00& 10 & 20 & 3 & 13 & 3.54 & 0  & 3.22 & 0   &6&20&3&0.05 &12&75\\
         & PSM01$^{\dagger}$  & 10 & 20 & 3 & 13 & --   & 0  & --   & 0   &1&2 &0&--   &0 &45\\
         & WL04$^{*}$   & 10 & 20 & 4 & 15 & 3.02 & 9  & --   & 0   &6& 20& 3&--   &6 &80\\
NGC 6565 & EBW04  & 10 & 20 & 2 & 11 & 2.97 & 8  & 2.41 & 0   &3& 10& 3&--   &6 &65\\
         & MKHC02$^{*}$ & 10 & 20 & 2 & 11 & 2.92 & 5  & 3.02 & 9   &4& 12& 3&-0.02&15&82\\
         & WL07   & 10 & 20 & 3 & 13 & 3.19 & 3  & 3.11 & 5   &6& 20& 3&--   &6 &77\\
NGC 6572 & F81	  & 0  & 20 & 2 & 11 & 2.69 & 0  & 2.26 & 0   &3& 10& 3&--   &6 &47\\
         & GKA07  & 10 & 20 & 2 & 11 & 3.07 & 8  & 2.97 & 9   &3& 11& 3&--   &6 &75\\
         & HAF94b$^{*}$ & 10 & 20 & 4 & 15 & 2.92 & 5  & 2.98 & 9   &6& 20& 3&0.01 &9 &88\\
         & KH01   & 10 & 20 & 2 & 11 & 2.99 & 8  & 2.94 & 7   &4& 13& 2&--   &6 &75\\
         & LLLB04 & 10 & 20 & 4 & 15 & 2.99 & 8  & --   & 0   &6& 20& 3&--   &6 &79\\
NGC 6720 & B80	  & 0  & 20 & 1 & 9  & --   & 0  & 3.21 & 1   &3& 11& 3&--   &6 &47\\
         & F81    & 0  & 20 & 2 & 11 & 2.64 & 0  & 2.76 & 0   &3& 9 & 3&--   &6 &46\\
         & GMC97$^{\dagger\ddagger}$  & 10 & 20 & 1 & 9  & 2.96 & 7  & 2.93 & 7   &6& 17& 3&--   &0 &70\\
         & HM77   & 0  & 20 & 1 & 9  & 3.43 & 0  & 3.18 & 2   &3& 11& 3&--   &6 &48\\
         & KH98$^{\dagger\ddagger}$   & 10 & 20 & 2 & 11 & 3.14 & 5  & 3.03 & 9   &4& 12& 2&--   &6 &73\\
         & LLLB04$^{*}$ & 10 & 20 & 4 & 15 & 3.07 & 8  & --   & 0   &6& 19& 3&--   &6 &78\\
NGC 6741 & HA97a$^{*}$  & 10 & 20 & 4 & 15 & 2.96 & 7  & 2.84 & 2   &6& 15& 3&-0.01&15&84\\
         & LLLB04$^{\dagger\ddagger}$ & 10 & 20 & 4 & 15 & 3.06 & 9  & --   & 0   &6& 17& 3&--   &6 &77\\
         & MKHS10 & 10 & 20 & 3 & 13 & 3.06 & 8  & 3.05 & 8   &4& 13& 3&--   &6 &78\\
NGC 6751 & AC79   & 0  & 20 & 2 & 11 & 2.95 & 7  & 2.95 & 8   &3& 11& 3&--   &6 &63\\
         & CMJK91 & 10 & 20 & 2 & 11 & 3.02 & 10 & 2.93 & 7   &3& 11& 3&--   &6 &75\\
         & KB94   & 0  & 20 & 1 & 9  & 3.43 & 0  & --   & 0   &4& 11& 3&--   &6 &46\\
         & MKHS10$^{*}$ & 10 & 20 & 2 & 11 & 3.09 & 7  & 3.03 & 9   &5& 17& 3&--   &6 &80\\
         & PSM01  & 10 & 20 & 2 & 11 & --   & 0  & --   & 0   &1& 2 & 0&--   &0 &43\\
NGC 6790 & AC79   & 0  & 20 & 2 & 11 & --   & 0  & 3.09 & 6   &6& 16& 3&--   &6 &59\\
         & AHF96  & 10 & 20 & 3 & 13 & 3.42 & 0  & 3.21 & 1   &6& 16& 3&0.08 &6 &66\\
         & F81    & 0  & 20 & 2 & 11 & 2.55 & 0  & 2.58 & 0   &3& 9 & 3&--   &6 &46\\
         & KH01$^{*}$   & 10 & 20 & 1 & 9  & 2.87 & 3  & 3.00 & 10  &5& 16& 3&--   &6 &74\\
         & LLLB04$^{\ddagger}$ & 10 & 20 & 4 & 15 & 2.98 & 8  & --   & 0   &6& 16& 3&--   &0 &69\\
NGC 6884 & AC79   & 0  & 20 & 3 & 13 & --   & 0  & 3.09 & 6   &6& 20& 3&--   &6 &65\\
         & HAF97  & 10 & 20 & 3 & 13 & 3.28 & 0  & 3.06 & 7   &6& 17& 3&-0.23&6 &73\\
         & KH01$^{*}$   & 10 & 20 & 2 & 11 & 2.97 & 7  & 3.07 & 7   &4& 14& 3&--   &6 &75\\
         & LLLB04$^{\ddagger}$ & 10 & 20 & 4 & 15 & 3.02 & 10 & --   & 0   &6& 17& 3&--   &0 &72\\
NGC 7009 & CA79   & 0  & 20 & 4 & 15 & 2.95 & 6  & 3.11 & 5   &6& 19& 3&0.03 &15&80\\
         & F81    & 0  & 20 & 2 & 11 & --   & 0  & 2.58 & 0   &3& 10& 3&--   &6 &47\\
         & FL11   & 10 & 20 & 4 & 15 & 2.78 & 0  & 2.99 & 9   &6& 20& 3&-0.10&9 &83\\
         & HA95a$^{\dagger}$  & 10 & 20 & 3 & 13 & 3.49 & 0  & 2.78 & 0   &6& 17& 3&-0.01&15&75\\
         & HA95b  & 10 & 20 & 4 & 15 & 3.47 & 0  & 3.17 & 2   &6& 19& 3&0.00 &15&81\\
         & KC06$^{*}$   & 10 & 20 & 2 & 11 & 3.03 & 10 & 3.00 & 10  &6& 18& 3&--   &6 &85\\
         & KH98   & 10 & 20 & 2 & 11 & 3.93 & 0  & 3.10 & 5   &5& 16& 2&--   &6 &68\\
NGC 7662 & AC83   & 0  & 20 & 3 & 13 & --   & 0  & 2.97 & 9   &5& 17& 3&--   &6 &65\\
         & HA97b$^{*}$  & 10 & 20 & 4 & 15 & 3.79 & 0  & 3.32 & 0   &6& 15& 3&-0.08&12&72\\
         & LLLB04 & 10 & 20 & 4 & 15 & 2.70 & 0  & --   & 0   &6& 18& 3&--   &6 &69\\
PB 6     & GKA07$^{\dagger}$  & 10 & 20 & 1 & 9  & 3.08 & 8  & 2.98 & 9   &1& 2 & 3&--   &6 &64\\
         & GPP09$^{*}$  & 10 & 20 & 3 & 13 & 2.85 & 2  & 2.91 & 6   &4& 11& 3&-0.06&12&74\\
         & MKHS10 & 10 & 20 & 2 & 11 & 3.09 & 7  & 3.02 & 9   &3& 9 & 3&--   &6 &72\\
         & PSEK98 & 10 & 20 & 1 & 9  & --   & 0  & --   & 0   &1& 2 & 0&--   &0 &41\\
PC 14    & GKA07$^{\dagger\ddagger}$  & 10 & 20 & 2 & 11 & 2.98 & 8  & 3.05 & 8   &3& 11& 3&--   &6 &74\\
         & GPMDR12$^{*}$& 10 & 20 & 4 & 15 & 3.02 & 10 & 3.03 & 9   &6& 20& 3&0.02 &15&99\\
         & MKHC02$^{\dagger\ddagger}$ & 10 & 20 & 2 & 11 & 2.50 & 0  & 3.04 & 8   &4& 13& 3&-0.09&12&74\\
Pe 1-1   & BK13   & 10 & 20 & 2 & 11 & 3.14 & 5  & 3.03 & 8   &3& 10& 3&--   &0 &64\\
         & G14    & 10 & 20 & 2 & 11 & 3.14 & 5  & 2.97 & 9   &3& 11& 3&--   &6 &72\\
         & GKA07$^{\dagger}$  & 10 & 20 & 2 & 11 & 3.23 & 1  & 3.03 & 9   &3& 11& 3&--   &6 &68\\
         & GPMDR12$^{*}$& 10 & 20 & 4 & 15 & 2.95 & 7  & 3.00 & 10  &6& 20& 3&0.06 &12&94\\
\hline
\end{tabular}
\\
$^\dagger$ O/H, N/H, and S/H are within 0.1~dex of the reference spectrum for calculations based on the blue [\ion{O}{ii}] lines.\\
$^\ddagger$ O/H, N/H, and S/H are within 0.1~dex of the reference spectrum for calculations based on the red [\ion{O}{ii}] lines.\\
References for the line intensities: AC79, \citet{AC79}; AC83, \citet{AC83}; AHF96, \citet{AHF96}; AK87, \citet{AK87}; AKRO81, \citet{AKRO81}; B78, \citet{B78}; B80, \citet{B80}; B91, \citet{B91}; BERD15, \citet{BERD15}; BK13, \citet{BK13}; BC84, \citet{BC84}; CA79, \citet{CA79}; CMJK91, \citet{CMJK91}; CPT87, \citet{CPT87}; CSPT83, \citet{CSPT83}; DASNA17, \citet{DASNA17}; DKSH15, \citet{DKSH15}; DRMV09, \citet{DRMV09}; EBW04, \citet{EBW04}; F81, \citet{F81}; FGMR15, \citet{FGMR15}; FL11, \citet{FL11}; G14, \citet{G14};
\end{minipage}
\end{table*}

\begin{table*}
\begin{minipage}{180mm}
\contcaption{Partial and final scores for the sample spectra.}
\begin{tabular}{llrrrrrrrr}
\hline
\end{tabular}
\hspace{-0.5em}GCSC09, \citet{GCSC09}; GDGD18, \citet{GDGD18}; GKA07, \citet{GKA07}; GMC97, \citet{GMC97}; GPMDR12, \citet{GPMDR12}; GPP09, \citet{GPP09}; HA95a, \citet{HA95a}; HA95b, \citet{HA95b}; HA97a, \citet{HA97a}; HA97b, \citet{HA97b}; HA98, \citet{HA98}; HAF94a, \citet{HAF94a}; HAF94b, \citet{HAF94b}; HAF97, \citet{HAF97}; HAFL01, \citet{HAFL01}; HAFLK00, \citet{HAFLK00}; HAL01, \citet{HAL01}; HKB00, \citet{HKB00}; HM77, \citet{HM77}; HPF04, \citet{HPF04}; KB94, \citet{KB94}; KC06, \citet{KC06}; KH98, \citet{KH98}; KHM03, \citet{KHM03}; KZGP08, \citet{KZGP08}; LLLB04, \citet{LLLB04}; KH01, \citet{KH01}; MKHC02, \citet{MKHC02}; MKHS10, \citet{MKHS10}; OHLIT09, \citet{OHLIT09}; OT13, \citet{OT13}; OTHI10, \citet{OTHI10}; PSEK98, \citet{PSEK98}; PSM01, \citet{PSM01}; PT87, \citet{PT87}; RCK14, \citet{RCK14}; SAKC81, \citet{SAKC81}; SCT87, \citet{SCT87}; SWBvH03, \citet{SWBvH03}; TBLDS03, \citet{TBLDS03}; TP79, \citet{TP79}; WL04, \citet{WL04}; WL07, \citet{WL07}; WLB05, \citet{WLB05}; Z76, \citet{Z76}; ZFCH12, \citet{ZFCH12}.\\
\end{minipage}
\end{table*}


\bsp	
\label{lastpage}
\end{document}